\documentclass[aps,prd,10pt,notitlepage,nofootinbib,superscriptaddress,showkeys,showpacs]{revtex4-2}
\pdfoutput=1
\usepackage{amstext,amsmath,amssymb,amsfonts,bbm,euscript,color,dsfont}
\usepackage[latin1]{inputenc}
\usepackage{epsfig}
\usepackage{hyperref}
\usepackage{amsthm}
\usepackage{subfigure}
\usepackage{color}
\usepackage{multirow}
\usepackage{bbold}
\usepackage{tikz}
\usetikzlibrary{arrows,backgrounds}
\usepackage{verbatim} 
\usepackage{graphicx}
\usepackage{latexsym}

\makeatother

\begin{document}

\title{An all order renormalizable Refined-Gribov-Zwanziger
model with BRST invariant fermionic horizon function in linear covariant gauges}

\author{M.~A.~L.~Capri}\email{caprimarcio@gmail.com}
\affiliation{Universidade do Estado do Rio de Janeiro (UERJ),
Instituto de F\'{i}sica, Departamento de F\'{i}sica Te\'{o}rica,
Rua S\~ao Francisco Xavier 524, Maracan\~{a}, Rio de Janeiro, Brasil, CEP 20550-013}

\author{S.~P.~Sorella} \email{silvio.sorella@gmail.com}
\affiliation{Universidade do Estado do Rio de Janeiro (UERJ),
Instituto de F\'{i}sica, Departamento de F\'{i}sica Te\'{o}rica,
Rua S\~ao Francisco Xavier 524, Maracan\~{a}, Rio de Janeiro, Brasil, CEP 20550-013}

\author{R.~C.~Terin} \email{rodrigorct@ita.br}
\affiliation{Instituto Tecnol\'{o}gico de Aeron\'{a}utica (ITA), DCTA, Departamento de F\'{i}sica, Pra\c{c}a Marechal Eduardo Gomes, 50 - Vila das Ac\'{a}cias, S\~{a}o Jos\'{e} dos Campos, S\~{a}o Paulo, Brasil, CEP 12228-900}

\begin{abstract}
We introduce, within the  Refined-Gribov-Zwanziger setup,  a composite BRST invariant fermionic operator coupled to the inverse of the Faddeev-Popov operator. As a result, an effective BRST invariant action in Euclidean space-time is constructed, enabling us to pave the first step towards the study of the behaviour of the fermion propagator in the infrared region in the class of the linear  covariant gauges. The aforementioned action is proven to be renormalizable to all orders by means of the algebraic renormalization procedure.  
\end{abstract}

\maketitle
\section{Introduction}
\label{RGZLCGreview} 

Despite the non-trivial progress done in the last decades, see \cite{Vandersickel:2012tz} for a general overview, a satisfactory solution of the  Gribov problem \cite{Gribov:1977wm} is still lacking. As it is well known, the standard Faddeev-Popov gauge fixing quantization procedure of non-abelian asymptotically free gauge theories yields remarkable results in the deep ultraviolet region. Though, its naive extension to the non-perturbative infrared region fails, due to the existence of the Landau pole. Such a non-perturbative region is  deeply affected by the existence of the so called Gribov copies \cite{Gribov:1977wm},
{\it i.e.} by equivalent field configurations: configurations which are related by a gauge transformation while obeying the same gauge-fixing condition. The  whole issue arises
from the observation \cite{Gribov:1977wm} that a local covariant gauge-fixing is unable to account in a  complete manner for the gauge freedom. 
It was soon realized that the Gribov problem  is not a particular problem of some
specific gauge-fixing, but an intrinsic problem related to the nontrivial
geometrical structure of the space of the gauge orbits of non-abelian gauge theories \cite{Singer:1978dk}. \\\\The Gribov problem has been faced  under distinct viewpoints as, for example,  the Refined-Gribov-Zwanziger (RGZ) approach \cite{Zwanziger:1989mf,Zwanziger:1990tn,Dudal:2008sp,Dudal:2007cw,Dudal:2011gd,Dudal:2019ing}\footnote{See \cite{Vandersickel:2011zc} for a pedagogical introduction.} which has been object of intensive investigations in last years,  including the important result of the existence of an exact   Becchi-Rouet-Stora-Tyutin (BRST) invariance, see \cite{Capri:2015ixa,Capri:2016gut,Capri:2017bfd,Capri:2017abz}, which has allowed to extend the Gribov-Zwanziger original construction  \cite{Gribov:1977wm,Zwanziger:1989mf,Zwanziger:1990tn} from the Landau to the linear covariant gauges. In the RGZ approach, the domain of integration in the functional integral is restricted to the so-called first Gribov region \cite{Gribov:1977wm} defined as the set of all gauge field configurations obeying the Landau gauge fixing condition, $\partial_\mu A^a_\mu=0$  and for which the Faddeev-Popov operator, ${\cal M}^{ab}(A) = -\partial_\mu D^{ab}_\mu(A)$ is strictly positive, {\it i.e} ${\cal M}^{ab}(A)>0$. For other recent alternative approaches to the Gribov problem we quote  \cite{Serreau:2012cg,Serreau:2015yna,Reinosa:2020skx}, where the gauge-fixing procedure is supplemented by an averaging over all Gribov regions by introducing a Boltzmann weight aiming at favouring field configurations close to the Fundamental Modular Region (FMR), a region which is contained within the Gribov region and which is known to be free from Gribov copies. \\\\It is worth mentioning here that both the Refined-Gribov-Zwanziger approach as well as other approaches based on the study of the Schwinger-Dyson equations \cite{Alkofer:2000wg,Binosi:2009qm,Brambilla:2014jmp,Huber:2018ned}, Renormalization Group Equations \cite{Pawlowski:2005xe}  and effective massive gluon models \cite{Tissier:2010ts,Tissier:2011ey,Pelaez:2013cpa,Siringo:2014lva,Siringo:2015gia,Frasca:2007uz}, have already provided  good results for the two-point gluon correlation functions in the infrared region, which are in good agreement with the  lattice numerical simulations see, for example, 
\cite{Cucchieri:2007md,Cucchieri:2007rg,Dudal:2018cli,Dudal:2019gvn,Binosi:2019ecz} and refs therein. \\\\In the present work, we shall  focus  on the RGZ model and its exact BRST invariance in the class of the linear covariant gauges (LCG). As mentioned before, an exact BRST symmetry   \cite{Capri:2015ixa,Capri:2016gut,Capri:2017bfd,Capri:2017abz} of the RGZ action has been constructed out of the non-local,
transverse and gauge-invariant composite field $(A_\mu^h=A_{\mu}^{h,a}T^a)$\footnote{The matrices $\{T^a, a=1,.., N^2-1\}$ denote the Hermitean generators of the gauge group $SU(N)$, $[T^a,T^b]=if^{abc}T^c $.}, introduced in \cite{Zwanziger:1990tn,Lavelle:1995ty}, namely: 
\begin{equation}
A_{\mu}^{h}=\left(\delta_{\mu\nu}-\frac{\partial_{\mu}\partial_{\nu}}{\partial^{2}}\right)\left(A_{\nu}-ig\left[\frac{1}{\partial^{2}}\partial A,A_{\nu}\right]+\frac{ig}{2}\left[\frac{1}{\partial^{2}}\partial A,\partial_{\nu}\frac{1}{\partial^{2}}\partial A\right]+\mathcal{O}(A^{3})\right)\,.\label{intro13-1}
\end{equation}
As it is easily checked, expression \eqref{intro13-1} is left invariant, order by order in powers of the coupling constant $g$,  by the BRST transformation, {\it  i.e.} 
\begin{equation}
sA_{\mu}^{h,a}=0\,,\qquad sA_{\mu}^{a}=-D_{\mu}^{ab}(A)c^{b}\,,\label{intro14-1}
\end{equation}
where $s$ is the nilpotent BRST operator, $(D^{ab}_{\mu}=\delta^{ab}\partial_{\mu}-gf^{abc}A^{c}_{\mu})$ is the covariant derivative in the adjoint representation of the $SU(N)$ gauge group and $c^{a}(x)$ is the  Faddeev-Popov ghost field. Let us also remind, for further use, the expression of the gauge-fixed Yang-Mills action in the linear covariant gauges parametrized by the gauge parameter $\alpha$:  
\begin{equation}
S_{\mathrm{FP}}=\int d^{4}x\,\bigg[\frac{1}{4}F_{\mu\nu}^{a}F_{\mu\nu}^{a}+ib^{a}\partial_{\mu}A_{\mu}^{a}+\frac{\alpha}{2}b^{a}b^{a}+\bar{c}^{a}\partial_{\mu}D_{\mu}^{ab}c^{b}\bigg]=\int d^{4}x\,\bigg[\frac{1}{4}F_{\mu\nu}^{a}F_{\mu\nu}^{a}+s\left(\bar{c}^{a}\partial_{\mu}A_{\mu}^{a}-\frac{i\alpha}{2}\bar{c}^{a}b^{a}\right)\bigg]\,,\label{rgzlcg1}
\end{equation}
where $b^{a}$ is the auxiliary Nakanishi-Lautrup field, while $\bar{c}^{a}$ is the Faddeev-Popov antighost. The field strength $F_{\mu\nu}^{a}$ is given by
\begin{equation}
F_{\mu\nu}^{a}=\partial_{\mu}A_{\nu}^{a}-\partial_{\nu}A_{\mu}^{a}+gf^{abc}A_{\mu}^{b}A_{\nu}^{c}\,.\label{intro6-1}
\end{equation}
Following \cite{Capri:2015ixa,Capri:2016gut,Capri:2017bfd,Capri:2017abz}, for the BRST invariant partition function in the linear covariant gauges implementing the restriction to the Gribov region, we have 
\begin{equation}
\mathcal{Z}_{LCG}=\int\left[\mathcal{D}\Phi\right]\mathrm{e}^{-(S_{\mathrm{FP}}+\gamma^{4}H(A^{h})-4V\gamma^{4}(N^{2}-1))}\,,\label{intro7-1}
\end{equation}
where $\left[\mathcal{D}\Phi\right]$ stands for integration over all fields $(A^a_\mu, b^a, {\bar c}^a,c^a)$ and 
\begin{equation}
H(A^{h})=g^{2}\int d^{4}xd^{4}y~f^{abc}A_{\mu}^{h,b}(x)\left[\mathcal{M}^{-1}(A^{h})\right]^{ad}(x,y)f^{dec}A_{\mu}^{h,e}(y)\,,\label{intro8-1}
\end{equation}
is the Gribov-Zwanziger horizon function \cite{Gribov:1977wm,Zwanziger:1989mf,Zwanziger:1990tn}, with $[\mathcal{M}^{ab}(A^{h})]^{-1}=[-\partial_{\mu}D_{\mu}^{ab}(A^{h})]^{-1}$, $V$
 the space-time volume and $N$ the number of colors. The parameter
$\gamma$, known as the Gribov parameter,   has mass dimension one. It is 
not a free parameter, being determined by the Gribov-Zwanziger \cite{Gribov:1977wm,Zwanziger:1989mf,Zwanziger:1990tn}  gap equation,
\begin{equation}
\langle H(A^{h})\rangle=4V(N^{2}-1)\,,\label{intro9-1}
\end{equation}
where the expectation value is taken with respect to the measure \eqref{intro7-1}. \\\\As it is apparent from \eqref{intro8-1}, the horizon function is a
nonlocal expression. Though, the whole partition function \eqref{intro7-1} can be cast in local form following the two steps outlined in \cite{Capri:2015ixa,Capri:2016gut,Capri:2017bfd,Capri:2017abz}. First, we introduce a pair of bosonic fields $(\bar{\varphi},\varphi)_{\mu}^{ab}$
as well as a pair of anticommuting ones $(\bar{\omega},\omega)_{\mu}^{ab}$, known 
 as the Zwanziger fields. These fields enable us to localize the horizon function, yielding the so-called Gribov-Zwanziger action 
\begin{eqnarray}
S_{\mathrm{GZ}} & = & S_{FP}-\int d^{4}x\left(\bar{\varphi}_{\mu}^{ac}\mathcal{M}^{ab}(A^{h}){\varphi}_{\mu}^{bc}-\bar{\omega}_{\mu}^{ac}\mathcal{M}^{ab}(A^{h})\omega_{\mu}^{bc}\right)-\gamma^{2}\int d^{4}x~gf^{abc}(A^{h})_{\mu}^{a}(\varphi+\bar{\varphi})_{\mu}^{bc}\,.
\end{eqnarray}
Further, we perform a second localization procedure in order to cast in local form the non-local quantity $A^h_\mu$. As shown in \cite{Capri:2015ixa,Capri:2016gut,Capri:2017bfd,Capri:2017abz}, this is done by introducing an auxiliary localizing Stueckelberg field  \cite{Lavelle:1995ty,Delbourgo:1986wz,Delbourgo:1987np,Dragon:1996tk,Ruegg:2003ps} $\xi^a$ by means of 
\begin{eqnarray}
h & =e^{ig\xi} & =e^{ig\xi^{a}T^{a}}\,,\label{hxi-1} \\ 
A_{\mu}^{h} & \equiv A_{\mu}^{h,a}\,T^{a} & =h^{\dagger}A_{\mu}h+\frac{i}{g}h^{\dagger}\partial_{\mu}h\,.\label{local_Ah-1}
\end{eqnarray}
Under a gauge transformation with group element $U$, one has
\begin{equation}
    A_{\mu}\rightarrow A_{\mu}^{U}=U^{\dagger}A_{\mu}U + \frac{i}{g}U^{\dagger}\partial_{\mu}U\,,\quad h\rightarrow h^{U}= U^{\dagger}h\,,\quad h^{\dagger}\rightarrow (h^{U})^{\dagger}= h^{\dagger}U\,,
    \label{A_and_h_transf}
\end{equation}
from which the gauge invariance of $A^h_\mu$ follows immediately: 
\begin{equation}
    A_{\mu}^{h}\,\rightarrow (A_{\mu}^{U})^{h}= A_{\mu}^{h}\,.
\end{equation}
Expanding expression \eqref{local_Ah-1} in powers of $\xi^a$, we have a non-polynomial infinite series whose first terms read 
\begin{eqnarray}
(A^{h})_{\mu}^{a} & = & A_{\mu}^{a}-\partial_{\mu}\xi^{a}+gf^{abc}A_{\mu}^{b}\xi^{c}-\frac{g
^{2}}{2}f^{abc}\xi^{b}\partial_{\mu}\xi^{c}+{\rm higher\;orders}\,.
\label{gfexpansion}
\end{eqnarray}
Solving now the transversality condition, $\partial_\mu A^{h}_\mu=0$, for the auxiliary Stueckelberg field $\xi^a$, we get back the non-local expression \eqref{intro13-1}. It remains now to write down the starting action in its local form. This is done by introducing a new Lagrange multiplier $\tau^{a}$ which implements the transversality condition, $\partial_\mu A^{h}_\mu=0$, as well as the corresponding Jacobian which, as much as the Faddeev-Popov determinant, can be rewritten in a local exponential way  by means of a  new pair of ghosts $(\eta,\bar{\eta})$. Finally \cite{Capri:2015ixa,Capri:2016gut,Capri:2017bfd,Capri:2017abz}, for the 
 the localized  BRST-invariant form of the Gribov-Zwanziger action in LCG we obtain 
\begin{eqnarray}
S_{\mathrm{GZ}} & = & S_{FP}-\int d^{4}x\left(\bar{\varphi}_{\mu}^{ac}\mathcal{M}^{ab}(A^{h}){\varphi}_{\mu}^{bc}-\bar{\omega}_{\mu}^{ac}\mathcal{M}^{ab}(A^{h})\omega_{\mu}^{bc}\right)-\gamma^{2}\int d^{4}x~gf^{abc}(A^{h})_{\mu}^{a}(\varphi+\bar{\varphi})_{\mu}^{bc}\nonumber \\
 & + & \int d^{4}x~\tau^{a}\partial_{\mu}(A^{h})_{\mu}^{a}-\int d^{4}x~\bar{\eta}^{a}\mathcal{M}^{ab}(A^{h})\eta^{b}\,.\label{rgzlcg2-1}
\end{eqnarray}
The final step in now that of moving from the Gribov-Zwanziger action to its Refined version, a task easily achieved by adding the BRST invariant dimension two operators  $( A_{\mu}^{h,a}A_{\mu}^{h,a})$
and $(\bar{\varphi}_{\mu}^{ab}\varphi_{\mu}^{ab}-\bar{\omega}_{\mu}^{ab}\omega_{\mu}^{ab})$. Therefore, for the local BRST invariant Refined-Gribov-Zwanziger action in the LCG gauge one gets
\begin{equation}
S_{\mathrm{RGZ}}=S_{\mathrm{GZ}}+\frac{m^{2}}{2}\int d^{4}x~A_{\mu}^{h,a}A_{\mu}^{h,a}-M^{2}\int d^{4}x\left(\bar{\varphi}_{\mu}^{ab}\varphi_{\mu}^{ab}-\bar{\omega}_{\mu}^{ab}\omega_{\mu}^{ab}\right)\,.\label{intro11-1}
\end{equation}
where,  analogously to the Gribov parameter $\gamma$, the two new mass parameters $(m,M)$ are determined by their own gap equations, see \cite{Dudal:2011gd,Dudal:2019ing}, from which it turns out that the Refined-Gribov-Zwanziger setup is energetically favoured with respect to its Gribov-Zwanziger version, eq.\eqref{rgzlcg2-1}. \\\\Both Gribov-Zwanziger and Refined-Gribov-Zwanziger actions enjoy the following exact nilpotent BRST symmetry defined as 
\begin{align}
sA_{\mu}^{a} & =-D_{\mu}^{ab}c^{b}\,, &  & sc^{a}=\frac{g}{2}f^{abc}c^{b}c^{c}\,,\nonumber \\
s\bar{c}^{a} & =ib^{a}\,, &  & sb^{a}=0\,,\nonumber \\
s\varphi_{\mu}^{ab} & =0\,, &  & s\omega_{\mu}^{ab}=0\,,\nonumber \\
s\bar{\omega}_{\mu}^{ab} & =0\,, &  & s\bar{\varphi}_{\mu}^{ab}=0\,,\nonumber \\
sh^{ij} & =-igc^{a}(T^{a})^{ik}h^{kj}\;, &  & sA_{\mu}^{h,a}=0\,,\nonumber \\
s\tau^{a} & =0\,, &  & s\bar{\eta}^{a}=0\,,\nonumber \\
s\eta^{a} & =0\,, &  & s^{2}=0\,,\label{intro22-1-1}
\end{align}
\begin{equation} 
s S_{\mathrm{GZ}} = s S_{\mathrm{RGZ}} = 0 \;. \label{binv}
\end{equation} 
The BRST transformation  of the Stueckelberg field $\xi^{a}$ can be obtained iteratively by expanding in power series the BRST transformation of $h^{ij}$ in equations \eqref{intro22-1-1}, yielding 
\begin{equation}
s\xi^{a}=g^{ab}(\xi)c^{b}\,,
\label{s_xi}
\end{equation}
with $g^{ab}$ given by 
\begin{equation}
g^{ab}(\xi)=-\delta^{ab}+\frac{g}{2}f^{abc}\xi^{c}-\frac{g^{2}}{12}f^{amr}f^{mbq}\xi^{q}\xi^{r}+O(\xi^{3})\,.\label{intro22a-1-1}
\end{equation}
As shown in \cite{Capri:2015ixa,Capri:2016gut,Capri:2017bfd,Capri:2017abz} the Refined-Gribov-Zwanziger action, eq.\eqref{intro11-1}, gives rise to a set of non-trivial results which we enlist below: 
\begin{itemize} 

\item despite its non-polynomial character, expression \eqref{intro11-1} turns out to be renormalizable to all orders, 

\item the BRST invariance ensures that the parameters $(\gamma,m,M)$ will be not affected by the gauge parameter $\alpha$. As such, $(\gamma,m,M)$ are physical quantities entering the expression of the correlation functions of gauge invariant operators, 

\item due to the Nielsen identities following from the BRST invariance, the pole-mass of the transverse component of the gluon propagator turns out to be independent from the gauge parameter $\alpha$ to all orders,  

\item the Nielsen identities also imply that the longitudinal component of the gluon propagator does not get any quantum correction to all orders, a property shared also by the studies of the LCG gauge within the Schwinger-Dyson framework 
\cite{Aguilar:2016ock,Napetschnig:2021ria} as well as by lattice numerical simulations \cite{Cucchieri:2011aa,Cucchieri:2018doy,Bicudo:2015rma},

\item  when specializing to the Landau gauge, $\alpha=0$, the transverse gluon propagator computed from the RGZ action \eqref{intro11-1} is in very good  agreement with the most recent lattice simulations, see for example the analysis  performed in  \cite{Dudal:2018cli}  where the agreement is shown to remain valid from the very deep infrared, $p\sim 0$, till the UV region,  $p\sim 10 GeV$,

\item  so far, the Refined-Gribov-Zwanziger setup has already been employed in a variety of physical applications as, for example: study of the spectrum of the glueballs \cite{ Dudal:2010cd,Dudal:2013wja}, $\rho$-meson mass estimate \cite{Dudal:2013vha}, study of the topological susceptibility \cite{Dudal:2017kxb}, pomeron physics \cite{Canfora:2017pfh}, thermodynamics of the Polyakov loop \cite{Canfora:2015yia}.

\end{itemize}
The aim of the present work is that of establishing the all orders renormalizability of the Refined-Gribov-Zwanziger action, eq.\eqref{intro11-1}, when the BRST invariant composite fermionic quark matter operator $\psi^h$
\begin{eqnarray}
\psi^{h} =  h^{\dagger}\psi\,,\label{psi}
\end{eqnarray}
and its corresponding  non-local  matter horizon function are included in the starting action. As we shall see, the addition of the aforementioned BRST invariant fermionic term will enable us to pave the first steps towards the non-perturbative study of the quark propagator in the class of the linear covariant gauges, a topic which is under current intensive investigations.\\\\ The paper is organized as follows: in Section \ref{Rev} we introduce the BRST-invariant composite fermionic field $\psi^h$. In Section \ref{RGZLCGMF}
we present the fermionic matter horizon function. Section \ref{ext} is devoted to the derivation of the Slavnov-Taylor identities, translating in functional form the BRST exact symmetry of the starting action.  In  Sections \ref{ExtBRST}, \ref{Wardid} we establish the so-called extended BRST symmetry as well as the whole set of Ward identities satisfied by our model.  In Section \ref{REn} 
we prove the all order renormalizability of the model by means of the algebraic renormalization procedure.  Finally, in Section \ref{conclu} we present our conclusions and future perspectives.

\section{The BRST-invariant composite fermionic field}
\label{Rev}
\hspace{0.5cm} Within the  BRST framework reviewed in the Introduction, we proceed by exhibiting the construction of a BRST-invariant local, non-polynomial, fermionic composite fields  $(\psi^{h}(x),\bar{\psi}^{h}(x))$  in the same way as we have constructed $A^{h}_{\mu}(x)$, see eqs.\,\eqref{hxi-1}, \eqref{local_Ah-1} and \eqref{gfexpansion}.
Let us start by introducing the following spinor quantity\footnote{According to the notations adopted here, the Greek indices $\{\mu,\nu,\rho,\sigma\}$ are the vector indices of the Euclidean space, while the Greek indices $\{\alpha,\beta,\gamma,\delta\}$ to spinor indices. The Latin indices $\{a,b,c,d,e\}$, running from $1$ to $N^{2}-1$, are the indices of the adjoint representation of $SU(N)$; while the Latin indices $\{i,j,k,l\}$, running from $1$ to $N$, represent the indices of the fundamental representation of $SU(N)$.  The adopted convention for the Dirac gamma matrices $\gamma_{\mu}$ in Euclidean space is given by
$$
\gamma_{4}=\left(
\begin{tabular}{cc}
    $0$ & $\openone$ \\
     $\openone$&$0$ 
\end{tabular}\right)\,,
\qquad
\gamma_{k}=-i\left(
\begin{tabular}{cc}
    $0$ & $\sigma_{k}$ \\
     $-\sigma_{k}$&$0$ 
\end{tabular}\right)\,,\qquad
\gamma_{5}=\gamma_{4}\gamma_{1}\gamma_{2}\gamma_{3}=\left(
\begin{tabular}{cc}
    $\openone$ & $0$ \\
     $0$&$-\openone$ 
\end{tabular}\right)\,,
$$
where $\openone$ is the $2\times2$ identity matrix, $k=1,2,3$ and $\sigma_{k}$ are the  Pauli matrices. }
\begin{eqnarray}
\psi_{\alpha}^{h,i} & \equiv & h^{\dagger}\psi_{\alpha}^{i}\,,\label{local_psih}
\end{eqnarray}
where $\psi_{\alpha}^{i}$ is the complex spinor field and $h$ is given by
\eqref{hxi-1}. As the  field $\psi$ transforms as $\psi \rightarrow  U^{\dagger}\psi$ and $h^{\dagger}$ as $h^{\dagger}\rightarrow h^{\dagger} U$,  for a finite gauge transformation $U$, it is immediate to realize that $\psi^{h}$ is gauge-invariant. Of course, the same procedure can be done for the field $\bar{\psi}$, giving rise to the invariant composite field $\bar{\psi}^{h}\equiv\bar{\psi}h$. Expanding $h$ in powers of the Stueckelberg field $\xi^a$, one gets the infinite power series
\begin{eqnarray}
\psi^{h,i}_{\alpha} & = & \psi_{\alpha}^{i}-ig\xi^{a}(T^{a})^{ij}\psi_{\alpha}^{j}-\frac{g^{2}}{2}\xi^{a}\xi^{b}(T^{a})^{ij}(T^{b})^{jk}\psi_{\alpha}^{k}+\mathcal{O}(\xi^{3})\,.\label{psih_expansion}
\end{eqnarray}
It is worth mentioning here that a study of the fermionic quantity $\psi^h$ as a composite field in Euclidean Yang-Mills theory  was already done in \cite{Capri:2019drm}, where its renormalizability was established to all orders in the loop expansion. As a recent applications of the BRST invariant operators 
$({\bar \psi}^h,\psi^h)$, we quote the refs. \cite{DeMeerleer:2018txc,DeMeerleer:2019kmh,DallOlio:2021njq} where the renormalizable  non-abelian Landau-Khalatnikov-Fradkin (LKF) transformations in the class of the linear covariant gauges where derived by direct use of  $(A^{h}_{\mu}, \psi^{h},\bar{\psi}^{h})$, from which it follows that the correlation functions 
$$\langle A^{h}_{\mu_{1}}(x_1)\dots A^{h}_{\mu_i}(x_{i})
\psi^{h}(y_1)\dots\bar{\psi}^{h}(y_j)\rangle$$ are independent from the gauge parameter $\alpha$, namely
\begin{equation}
\langle A^{h}_{\mu_{1}}(x_1)\dots A^{h}_{\mu_i}(x_{i})
\psi^{h}(y_1)\dots\bar{\psi}^{h}(y_j)\rangle_{\alpha\neq0}=
\langle A^{h}_{\mu_{1}}(x_1)\dots A^{h}_{\mu_i}(x_{i})
\psi^{h}(y_1)\dots\bar{\psi}^{h}(y_j)\rangle_{\alpha=0}
\,.
\label{sei_lah}
\end{equation}
Following \cite{DeMeerleer:2018txc,DeMeerleer:2019kmh,DallOlio:2021njq}, once expanded in powers of the Stueckelberg field $\xi^a$, eq.\,\eqref{sei_lah} enables one to evaluate the
Green function $$\langle A^{h}_{\mu_{1}}(x_1)\dots A^{h}_{\mu_i}(x_{i})
\psi^{h}(y_1)\dots\bar{\psi}^{h}(y_j)\rangle_{\alpha\neq0}$$ for a  given non-vanishing value of the gauge parameter $\alpha$ from the knowledge of the corresponding Green function evaluated in the Landau gauge, {\it i.e.} $\alpha = 0$, yielding
thus the desired LKF transformations. \\\\As mentioned in the Introduction the composite operators $({\bar \psi}^h,\psi^h)$ will be employed to built up an  effective BRST invariant horizon fermionic function which can allow to investigate the non-perturbative behaviour of the quark propagator in the linear covariant gauges, a topic on which we hope to report soon.

\section{The horizon function for the matter}
\label{RGZLCGMF}

As outlined  in \cite{Capri:2014fsa,Capri:2014bsa,Capri:2017abz}, in analogy with the Gribov-Zwanziger gauge field horizon function, eq.\eqref{intro8-1}, we have provided a rationale for the introduction of an effective  generalized horizon matter function, namely 
\begin{eqnarray}
H_\Psi(A^{h},\Psi)=-\int d^{4}xd^{4}y\,\,(T^{a})^{ij}\,\bar{\Psi}^{i}(x)\left[\mathcal{M}^{-1}(A^{h})\right]^{ab}(x,y)\,(T^{b})^{jk}\,\Psi^{k}(y)\,,
\label{H_geral}
\end{eqnarray}
where $\Psi$ can be either a spinor  field or a scalar field in the representation $T^a$. The introduction of such  matter horizon function can be understood by observing that, once the Gribov horizon is introduced into the theory via the Gribov-Zwanziger horizon function \eqref{intro8-1}, its presence is, somehow, transmitted to the other matter fields present in the theory. As one easily figure out  from expression \eqref{intro8-1}, the Gribov horizon is encoded in  the inverse of the  Fadeev-Popov operator, $\left[\mathcal{M}^{-1}(A^{h})\right]^{ab}$, which becomes more and more singular as we are approaching the boundary of the Gribov region, {\it i.e.} the first Gribov horizon. \\\\ Though, unlike the Gribov-Zwanziger horizon function, eq.\eqref{intro8-1}, for which a well established derivation based on the properties of the Gribov region has been provided \cite{Gribov:1977wm,Zwanziger:1989mf,Zwanziger:1990tn}, an analogous construction for the horizon matter function $H_\Psi(A^{h},\Psi)$, eq.\eqref{H_geral}, is not yet at our disposal. Nevertheless, it is worth pointing out that the requirements of BRST invariance and of renormalizability, see the following sections, seem to select in a unique way the expression \eqref{H_geral} as a non-perturbative matter term which can be introduced in order to investigate the behavior of the quark propagator in the infrared region. From that point of view, it is interesting to observe that expression \eqref{H_geral} yields a propagator, eq.\eqref{two-point-quark}, exhibiting a quark mass function $\mathcal{A}(p^{2})$, eq.\eqref{mass-function}, in agreement with the available numerical lattice simulations. As such, expression \eqref{H_geral} can be regarded as an useful effective non-perturbative matter term which enjoys the non-trivial properties of preserving both BRST invariance and all orders renormalizability, while providing a helpful quark mass function $\mathcal{A}(p^{2})$.  \\\\Willing thus investigating the  procedure of \cite{Capri:2014fsa,Capri:2014bsa,Capri:2017abz} in a BRST invariant environment for the fermionic field in the linear covariant gauges, we start by considering the following starting  action
\begin{eqnarray}
S & = & S_{RGZ}+S_{matter}+S_{\sigma}\,,
\label{35}
\end{eqnarray}
where $S_{RGZ}$ stands for the  the RGZ action in linear covariant gauges, eq.\eqref{intro11-1}, while $S_{matter}$ and $S_{\sigma}$ are given by
\begin{eqnarray}
S_{matter} & = & \int d^{4}x\,\left[i\bar{\psi}^{i\alpha}(\gamma_{\mu})_{\alpha\beta}D_{\mu}^{ij}\psi^{j\beta}-m_{\psi}\,\bar{\psi}^{i\alpha}\psi_{\alpha}^{i}\right]\,,
\label{act1}\\[2mm]
S_{\sigma} & = & -\sigma^{3}H_{\psi}(A^{h},\psi^{h})
=-\sigma^{3}\int d^{4}xd^{4}y\,\bar{\psi}^{h,i}_{\alpha}(x)T^{a,ij}\left[\mathcal{M}^{-1}(A^{h})\right]^{ab}(x,y)T^{b,jk}\psi^{h,k}_{\alpha}(y)\,.
\label{horizon}
\end{eqnarray}
The term \eqref{act1} is the usual fermionic matter term coupled to the  Yang-Mills field through the covariant derivative in the fundamental representation, $D_{\mu}^{ij} =  \delta^{ij}\partial_{\mu}-igT^{a,ij}\,A_{\mu}^{a}$. The parameter $m_\psi$ is the fermion mass.  The term \eqref{horizon} stands for the effective fermionic horizon term written in an explicitly BRST invariant fashion by means of the variables $({\bar \psi}^h, \psi^h, A^h_\mu)$. Here, the parameter $\sigma$ is an effective massive parameter which plays a role akin to that of the Gribov parameter $\gamma$.  Notice also that the horizon function \eqref{intro8-1} has mass dimension $-4$, therefore it has to be introduced in the action together with a power four mass parameter, {\it i.e.} $\gamma^{4}$.  On the other hand, the fermionic horizon function $H_{\psi}(A^{h},\psi^{h})$ has mass dimension $-3$, justifying thus the presence of the term $\sigma^3$. \\\\Being $S_{\sigma}$ a non-local expression, the first issue  to be faced is that of its localization. Proceeding in
a very similar way as in the case of the horizon function of the gauge-invariant  composite field $A^h_\mu$ reviewed in the introduction \eqref{RGZLCGreview}, for the local version of $S_{\sigma}$  we get 
\begin{eqnarray}
S_{\sigma}^{local} & = & \int d^{4}x\Bigg[\bar{\lambda}_{\alpha}^{ai}\left(-\partial_{\mu}D_{\mu}^{ab}(A^{h})\right)\lambda^{\alpha,bi}+\bar{\zeta}_{\alpha}^{ai}\left(-\partial_{\mu}D_{\mu}^{ab}(A^{h})\right)\zeta^{\alpha,bi}\nonumber \\
 & + & \sigma^{\frac{3}{2}}\left(\bar{\lambda}_{\alpha}^{ai}T^{a,ij}\psi^{h,j\alpha}+\bar{\psi}_{\alpha}^{h,i}T^{a,ij}\lambda^{aj\alpha}\right)\Bigg]\,,
 \label{localfermion}
\end{eqnarray}
where $(\lambda_{\alpha}^{ai},\bar{\lambda}_{\alpha}^{ai})$ are anti-commuting
localizing auxiliary spinor fields while  $(\zeta_{\alpha}^{ai},\bar{\zeta}_{\alpha}^{ai})$
are  commuting ones. It is easy to check out that  integration over
the auxiliary fields $(\lambda_{\alpha}^{ai},\bar{\lambda}_{\alpha}^{ai},\zeta_{\alpha}^{ai},\bar{\zeta}_{\alpha}^{ai})$
gives back the original nonlocal expression \eqref{horizon}.
The action \eqref{35} gives place to an equivalent local starting action:  
\begin{equation}
    S^{local}=S_{RGZ}+S_{matter}+S_{\sigma}^{local}\,,
    \label{S_local}
\end{equation}
where expression \eqref{horizon} has been replaced by \eqref{localfermion}. The new local action \eqref{S_local} is left invariant by the BRST transformations \eqref{intro22-1-1} together with the BRST transformations of $(\psi,\bar{\psi},\lambda,\bar{\lambda},\zeta, \bar{\zeta})$,
\begin{align}
s\psi_{\alpha}^{i} & =-i(T^{a})^{ij}c^{a}\psi_{\alpha}^{j}\,, & s\bar{\psi}_{\alpha}^{i} & =-i\bar{\psi}_{\alpha}^{j}(T^{a})^{ji}c^{a}\,,\nonumber\\
s\bar{\zeta}_{\alpha}^{ai} & =0\,, & s\lambda_{\alpha}^{ai} & =0\,,\nonumber\\
s\bar{\lambda}_{\alpha}^{ai} & =0\,, & s\zeta_{\alpha}^{ai} & =0\,,\nonumber\\
\label{s}
\end{align}
with 
\begin{equation} 
s S^{local} = 0 \;. \label{ssloc} 
\end{equation}

\section{Introduction of external sources}
\label{ext}
In order to prove the all order renormalizability of the action $S^{local}$ we follow the original procedure devised in \cite{Zwanziger:1989mf,Zwanziger:1990tn} and embed $S^{local}$ into a more general action by means of the introduction of a suitable set of external sources. For the benefit of the reader, let us give a very brief overview of how this is done. 

\subsection{Embedding the model into a more general one}

\hspace{0.5cm} Following \cite{Zwanziger:1989mf,Zwanziger:1990tn}, let us consider the RGZ action \eqref{intro11-1} and take a look at the term proportional to Gribov's parameter $\gamma^{2}$, namely
\begin{equation}
    S_{\gamma^{2}}=\int d^{4}x~\gamma^{2}f^{abc}\,A^{h,a}_{\mu}\,(\varphi+\bar{\varphi})_{\mu}^{bc}\,.
\end{equation}
As shown in \cite{Zwanziger:1989mf,Zwanziger:1990tn} this term might  be seen as a  particular case of a more general term depending on a set of external sources, {\it i.e.}
\begin{eqnarray}
S(M,N,U,V)&=&\int d^{4}x\bigg[M_{\mu\nu}^{ac}D_{\mu}^{ab}(A^{h})\varphi^{bc}_{\nu}+V_{\mu\nu}^{ac}D_{\mu}^{ab}(A^{h})\bar{\varphi}^{bc}_{\nu}
-N_{\mu\nu}^{ac}D_{\mu}^{ab}(A^{h})\omega^{bc}_{\nu}
\nonumber \\ &+& 
 U_{\mu\nu}^{ac}D_{\mu}^{ab}(A^{h})\bar{\omega}^{bc}_{\nu}-M_{\mu\nu}^{ab}V_{\mu\nu}^{ab}
 +N_{\mu\nu}^{ab}U_{\mu\nu}^{ab}\bigg]\,,
 \label{S_MNUV}
\end{eqnarray}
with $(M,V)$ being commuting sources and $(N,U)$ anti-commuting ones. In fact, when the sources are set to their physical values:   
\begin{eqnarray}
M_{\mu\nu}^{ab}\Big|_{phys} & = & V_{\mu\nu}^{ab}\Big|_{phys}=\gamma^{2}\delta^{ab}\delta_{\mu\nu}\,,\nonumber \\
N_{\mu\nu}^{ab}\Big|_{phys} & = & U_{\mu\nu}^{ab}\Big|_{phys}=0\,,
\label{phys_values_gamma}
\end{eqnarray}
we have
\begin{equation}
S(M,N,U,V)\Bigl|_{phys}= S_{\gamma^{2}}\,,    
\end{equation}
modulo a vacuum term, $4V(N^2-1)$, coming from the  product $MV$ in \eqref{S_MNUV}, which is allowed by power-counting. The introduction of this set of external sources allows us to write the following symmetry transformations 
\begin{eqnarray}
\delta \varphi^{ab}_{\mu} &=&\omega^{ab}_{\mu}\,,\qquad\delta\omega^{ab}_{\mu}\,=\,0\,,
\nonumber\\
\delta\bar{\omega}^{ab}_{\mu}&=&\bar{\varphi}^{ab}_{\mu}\,,\qquad\delta\bar{\varphi}^{ab}_{\mu}\,=\,0\,,
\nonumber\\
\delta N^{ab}_{\mu\nu}&=&M^{ab}_{\mu\nu}\,,\qquad \delta M^{ab}_{\mu\nu}\,=\,0\,,
\nonumber\\
\delta V^{ab}_{\mu\nu}&=& U^{ab}_{\mu\nu}\,,\qquad
\delta U^{ab}_{\mu\nu}\,=\,0\,.
\end{eqnarray}
As much as the BRST operator $s$, the operator $\delta$ is nilpotent too. It acts non-trivially only on the variables $(\varphi,\bar\varphi,\omega,\bar\omega,M,N,U,V)$.  Moreover, as we shall see later, the operators $s$ and $\delta$ can be  combined together in order to write down a helpful extended generalized nilpotent BRST operator. Let us notice that  the term \eqref{S_MNUV} can be rewritten as a total $\delta$-variation,
\begin{eqnarray}
S(M,N,U,V)&=&\delta\int d^{4}x~
\bigg[N_{\mu\nu}^{ac}D_{\mu}^{ab}(A^{h})\varphi^{bc}_{\nu}+V_{\mu\nu}^{ac}D_{\mu}^{ab}(A^{h})\bar{\omega}^{bc}_{\nu}-N_{\mu\nu}^{ab}V_{\mu\nu}^{ab}\bigg]\,,
\end{eqnarray}
as well as the RGZ dimension two operator
\begin{equation}
\bar{\varphi}^{ab}_{\mu}\varphi^{ab}_{\mu} - 
\bar{\omega}^{ab}_{\mu}\omega^{ab}_{\mu} = \delta\left( \bar{\omega}^{ab}_{\mu}\varphi^{ab}_{\mu}\right)\,.
\end{equation}
In fact, making use of $(M,N,U,V)$ the whole local version of the Gribov-Zwanziger horizon function can be  rewritten as a $\delta$-variation. As a consequence, once $\delta$ will be combined together with $s$ to give rise  to an extended generalized BRST operator, the horizon function as well as the RGZ operator $(\bar{\varphi}^{ab}_{\mu}\varphi^{ab}_{\mu} - 
\bar{\omega}^{ab}_{\mu}\omega^{ab}_{\mu})$ will turn out to belong to the trivial sector of the cohomology of such an extended BRST operator. With respect to the $s$ operator, the sources $(M,N,U,V)$ transform as singlets, {\it i.e.} 
\begin{equation}
    sM^{ab}_{\mu\nu} = sN^{ab}_{\mu\nu} = sU^{ab}_{\mu\nu} = sV^{ab}_{\mu\nu} =0\,.
\end{equation}
As shown in \cite{Zwanziger:1989mf,Zwanziger:1990tn}, we can introduce an useful  $U(4(N^{2}-1))$ symmetry for $S(M,N,U,V)$, given by 
\begin{eqnarray}
Q^{ab}_{\mu\nu}(S(M,N,U,V))=0\,,
\end{eqnarray}
with
\begin{eqnarray}
Q^{ab}_{\mu\nu}&=&\int d^{4}x~\bigg(
\varphi^{ca}_{\mu}\frac{\delta}{\delta\varphi^{cb}_{\nu}}
-\bar{\varphi}^{cb}_{\nu}\frac{\delta}{\delta\bar{\varphi}^{ca}_{\mu}}
+\omega^{ca}_{\mu}\frac{\delta}{\delta\omega^{cb}_{\nu}}
-\bar{\omega}^{cb}_{\nu}\frac{\delta}{\delta\bar{\omega}^{ca}_{\mu}}
\nonumber\\
&+&
V^{ca}_{\sigma\mu}\frac{\delta}{\delta V^{cb}_{\sigma\nu}}
-M^{cb}_{\sigma\nu}\frac{\delta}{\delta{M}^{ca}_{\sigma\mu}}
+U^{ca}_{\sigma\mu}\frac{\delta}{\delta U^{cb}_{\sigma\nu}}
-N^{cb}_{\sigma\nu}\frac{\delta}{\delta{N}^{ca}_{\sigma\mu}}\bigg)\,.
\label{Op_Q}
\end{eqnarray}
The trace of this operator can be used to define a quantum charge for the  variables $(\varphi,\bar\varphi,\omega,\bar\omega,M,N,U,V)$. Also, this symmetry enables us to introduce the so-called multi-index notation, in which a pair of indices is combined into a single index: 
\begin{equation}
(\varphi^{ab}_{\nu},\bar\varphi^{ab}_{\nu},\omega^{ab}_{\nu},\bar\omega^{ab}_{\nu},M^{ab}_{\mu\nu},N^{ab}_{\mu\nu},U^{ab}_{\mu\nu},V^{ab}_{\mu\nu})\equiv
(\varphi^{aI},\bar\varphi^{aI},\omega^{aI},\bar\omega^{aI},M^{aI}_{\mu},N^{aI}_{\mu},U^{aI}_{\mu},V^{aI}_{\mu})\,.
\end{equation}
where  $I\equiv\{b,\nu\}$ is a combination of the color and Lorentz indices  $b$ and $\nu$. We can make use  thus of  a new set of indices $\{I,J,K,L,\dots\}$  which  run from $1$ to $4\times(N^{2}-1)$. In terms of the multi-index notation, the term \eqref{S_MNUV} takes the form
\begin{eqnarray}
S(M,N,U,V)&=&\delta\int d^{4}x~
\bigg[N_{\mu}^{aI}D_{\mu}^{ab}(A^{h})\varphi^{bI}+V_{\mu}^{aI}D_{\mu}^{ab}(A^{h})\bar{\omega}^{bI}-N_{\mu}^{aI}V_{\mu}^{aI}\bigg]\nonumber\\
&=&
\int d^{4}x\bigg[M_{\mu}^{aI}D_{\mu}^{ab}(A^{h})\varphi^{bI}+V_{\mu}^{aI}D_{\mu}^{ab}(A^{h})\bar{\varphi}^{bI}
-N_{\mu}^{aI}D_{\mu}^{ab}(A^{h})\omega^{bI}
\nonumber \\ &+& 
 U_{\mu}^{aI}D_{\mu}^{ab}(A^{h})\bar{\omega}^{bI}-M_{\mu}^{aI}V_{\mu}^{aI}
 +N_{\mu}^{aI}U_{\mu}^{aI}\bigg]\,.
\end{eqnarray}
As expected, a similar procedure  can be repeated for the local version of the horizon function for the fermionic matter term, which can be embedded into a more general action. For such a purpose,  we shall employ the set of external sources $(\Pi,\bar{\Pi},\Lambda,\bar{\Lambda})_{\alpha\beta}^{ij}$, with $(\Lambda,\bar{\Lambda})$ being commuting variables and $(\Pi,\bar\Pi)$ anti-commuting ones, and write the following expression:
\begin{eqnarray}
S(\Pi,\bar{\Pi},\Lambda,\bar{\Lambda})&=&\int d^{4}x\,\bigg[\bar{\Lambda}_{\alpha\beta}^{jk}\bar{\psi}^{h,i\alpha}T^{a,ij}\lambda^{ak\beta}
+\bar{\Pi}_{\alpha\beta}^{jk}\bar{\psi}^{h,i\alpha}T^{a,ij}\zeta^{ak\beta}
\nonumber\\
&+&
\Pi^{ik\beta}_{\alpha}\bar{\zeta}_{\beta}^{ak}T^{a,ij}\psi^{h,j\alpha}
+\Lambda^{ik\beta}_{\alpha}\bar{\lambda}_{\beta}^{ak}T^{a,ij}\psi^{h,j\alpha}\bigg]\,.
\label{S_lambda_pi}
\end{eqnarray}
When the sources attain their physical values
\begin{eqnarray}
\Lambda_{\alpha\beta}^{ij}\Big|_{phys} & = & \bar{\Lambda}_{\alpha\beta}^{ij}\Big|_{phys}=\sigma^{\frac{3}{2}}\delta^{ij}\delta_{\alpha\beta}\,,\nonumber \\
\Pi_{\alpha\beta}^{ij}\Big|_{phys} & = & \bar{\Pi}{}_{\alpha\beta}^{ij}\Big|_{phys}=0\,,\label{phys_values_sigma}
\end{eqnarray}
the term $S(\Pi,\bar{\Pi},\Lambda,\bar{\Lambda})$ yields 
\begin{equation}
    S(\Pi,\bar{\Pi},\Lambda,\bar{\Lambda})\Big|_{phys} = \int d^{4}x~\sigma^{\frac{3}{2}}\left(\bar{\lambda}_{\alpha}^{ai}T^{a,ij}\psi^{h,j\alpha}+\bar{\psi}_{\alpha}^{h,i}T^{a,ij}\lambda^{aj\alpha}\right)\,,
\end{equation}
which is exactly the term proportional to $\sigma^{\frac{3}{2}}$ in eq.\eqref{localfermion}. The sources $(\Pi,\bar{\Pi},\Lambda,\bar{\Lambda})_{\alpha\beta}^{ij}$ are BRST singlets,
\begin{equation}
    s\Lambda^{ij}_{\alpha\beta}=s\bar{\Lambda}^{ij}_{\alpha\beta}=s\Pi^{ij}_{\alpha\beta}
    =s\bar{\Pi}^{ij}_{\alpha\beta}=0\,.
\end{equation}
from which it immediately follows  that $sS(\Pi,\bar{\Pi},\Lambda,\bar{\Lambda})=0$. On the other hand, as done in the case of $(M,N,U,V)$, a second nilpotent operator $\hat{\delta}$, $\hat{\delta}^2=0$, acting  only on $(\lambda,\bar\lambda,\zeta,\bar\zeta,\Pi,\bar{\Pi},\Lambda,\bar{\Lambda})$ can be introduced as:
\begin{eqnarray}
\hat{\delta}\lambda^{ai}_{\alpha}&=&\zeta^{ai}_{\alpha}\,,\qquad \hat{\delta}\zeta^{ai}_{\alpha}\,=\,0\,,\nonumber\\
\hat{\delta}\bar{\zeta}^{ai}_{\alpha}&=&\bar{\lambda}^{ai}_{\alpha}\,,\qquad \hat{\delta}\bar{\lambda}^{ai}_{\alpha}\,=\,0\,,\nonumber\\
\hat{\delta}\Lambda^{ij}_{\alpha\beta}&=&\Pi^{ij}_{\alpha\beta}\,,\qquad \hat{\delta}\Pi^{ij}_{\alpha\beta}\,=\,0\,,\nonumber\\
\hat{\delta}\bar{\Pi}^{ij}_{\alpha\beta}&=&\bar{\Lambda}^{ij}_{\alpha\beta}\,,\qquad \hat{\delta}\bar{\Lambda}^{ij}_{\alpha\beta}\,=\,0\,.
\end{eqnarray}
Again, the term  $S(\Pi,\bar{\Pi},\Lambda,\bar{\Lambda})$ can be re-written as an exact $\hat{\delta}$-variation,
\begin{equation}
    S(\Pi,\bar{\Pi},\Lambda,\bar{\Lambda}) = \hat{\delta}\int d^{4}x~\bigg[\bar{\Pi}_{\alpha\beta}^{jk}\bar{\psi}^{h,i\alpha}T^{a,ij}\lambda^{ak\beta}
    +
    \Lambda^{ik\beta}_{\alpha}\bar{\zeta}_{\beta}^{ak}T^{a,ij}\psi^{h,j\alpha}\bigg]\,.
\end{equation}
We can also define a new $\hat{\delta}$-invariant dimension two operator
\begin{equation}
    \bar{\lambda}^{ai}_{\alpha}\lambda^{ai\alpha}
    +
    \bar{\zeta}^{ai}_{\alpha}\zeta^{ai\alpha}
    =\hat{\delta}\left(\bar{\zeta}^{ai}_{\alpha}\lambda^{ai\alpha}\right)
    \,,  \label{f2op}
\end{equation}
which will be considered later on. Furthermore, $S(\Pi,\bar{\Pi},\Lambda,\bar{\Lambda})$ displays  an exact  $U(4N)$ symmetry:   
\begin{equation}
    \hat{Q}^{ij}_{\alpha\beta}(S(\Pi,\bar{\Pi},\Lambda,\bar{\Lambda})) = 0\,,
\end{equation}
with
\begin{eqnarray}
\hat{Q}^{ij}_{\alpha\beta} &=& \int d^{4}x~\Bigg(
\lambda^{ai}_{\alpha}\,
\frac{\delta}{\delta \lambda^{aj\beta}}
-\bar{\lambda}^{aj}_{\beta}\,
\frac{\delta}{\delta\bar{\lambda}^{ai\alpha}}
+\zeta^{ai}_{\alpha}\,
\frac{\delta}{\delta \zeta^{aj\beta}}
-\bar{\zeta}^{aj}_{\beta}\,
\frac{\delta}{\delta\bar{\zeta}^{ai\alpha}}
\nonumber\\
&+&
\Lambda^{ki}_{\gamma\alpha}\,
\frac{\delta}{\delta \Lambda^{kj\beta}_{\gamma}}
-\bar{\Lambda}^{kj}_{\gamma\beta}\,
\frac{\delta}{\delta\bar{\Lambda}^{ki\alpha}_{\gamma}}
+
\Pi^{ki}_{\gamma\alpha}\,
\frac{\delta}{\delta \Pi^{kj\beta}_{\gamma}}
-\bar{\Pi}^{kj}_{\gamma\beta}\,
\frac{\delta}{\delta\bar{\Pi}^{ki\alpha}_{\gamma}}
\Bigg)\,.
\label{Op_Q_hat}
\end{eqnarray}
As in the case of \eqref{Op_Q}, the trace of $\hat{Q}^{ij}_{\alpha\beta}$ defines a charge for $(\lambda,\bar\lambda,\zeta,\bar\zeta,\Pi,\bar{\Pi},\Lambda,\bar{\Lambda})$ and a new multi-index $\hat{I}\equiv\{j,\beta\}$ can be established:
\begin{equation}
    (\lambda^{aj}_{\beta},\bar{\lambda}^{aj}_{\beta},\zeta^{aj}_{\beta},\bar{\zeta}^{aj}_{\beta},\Pi^{ij}_{\alpha\beta},\bar{\Pi}^{ij}_{\alpha\beta},\Lambda^{ij}_{\alpha\beta},\bar{\Lambda}^{ij}_{\alpha\beta})
    \equiv
    (\lambda^{a}_{\hat{I}},\bar{\lambda}^{a}_{\hat{I}},\zeta^{a}_{\hat{I}},\bar{\zeta}^{a}_{\hat{I}},\Pi^{i}_{\alpha\hat{I}},\bar{\Pi}^{i}_{\alpha\hat{I}},\Lambda^{i}_{\alpha\hat{I}},\bar{\Lambda}^{i}_{\alpha\hat{I}})\,.
\end{equation}
where the indices $\{\hat{I},\hat{J},\hat{K}, \dots\}$ vary from $1$ to $4\times N$. In terms of the new multi-indices, expression \eqref{S_lambda_pi} can be  re-written as
\begin{eqnarray}
S(\Pi,\bar{\Pi},\Lambda,\bar{\Lambda}) 
&=& 
\hat{\delta}\int d^{4}x~\bigg[\bar{\Pi}_{\alpha\hat{I}}^{j}\,\bar{\psi}^{h,i\alpha}\,T^{a,ij}\,\lambda^{a\hat{I}}
    +
    \Lambda^{i\hat{I}}_{\alpha}\,\bar{\zeta}_{\hat{I}}^{a}\,T^{a,ij}\,\psi^{h,j\alpha}\bigg]
\nonumber\\
&=&\int d^{4}x\,\bigg[\bar{\Lambda}_{\alpha\hat{I}}^{j}\,\bar{\psi}^{h,i\alpha}\,T^{a,ij}\,\lambda^{a\hat{I}}
+\bar{\Pi}_{\alpha\hat{I}}^{j}\,\bar{\psi}^{h,i\alpha}\,T^{a,ij}\,\zeta^{a\hat{I}}
\nonumber\\
&+&
\Pi^{i\hat{I}}_{\alpha}\,\bar{\zeta}_{\hat{I}}^{a}\,T^{a,ij}\,\psi^{h,j\alpha}
+\Lambda^{i\hat{I}}_{\alpha}\,\bar{\lambda}_{\hat{I}}^{a}\,T^{a,ij}\,\psi^{h,j\alpha}\bigg]\,.
\end{eqnarray}
Finally, we can replace the full action $S^{local}$, eq.\,\eqref{S_local}, by a more general one or, equivalently, we can also state that  $S^{local}$ is embedded in the following action
\begin{eqnarray}
S_{1}&=&
\int d^{4}x~\Bigg[\,\frac{1}{4}\,F^{a}_{\mu\nu}F^{a}_{\mu\nu}
+\frac{\alpha}{2}\,b^{a}b^{a}
+ib^{a}\,\partial_{\mu}A^{a}_{\mu}
+\bar{c}^{a}\,\partial_{\mu}D^{ab}_{\mu}(A)c^{b}
+i\bar{\psi}^{i\alpha}(\gamma_{\mu})_{\alpha\beta}D_{\mu}^{ij}\psi^{j\beta}\Bigg]\nonumber\\
&+&
\int d^{4}x~\Bigg[\tau^{a}\,\partial_{\mu}A_{\mu}^{h,a}+\bar{\eta}^{a}\,\partial_{\mu}D_{\mu}^{ab}(A^{h})\eta^{b}\Bigg]
\nonumber \\
 &+&
\int d^{4}x~\Bigg[ \bar{\varphi}^{aI}\,\partial_{\mu}D_{\mu}^{ab}(A^{h})\varphi^{bI}
 -\bar{\omega}^{aI}\,\partial_{\mu}D_{\mu}^{ab}(A^{h})\omega^{bI}
 + M_{\mu}^{aI}D_{\mu}^{ab}(A^{h})\varphi^{bI}
 \nonumber\\
 &+&
V_{\mu}^{aI}D_{\mu}^{ab}(A^{h})\bar{\varphi}^{bI}
-N_{\mu}^{aI}D_{\mu}^{ab}(A^{h})\omega^{bI}
+U_{\mu}^{aI}D_{\mu}^{ab}(A^{h})\bar{\omega}^{bI}-M_{\mu}^{aI}V_{\mu}^{aI}
 +N_{\mu}^{aI}U_{\mu}^{aI}\Bigg]
 \nonumber\\
 &+&
 \int d^{4}x~\Bigg[\bar{\lambda}_{\hat{I}}^{a}\left(-\partial_{\mu}D_{\mu}^{ab}(A^{h})\right)\lambda^{b\hat{I}}+\bar{\zeta}_{\hat{I}}^{a}\left(-\partial_{\mu}D_{\mu}^{ab}(A^{h})\right)\zeta^{b\hat{I}}\nonumber \\
 &+&
 \bar{\Lambda}_{\alpha\hat{I}}^{j}\,\bar{\psi}^{h,i\alpha}\,T^{a,ij}\,\lambda^{a\hat{I}}
+\bar{\Pi}_{\alpha\hat{I}}^{j}\,\bar{\psi}^{h,i\alpha}\,T^{a,ij}\,\zeta^{a\hat{I}}
+
\Pi^{i\hat{I}}_{\alpha}\,\bar{\zeta}_{\hat{I}}^{a}\,T^{a,ij}\,\psi^{h,j\alpha}
+\Lambda^{i\hat{I}}_{\alpha}\,\bar{\lambda}_{\hat{I}}^{a}\,T^{a,ij}\,\psi^{h,j\alpha}\Bigg]
\nonumber\\
&+&
\int d^{4}x~\Bigg[
\frac{m^{2}}{2}\,A^{h,a}_{\mu}A^{h,a}_{\mu}
-M^{2}\,(\bar{\varphi}^{aI}\varphi^{aI}
-\bar{\omega}^{aI}\omega^{aI})
-m_{\psi}\,\bar{\psi}^{i\alpha}\psi^{i}_{\alpha}
\nonumber\\
&+&
w^{2}\,(\bar{\lambda}^{a\hat{I}}\lambda^{a}_{\hat{I}}
+\bar{\zeta}^{a\hat{I}}\zeta^{a}_{\hat{I}})\Bigg]\,,
\label{S1}
\end{eqnarray}
where, in the last line,  we have also included the dimension two invariant fermionic operator of eq.\eqref{f2op} trough the new mass parameter $w^2$, this term being allowed by both power counting and symmetry content. \\\\It is quickly checked that, when the externals sources $(M,N,U,V)$ and $(\Pi,\bar{\Pi},\Lambda,\bar{\Lambda})$ attain their physical values, eqs.\eqref{phys_values_gamma},\eqref{phys_values_sigma}, expression \eqref{S1} yields back the local action $S^{local}$ of equation \eqref{S_local}, with the addition of the invariant and power counting allowed quantity $(\bar{\lambda}^{a\hat{I}}\lambda^{a}_{\hat{I}}
+\bar{\zeta}^{a\hat{I}}\zeta^{a}_{\hat{I}})$. As we have seen before, expression \eqref{S1} enjoys the following exact symmetries
\begin{equation}
sS_{1}=\delta S_{1}=\hat{\delta}S_{1}=Q_{IJ}(S_{1})=\hat{Q}_{\hat{I}\hat{J}}(S_{1})=0,    
\end{equation}
Evidently, being the action $S^{local}$ a particular case of $S_{1}$, the all order renormalizability of the latter will imply the renormalizability of $S^{local}$. Therefore, from now on, we shall focus on $S_{1}$.

\subsection{Establishing the Slavnov-Taylor identities}
Following the setup of the algebraic renormalization \cite{Piguet:1995er} a key step in the proof of the renormalizability is the establishment of the Slavnov-Taylor identities, translating at the functional quantum level the BRST invariance. As we have several non-linear transformations as well as several composite operators and symmetries, the task of writing down the Slavnov-Taylor identities for the present model, {\it i.e.} for the action $S_{1}$, requires a few steps which will illustrated in the following subsections. \\\\First of all, from the BRST transformations of eqs.\,\eqref{s_xi}, \eqref{intro22a-1-1}, \eqref{intro22-1-1} and \eqref{s} one observes  that the transformations of the fields $(A^{a}_{\mu}$, $c^{a}$, $\psi^{i}$, $\bar{\psi}^{i})$ and $\xi^{a}$ are nonlinear in the quantum fields, so that they define composite operators which need to be properly defined at the quantum level  by means of a suitable set of external sources \cite{Piguet:1995er}. This leads to the following action    
\begin{equation}
    S_2=S_1+S_{BRST}\,,
    \label{S2}
\end{equation}
with $S_1$ being given by \eqref{S1} and $S_{BRST}$ by
\begin{eqnarray}
S_{BRST}&=&\int d^{4}x~\bigg[\Omega^{a}_{\mu}(sA^{a}_{\mu})
+L^{a}(sc^{a})+K^{a}(s\xi^{a})+
(s\bar{\psi}^{i\alpha})\varUpsilon_{\alpha}^{i}+\bar{\varUpsilon}_{\alpha}^{i}(s\psi^{i\alpha})\bigg]\nonumber\\
&=&\int d^{4}x~\bigg[\Omega^{a}_{\mu}\,D^{ab}_{\mu}(A)c^{b}
+\frac{g}{2}f^{abc}L^{a}c^{b}c^{c}
+K^{a}g^{ab}(\xi)c^{b}
\nonumber\\
&-&
i\bar{\psi}^{i\alpha}\,T^{a,ij}\,c^{a}\,\varUpsilon^{j}_{\alpha}
-i\bar{\varUpsilon}^{i}_{\alpha}\,T^{a,ij}\,c^{a}\,\psi^{j\alpha}\bigg]\,,
\end{eqnarray}
where $(\Omega^{a}_{\mu},L^{a},K^{a},\varUpsilon^{i}_{\alpha},\bar{\varUpsilon}^{i}_{\alpha})$ are a set of external BRST invariant sources 
\begin{equation}
    s\Omega^{a}_{\mu}=sL^{a}=sK^{a}=s\varUpsilon^{i}_{\alpha}=\bar{\varUpsilon}^{i}_{\alpha}=0\,.
\end{equation}
The quantum numbers of these sources can be found in tables I -- \ref{table2-1}. The BRST invariance can now be written as a functional identity known as the  Slavnov-Taylor identity:
\begin{equation}
    \mathcal{S}(S_2)=0\,,
\end{equation}
where $\mathcal{S}(F)$ being defined by 
\begin{eqnarray}
\mathcal{S}(F)&=&\int d^{4}x~\bigg(
\frac{\delta F}{\delta \Omega^{a}_{\mu}}\frac{\delta F}{\delta A^{a}_{\mu}}
+\frac{\delta F}{\delta L^{a}}\frac{\delta F}{\delta c^{a}}
+\frac{\delta F}{\delta K^{a}}\frac{\delta F}{\delta \xi^{a}}
+ib^{a}\frac{\delta F}{\delta \xi^{a}}\bigg)\,,
\end{eqnarray}
with $F$ standing for an arbitrary integrated functional of the fields and sources.

\subsection{Introducing the external sources for the composite operators $(A^{h}_{\mu}, \psi^{h}, \bar{\psi}^{h})$}

\hspace{0.5cm} Besides the sources $(\Omega^{a}_{\mu},L^{a},K^{a},\varUpsilon^{i}_{\alpha},\bar{\varUpsilon}^{i}_{\alpha})$ introduced before, we need to take into account that the quantities $(A^{h}_{\mu}, \psi^{h},\bar{\psi}^{h})$ are composite operators too and, as such, they do require to be introduced through suitable external 
sources. We get thus 
\begin{eqnarray}
S_{3}=S_{2}+\int d^{4}x~\bigg(
\mathcal{J}^{a}_{\mu}\,A^{h,a}_{\mu}
+\bar{\psi}^{h,i}_{\alpha}\,\Theta^{i\alpha}
+\bar{\Theta}^{i}_{\alpha}\,\psi^{h,i\alpha}\bigg)\,,
\label{S3}
\end{eqnarray}
where $S_2$ is given by $\eqref{S2}$ while the  sources $(\mathcal{J}^{a}_{\mu},\Theta^{i}_{\alpha},\bar{\Theta}^{i}_{\alpha})$ are coupled  to  $(A^{h}_{\mu},\psi^{h},\bar{\psi}^{h})$. The gauge invariance  of $(A^{h}_{\mu},\psi^{h},\bar{\psi}^{h})$ naturally  leads to  
\begin{equation}
s\mathcal{J}^{a}_{\mu}=s\Theta^{i}_{\alpha}=s\bar{\Theta}^{i}_{\alpha}=0\,.   \end{equation}
Furthermore, the BRST invariant mass terms built up with $(A^{h}_{\mu},\psi^{h},\bar{\psi}^{h})$ and with the two families of localizing auxiliary fields $(\varphi,\bar{\varphi},\omega,\bar{\omega})$ and $(\lambda,\bar{\lambda},\zeta,\bar{\zeta})$ which appear in the last two lines  of expression  \eqref{S1} can be treated as composite operators coupled to the sources $(J,J_\psi,J_\varphi,J_\lambda)$, which replace the mass parameters  $(m^{2},m_{\psi},M^{2},w^{2})$, namely 
\begin{eqnarray}
S(J,J_\psi,J_\varphi,J_\lambda)
&=&\int d^{4}x~\bigg[J\,A^{h,a}_{\mu}A^{h,a}_{\mu}
+J_{\psi}\,\bar{\psi}^{i}_{\alpha}\psi^{i\alpha}
+J_{\varphi}\,(\bar{\varphi}^{aI}\varphi^{aI}- \bar{\omega}^{aI}\omega^{aI})\nonumber\\
&+&
J_{\lambda}\,(\bar{\lambda}^{a\hat{I}}\lambda^{a}_{\hat{I}}
+\bar{\zeta}^{a\hat{I}}\zeta^{a}_{\hat{I}})\bigg]\,.
\label{Js}
\end{eqnarray}
At the end, we shall set 
\begin{eqnarray}
    J(x)\Big|_{phys} &=& \frac{m^{2}}{2}\,,\qquad
    J_{\psi}(x)\Big|_{phys}=- m_{\psi}\,,\nonumber\\
    J_{\varphi}(x)\Big|_{phys}&=&-M^{2}\,,\qquad
    J_{\lambda}(x)\Big|_{phys}=w^{2}\,.
    \label{J_phys}
\end{eqnarray}
thus recovering the  mass terms of $S_{1}$. From the invariance of the mass terms it follows 
\begin{equation}
    sJ=sJ_{\psi}=sJ_{\varphi}=sJ_{\lambda}=0\,.
\end{equation}
So far, the action we have constructed is given by the expression  $S_3$,  
eq.\eqref{S3},   with the masses replaced by the local sources $(J,J_\psi,J_\varphi,J_\lambda)$, {\it i.e.} with the last two lines  of eq.\eqref{S1} replaced by the term \eqref{Js}. Of course, at the end, when the sources attain theirs physical values, eq.\eqref{J_phys}, the action $S_3$ is recovered. In order to clarify the notation, let us define a new action 
\begin{equation}
    S_4\equiv S_{3}\Big|_{masses\,\rightarrow\,sources}\,.
    \label{S4}
\end{equation}
Though, the action $S_{4}$ is not yet the final expression for the complete tree level action which will be taken as the starting point. Two more steps are needed, as illustrated below. As we have seen before, besides the BRST invariance, we have additional global symmetries, namely $(\delta,\hat{\delta})$, which  will turn out to be very helpful for the algebraic proof of the renormalizability when translated into Ward identities. To that aim we shall introduce  extra local composite operators encoded in the term $S_{extra}$
\begin{equation}
    S_5=S_4+S_{extra}\,,
\end{equation}
with $S_{extra}$ being given by
\begin{eqnarray}
S_{extra} & = & \int d^{4}x~\bigg[\,\Xi_{\mu}^{a}\,D_{\mu}^{ab}(A^{h})\eta^{b}+\Gamma^{ab}\eta^{a}\eta^{b}-X^{I}\,\eta^{a}\bar{\omega}^{aI}-Y^{I}\,\eta^{a}\bar{\varphi}^{aI}-\bar{X}^{abI}\,\eta^{a}\omega^{bI}\nonumber \\&-&\bar{Y}^{abI}\,\eta^{a}\varphi^{bI}
- Z^{\hat{I}}\,\eta^{a}\bar{\lambda}_{\hat{I}}^{a}-W^{\hat{I}}\,\eta^{a}\bar{\zeta}_{\hat{I}}^{a}-\bar{Z}^{ab\hat{I}}\,\eta^{a}\lambda_{\hat{I}}^{b}
-\bar{W}^{ab\hat{I}}\,\eta^{a}\zeta_{\hat{I}}^{b}
\nonumber\\
&+&\Phi_{\alpha}^{i}\,\bar{\psi}^{h,j\alpha}\,T^{a,ji}\,\eta^{a}+\bar{\Phi}^{i\alpha}\,T^{a,ij}\,\eta^{a}\,\psi_{\alpha}^{h,j}
+K_{\varphi}\,\bar{\omega}^{aI}\varphi^{aI}
+K_{\lambda}\,\bar{\lambda}^{a\hat{I}}\zeta^{a}_{\hat{I}}\,
\bigg]\,.
\end{eqnarray}
This extra term is left invariant by the three operators  $(s,\delta,\hat{\delta})$ provided the new  external local sources transform as
\begin{equation}
    s(\Xi,\Gamma,\Phi,\bar{\Phi},X,\bar{X},Y,\bar{Y},Z,\bar{Z},W,\bar{W},K_{\varphi},K_{\lambda})=0\,;
\end{equation}
\begin{eqnarray}
    &&\delta(\Xi,\Gamma,\Phi,\bar{\Phi},Z,\bar{Z},W,\bar{W},K_{\lambda})=0\,,\nonumber\\
    &&\delta Y^{I}=X^{I}\,,\qquad \delta X^{I}=0\,,\nonumber\\
    &&
    \delta\bar{X}^{abI}=-\bar{Y}^{abI}\,,\qquad
    \delta\bar{Y}^{abI}=0\,,\nonumber\\
    &&
    \delta J_{\varphi}=K_{\varphi}\,,\qquad \delta K_{\varphi}=0\,;
    \label{more_delta_transf}
\end{eqnarray}
\begin{eqnarray}
    &&\hat{\delta}(\Xi,\Gamma,,\Phi,\bar{\Phi}X,\bar{X},Y,\bar{Y},K_{\varphi})=0\,,\nonumber\\
    &&\hat{\delta} Z^{\hat{I}}=-W^{\hat{I}}\,,\qquad \hat{\delta} W^{\hat{I}}=0\,,\nonumber\\
    &&
    \hat{\delta}\bar{W}^{ab\hat{I}}=\bar{Z}^{ab\hat{I}}\,,\qquad
    \hat{\delta}\bar{Z}^{ab\hat{I}}=0\,,\nonumber\\
    &&
    \hat{\delta} J_{\lambda}=K_{\lambda}\,,\qquad \hat{\delta} K_{\lambda}=0\,,
    \label{more_delta_hat_transf}
\end{eqnarray}
where  we have extended the $\delta$ and $\hat{\delta}$ transformations to the sources $J_{\varphi}$ and $J_{\lambda}$, respectively.\\\\The final step is now that of introducing pure vacuum terms in  the external sources, allowed  by the power counting, namely 
\begin{equation}
    S_{6}=S_5+S_{PC}\,,
    \label{S6}
\end{equation}
with
\begin{eqnarray}
S_{PC}&=&\int d^{4}x~\biggl[\frac{\kappa_{1}}{2}\,J^{2}
+\kappa_{2}\,JJ^{2}_{\psi}+\kappa_{3}\,J^{4}_{\psi}+\varrho\,J_{\psi}(\bar{\Lambda}^{i\alpha\hat{I}}\Lambda^{i}_{\alpha\hat{I}}
-\bar{\Pi}^{i\alpha\hat{I}}\Pi^{i}_{\alpha\hat{I}})\bigg]\,,
\end{eqnarray}
where $(\kappa_1,\kappa_2,\kappa_3,\varrho)$ are arbitrary coefficients needed to take into account  the UV divergences present in the  correlation functions:
\begin{equation*}
    \left\langle (A^{h}A^{h})_x (A^{h}A^{h})_y\right\rangle\,,
\end{equation*}
\begin{equation*}
    \left\langle (\bar{\psi}\psi)_x
    (\bar{\psi}\psi)_y(A^{h}A^{h})_z\right\rangle\,,
\end{equation*}
\begin{equation*}
    \left\langle (\bar{\psi}\psi)_x
    (\bar{\psi}\psi)_y
    (\bar{\psi}\psi)_z
    (\bar{\psi}\psi)_w\right\rangle\,,
\end{equation*}
\begin{equation*}
    \left\langle (\bar{\psi}\psi)_x
    \left(
    (\bar{\psi}^{h}T\lambda)_y
    (\bar{\lambda}T\psi^{h})_z
    -(\bar{\psi}^{h}T\zeta)_y
    (\bar{\zeta}T\psi^{h})_z
    \right)
    \right\rangle\,.
\end{equation*}
Other possible combinations in the external sources  like $J_{\varphi}^{2}$, $J_{\lambda}^{2}$, $J_{\psi}^{2}J_{\varphi}$, $J_{\psi}^{2}J_{\lambda}$, $J J_{\varphi}, J J_{\lambda}$ and $J_{\varphi}J_{\lambda}$ 
are forbidden, since are not left invariant  by the  transformations \eqref{more_delta_transf}, \eqref{more_delta_hat_transf} of $J_{\varphi}$ and $J_{\lambda}$,  meaning in fact  that the correlation functions
\begin{equation*}
    \left\langle(\bar{\varphi}\varphi-\bar{\omega}\omega)_{x}
    (\bar{\varphi}\varphi-\bar{\omega}\omega)_{y}\right\rangle\,,
\end{equation*}
\begin{equation*}
    \left\langle(\bar{\lambda}\lambda+\bar{\zeta}\zeta)_{x}
    (\bar{\lambda}\lambda+\bar{\zeta}\zeta)_{y}\right\rangle\,,
\end{equation*}
\begin{equation*}
    \left\langle(\bar{\psi}\psi)_{x}
    (\bar{\psi}\psi)_{y}
    (\bar{\varphi}\varphi-\bar{\omega}\omega)_{z}\right\rangle\,,
\end{equation*}
\begin{equation*}
    \left\langle(\bar{\psi}\psi)_{x}
    (\bar{\psi}\psi)_{y}
    (\bar{\lambda}\lambda+\bar{\zeta}\zeta)_{z}\right\rangle\,,
\end{equation*}
\begin{equation*}
    \left\langle(A^{h}A^{h})_{x}
        (\bar{\varphi}\varphi-\bar{\omega}\omega)_{y}\right\rangle\,,
\end{equation*}
\begin{equation*}
    \left\langle(A^{h}A^{h})_{x}
    (\bar{\lambda}\lambda+\bar{\zeta}\zeta)_{y}\right\rangle\,,
\end{equation*}
\begin{equation*}
    \left\langle(\bar{\varphi}\varphi-\bar{\omega}\omega)_{x}
    (\bar{\lambda}\lambda+\bar{\zeta}\zeta)_{y}
    \right\rangle
\end{equation*}
are not  UV divergent.

\section{Introduction of the extended generalized  BRST operator and of the complete tree-level action $\Sigma$}
\label{ExtBRST}

\hspace{0.5cm} In this section, we shall  introduce a unique generalized BRST operator encoding all three operators $(s, \delta, {\hat \delta})$. First of all, following  \cite{Piguet:1984js}, let us extend the action of the operator $s$ on the gauge parameter $\alpha$, a very powerful trick to control the  dependence of the invariant local counterterm as well as of the Green functions of the theory from $\alpha$. To that end \cite{Piguet:1984js}, we define the action of $s$ on the parameter $\alpha$ as 
\begin{eqnarray}
s\alpha=\chi\,,\qquad s\chi=0\,,\label{ra15}
\end{eqnarray}
where $\chi$ is a Grassmann parameter with ghost number $1$ which will be set to zero at the very end. According to  \cite{Piguet:1984js}, the linear covariant gauge-fixing term acquires now a $\chi$ dependent part, namely 
\begin{equation}
    s \int d^4x \left( - \frac{i \alpha}{2} {\bar c}^a b^a \right) = \int d^4x \left( - \frac{i \chi}{2} {\bar c}^a b^a +  \frac{ \alpha}{2} b^a b^a\right) \;, \label{chit}
\end{equation}
which is taken into account by introducing it in the final complete tree level action $\Sigma$ given by 
\begin{eqnarray}
\Sigma = S_{6}-\frac{i}{2}\int d^{4}x~\chi\,\bar{c}^{a}b^{a}\,, \label{SS}
\end{eqnarray}
where $S_6$ stands for the expression of eq.\eqref{S6}. As it is easily checked, the action  $\Sigma$ above is left invariant by all three operators\footnote{The parameters $(\alpha, \chi)$ do not transform under $(\delta, {\hat \delta})$
\begin{equation} 
\delta \alpha = \delta \chi = {\hat \delta} \alpha = {\hat \delta} \chi = 0 \;.\label{notr}
\end{equation} 
} $(s, \delta, {\hat \delta})$: 
\begin{equation}
    s\Sigma=\delta\Sigma=\hat{\delta}\Sigma=0\,.
\end{equation}
Therefore, noticing that 
\begin{equation}
    s^{2}=0\,,\qquad
    \delta^{2}=0\,,\qquad
    \hat{\delta}^{2}=0\,,\qquad
    \{s,\delta\}=0\,,\qquad
    \{s,\hat{\delta}\}=0\,,\qquad
    \{\delta,\hat{\delta}\}=0\,,
\end{equation}
it turns out to be very helpful to join all these three operators $(s, \delta, {\hat \delta})$ into a unique generalized BRST nilpotent operator $Q$ defined by 
\begin{equation}
    Q =s+\delta+\hat{\delta}\;, \qquad Q^2=0 \;. \label{genQ}
\end{equation}
For further use, let us enlist below the whole set of $Q$-transformations of all fields, sources and parameters introduced so far: 
\begin{itemize}
\item{ The nonlinear $Q$-transformations:
\begin{align}
QA^{a}_{\mu}&=-D^{ab}_{\mu}(A)c^{b}\,, 
\nonumber\\ 
Qc^{a}&=\frac{g}{2}f^{abc}c^{b}c^{c}\,, 
\nonumber\\
Q\psi^{i}_{\alpha}&=-iT^{a,ij}c^{a}\psi^{j}_{\alpha}\,, 
\nonumber\\
Q\bar{\psi}^{i}_{\alpha}&=-i\bar{\psi}^{j}_{\alpha}\,T^{a,ji}c^{a}\,, 
\nonumber\\
Q\xi^{a}&=g^{ab}(\xi)c^{b}\,;
\label{Q_nonlinear}
\end{align}
}
\item{ The $Q$-doublet transformations:
\begin{align}
Q\bar{c}^{a}&=ib^{a}\,,&
Qb^{a}&=0\,, &
Q\alpha&=\chi\,, &
Q\chi&=0\,,\nonumber\\
Q\varphi^{aI}&=\omega^{aI}\,,& 
Q\omega^{aI}&=0\,, &
Q\bar{\omega}^{aI}&=\bar{\varphi}^{aI}\,, &
Q\bar{\varphi}^{aI}&=0\,,\nonumber\\
QN^{aI}_{\mu}&=M^{aI}\,, &
QM^{aI}_{\mu}&=0\,, &
QV^{aI}_{\mu}&=U^{aI}\,, &
QU^{aI}_{\mu}&=0\,,\nonumber\\
Q\lambda^{a\hat{I}}&=\zeta^{a\hat{I}}\,, &
Q\zeta^{a\hat{I}}&=0\,, &
Q\bar{\zeta}^{a\hat{I}}&=\bar{\lambda}^{a\hat{I}}\,, &
Q\bar{\lambda}^{a\hat{I}}&=0\,,
\nonumber\\
Q\Lambda^{i\hat{I}}_{\alpha}&=\Pi^{i\hat{I}}_{\alpha}\,, &
Q\Pi^{i\hat{I}}_{\alpha}&=0\,, &
Q\bar{\Pi}^{i\hat{I}}_{\alpha}&=\bar{\Lambda}^{i\hat{I}}_{\alpha}\,, &
Q\bar{\Lambda}^{i\hat{I}}_{\alpha}&=0\,,
\nonumber\\
QJ_{\varphi}&=K_{\varphi}\,,&
QK_{\varphi}&=0\,, &
QJ_{\lambda}&=K_{\lambda}\,, &
QK_{\lambda}&=0\,,
\nonumber\\
QY^{I}&=X^{I}\,, &
QX^{I}&=0\,, &
Q\bar{X}^{abI}&=-\bar{Y}^{abI}\,, &
Q\bar{Y}^{abI}&=0\,,
\nonumber\\
QZ^{\hat{I}}&=-W^{\hat{I}}\,, &
QW^{\hat{I}}&=0\,, &
Q\bar{W}^{ab\hat{I}}&=\bar{Z}^{ab\hat{I}}\,, &
Q\bar{Z}^{ab\hat{I}}&=0\,;
\label{Q_doublets}
\end{align}
}
\item{ The $Q$-singlet transformations:
\begin{equation}
Q\,\left(\,\eta^{a},\bar{\eta}^{a},\tau^{a},J,J_{\psi},\Omega^{a}_{\mu},L^{a}, K^{a}, \varUpsilon^{i}_{\alpha},
\bar{\varUpsilon}^{i}_{\alpha},
\mathcal{J}^{a}_{\mu},
\Theta^{i}_{\alpha},
\bar{\Theta}^{i}_{\alpha},
\Xi^{a}_{\mu},\Gamma^{ab},
\Phi^{i}_{\alpha},
\bar{\Phi}^{i}_{\alpha}\,\right)=0\,.
\label{Q_singlets}
\end{equation}
}
\end{itemize}
Let us end this section by presenting  the explicit expression of the complete tree level starting action  $\Sigma$ of eq.\eqref{SS}: 
\begin{eqnarray}
\Sigma &=&\int d^{4}x~\Bigg[\,\frac{1}{4}\,F^{a}_{\mu\nu}F^{a}_{\mu\nu}
+i\bar{\psi}^{i\alpha}(\gamma_{\mu})_{\alpha\beta}D_{\mu}^{ij}\psi^{j\beta}+J\,A^{h,a}_{\mu}A^{h,a}_{\mu}+J_{\psi}\,\bar{\psi}^{i\alpha}\psi^{i}_{\alpha} 
\nonumber\\
\phantom{\bigg|}&+&(\mathcal{J}^{a}_{\mu}-\partial_{\mu}\tau^{a})A^{h,a}_{\mu}
+(\Xi^{a}_{\mu}-\partial_{\mu}\bar{\eta}^{a})D^{ab}_{\mu}(A^{h})\eta^{b} 
+\bar{\psi}^{h,i}_{\alpha}\Theta^{i\alpha}
+\bar{\Theta}^{i}_{\alpha}\psi^{h,i\alpha}
\nonumber\\
\phantom{\bigg|}&+&\Gamma^{ab}\eta^{a}\eta^{b}
+\Phi^{i}_{\alpha}\bar{\psi}^{h,j\alpha}\,T^{a,ji}\eta^{a}
+\bar{\Phi}^{i}_{\alpha}\,T^{a,ij}\eta^{a}\psi^{h,j\alpha}
+\frac{\kappa_{1}}{2}\,J^{2}
+\kappa_{2}\,JJ^{2}_{\psi}
+\kappa_{3}\,J^{4}_{\psi}
\nonumber\\
\phantom{\bigg|}
&-&\frac{i}{2}\,\chi\,\bar{c}^{a}b^{a}
+\frac{\alpha}{2}\,b^{a}b^{a}
+ib^{a}\,\partial_{\mu}A^{a}_{\mu}
-(\Omega^{a}_{\mu}+\partial_{\mu}\bar{c}^{a})D^{ab}_{\mu}(A)c^{b}
+\frac{g}{2}f^{abc}L^{a}c^{b}c^{c}
\nonumber\\
\phantom{\bigg|}
&+&K^{a}\,g^{ab}(\xi)c^{b}
-i\bar{\psi}^{i\alpha}\,T^{a,ij}c^{a}\varUpsilon^{j}_{\alpha}
-i\bar{\varUpsilon}^{i}_{\alpha}\,T^{a,ij}c^{a}\psi^{j\alpha}
+\bar{\varphi}^{aI}\,\partial_{\mu}D^{ab}_{\mu}(A^{h})\varphi^{bI}
\nonumber\\
\phantom{\bigg|}
&-&
\bar{\omega}^{aI}\,\partial_{\mu}D^{ab}_{\mu}(A^{h})\omega^{bI}
+M^{aI}_{\mu}\,D^{ab}_{\mu}(A^{h})\varphi^{bI}
-N^{aI}_{\mu}\,D^{ab}_{\mu}(A^{h})\omega^{bI}
+U^{aI}_{\mu}\,D^{ab}_{\mu}(A^{h})\bar{\omega}^{bI}
\nonumber\\
\phantom{\bigg|}
&+&V^{aI}_{\mu}\,D^{ab}_{\mu}(A^{h})\bar{\varphi}^{bI}
+ K_{\varphi}\,\bar{\omega}^{aI}\varphi^{aI}
+J_{\varphi}(\bar{\varphi}^{aI}\varphi^{aI}
-\bar{\omega}^{aI}\omega^{aI})
-M^{aI}_{\mu}V^{aI}_{\mu}
+N^{aI}_{\mu}U^{aI}_{\mu}
\nonumber\\
\phantom{\bigg|}
&-&
\bar{\lambda}^{a\hat{I}}\,\partial_{\mu}D^{ab}_{\mu}(A^{h})\lambda^{b}_{\hat{I}}
-\bar{\zeta}^{a\hat{I}}\,\partial_{\mu}D^{ab}_{\mu}(A^{h})\zeta^{b}_{\hat{I}}
+\bar{\Lambda}^{j\hat{I}}_{\alpha}\,\bar{\psi}^{h,i\alpha}\,T^{a,ij}\lambda^{a}_{\hat{I}}
-\bar{\Pi}^{j\hat{I}}_{\alpha}\,\bar{\psi}^{h,i\alpha}\,T^{a,ij}\zeta^{a}_{\hat{I}}
\nonumber\\
\phantom{\bigg|}
&+&
\Pi^{i\hat{I}}_{\alpha}\bar{\zeta}^{a}_{\hat{I}}\,T^{a,ij}\psi^{h,j\alpha}
+\Lambda^{i\hat{I}}_{\alpha}\bar{\lambda}^{a}_{\hat{I}}\,T^{a,ij}\psi^{h,j\alpha}
+K_{\lambda}\,\bar{\zeta}^{a\hat{I}}\lambda^{a}_{\hat{I}}
+J_{\lambda}\,(\bar{\lambda}^{a\hat{I}}\lambda^{a}_{\hat{I}}+\bar{\zeta}^{a\hat{I}}\zeta^{a}_{\hat{I}})
\nonumber\\
\phantom{\bigg|}
&+&
\varrho\,J_{\psi}(\bar{\Lambda}^{i\alpha\hat{I}}\Lambda^{i}_{\alpha\hat{I}}
-\bar{\Pi}^{i\alpha\hat{I}}\Pi^{i}_{\alpha\hat{I}})
-X^{I}\eta^{a}\bar{\omega}^{aI}
-Y^{I}\eta^{a}\bar{\varphi}^{aI}
-\bar{Y}^{abI}\eta^{a}{\varphi}^{abI}
-\bar{X}^{abI}\eta^{a}{\omega}^{abI}
\nonumber\\
\phantom{\bigg|}
&-&
\bar{Z}^{ab\hat{I}}\eta^{a}\lambda^{b}_{\hat{I}}
-\bar{W}^{ab\hat{I}}\eta^{a}\zeta^{b}_{\hat{I}}
-{W}^{\hat{I}}\eta^{a}\bar{\zeta}^{a}_{\hat{I}}
-{Z}^{\hat{I}}\eta^{a}\bar{\lambda}^{a}_{\hat{I}}\Bigg]\,,
\label{Sigma_full}
\end{eqnarray}
with 
\begin{equation} 
Q \Sigma = 0 \;. \label{QinvSS}
\end{equation} 
The expression \eqref{Sigma_full} as well as the generalized identity \eqref{QinvSS} will be taken as the starting point for the forthcoming proof of the algebraic all orders renormalization of the action  $S^{local}$, eq.\eqref{S_local}. We remind in fact that $S^{local}$ is obtained from $\Sigma$, eq.\eqref{Sigma_full},  by taking the particular values
\eqref{phys_values_gamma},\eqref{phys_values_sigma},\eqref{J_phys}, while   setting the remaining sources and the Grassmannian parameter $\chi$ to zero. $S^{local}$ can be thought thus as a particular case of the more general action $\Sigma$ which enjoys a very rich set of Ward identities, as we shall see in the next section. \\\\It is helpful here to provide the mass dimensions and the other quantum numbers of all fields and sources appearing in the complete action $\Sigma$. These quantum numbers are displayed in the tables below (up to table \ref{table60}), where  the commuting (C) or anti-commuting (A) nature of each variable is also shown, being determined as the  sum of the ghost charges and of the so called $e$-charge ({\it i.e.} the spinor index). When this sum is even, the corresponding field/source is considered a commuting variable, otherwise it is an anti-commuting one. 

\begin{table}[!ht]
{\small
\flushleft
\begin{tabular}{|c|c|c|c|c|c|c|c|c|c|c|c|c|c|c|}
\hline\hline
Fields  & $A$  & $b$  & $c$  & $\bar{c}$  & $\phantom{\Big|}\!\bar{\psi}\phantom{\Big|}\!$  & $\phantom{\Big|}\!\psi\phantom{\Big|}\!$  & $\xi$  & $\bar{\varphi}$  & $\varphi$  & $\bar{\omega}$  & $\omega$  & $\alpha$  & $\chi$  & $\tau$ \tabularnewline
\hline 
\hline 
Dimension  & 1  & 2  & 0  & 2  & $\frac{3}{2}$  & $\frac{3}{2}$  & 0  & 1  & 1  & 1  & 1  & 0  & 0  & 2 \tabularnewline
\hline 
$c$-ghost number  & 0  & 0  & 1  & $-1$  & $0$  & $0$  & 0  & 0  & 0  & $-1$  & 1  & 0  & 1  & 0 \tabularnewline
\hline 
$\eta$-ghost number  & 0  & 0  & 0  & 0  & $0$  & $0$  & 0  & 0  & 0  & 0  & 0  & 0  & 0  & 0 \tabularnewline
\hline 
$e$-charge  & 0  & 0  & 0  & 0  & $-1$  & $1$  & 0  & $0$  & 0  & $0$  & 0  & 0  & 0  & 0 \tabularnewline
\hline 
$U(4(N^{2}-1))$-charge  & 0  & 0  & 0  & 0  & $0$  & $0$  & 0  & $-1$  & 1  & $-1$  & 1  & 0  & 0  & 0 \tabularnewline
\hline 
$U(4N)$-charge  & 0  & 0  & 0  & 0  & 0  & $0$  & $0$  & 0  & 0  & 0  & 0  & 0  & 0  & 0 \tabularnewline
\hline 
Nature  & C  & C  & A  & A  & A  & A  & C  & C  & C  & A  & A  & C  & A  & C \tabularnewline
\hline 
\end{tabular}
\label{table10}
}
\caption{Quantum numbers of the fields.}
\end{table}

\begin{table}[!ht]
{\small
\flushleft
\begin{tabular}{|c|c|c|c|c|c|c|}
\hline\hline 
Fields  & $\eta$  & $\bar{\eta}$  & $\bar{\lambda}$  & $\lambda$  & $\phantom{\Big|}\!\bar{\zeta}\phantom{\Big|}\!$  & $\phantom{\Big|}\!\zeta\phantom{\Big|}\!$\tabularnewline
\hline 
\hline 
Dimension  & 0  & 2  & $1$  & $1$  & $1$  & $1$\tabularnewline
\hline 
$c$-ghost number  & 0  & 0  & $0$  & $0$  & $-1$  & $1$\tabularnewline
\hline 
$\eta$-ghost number  & 1  & $-1$  & $0$  & $0$  & $0$  & $0$\tabularnewline
\hline 
$e$-charge  & 0  & 0  & $-1$  & $1$  & $-1$  & $1$\tabularnewline
\hline 
$U(4(N^{2}-1))$-charge  & 0  & 0  & $0$  & $0$  & $0$  & $0$\tabularnewline
\hline 
$U(4N)$-charge  & $0$  & $0$  & $-1$  & 1  & $-1$  & 1 \tabularnewline
\hline 
Nature  & A  & A  & A  & A  & C  & C\tabularnewline
\hline 
\end{tabular}
\label{table50} 
}
\caption{Quantum numbers of the fields.}
\end{table}

\begin{table}[!ht]
{\small
\flushleft
\begin{tabular}{|c|c|c|c|c|c|c|c|c|c|c|c|c|c|c|c|}
\hline\hline  
Sources  & $\Omega$  & $L$  & $K$  & $J$ & $J_{\psi}$ & $\mathcal{J}$  & $M$  & $N$  & $U$  & $V$  & $\phantom{\Big|}\!{J_{\varphi}}\phantom{\Big|}\!$  & $K_{\varphi}$  & $\Xi$  & $X$  & $Y$ \tabularnewline
\hline 
\hline 
Dimension  & 3  & 4  & 4  & 2 & 1 & 3  & 2  & 2  & 2  & 2  & 2  & 2  & 3  & 3  & 3 \tabularnewline
\hline 
$c$-ghost number  & $-1$  & $-2$  & $-1$  & 0 & 0 & 0  & 0  & $-1$  & $1$  & 0  & 0  & 1  & 0  & 1  & 0 \tabularnewline
\hline 
$\eta$-ghost number  & 0  & 0  & 0 & 0  & 0  & 0  & 0  & 0  & 0  & 0  & 0  & 0  & $-1$  & $-1$  & $-1$ \tabularnewline
\hline 
$e$-charge  & 0  & 0  & 0  & 0 & 0 & $0$  & $0$  & 0  & $0$  & 0  & $0$  & 0  & 0  & 0  & 0 \tabularnewline
\hline 
$U(4(N^{2}-1))$-charge  & 0  & 0  & 0 & 0 & 0  & 0  & $-1$  & $-1$  & 1  & 1  & 0  & 0  & 0  & 1  & 1 \tabularnewline
\hline 
$U(4N)$-charge  & 0  & 0  & 0 & 0  & 0  & 0  & $0$  & $0$  & 0  & 0  & 0  & 0  & 0  & 0  & 0 \tabularnewline
\hline 
Nature  & A  & C  & A  & C & C  & C  & C  & A  & A  & C  & C  & A  & A  & C  & A \tabularnewline
\hline 
\end{tabular}
\label{table2} 
}
\caption{Quantum numbers of the sources.}
\end{table}

\begin{table}[!ht]
{\small
\flushleft
\begin{tabular}{|c|c|c|c|c|c|c|c|c|c|c|}
\hline 
Sources  & $\phantom{\Big|}\!\bar{X}\phantom{\Big|}\!$  & $\phantom{\Big|}\!\bar{Y}\phantom{\Big|}\!$  & $\phantom{\Big|}\!\bar{\varUpsilon}\phantom{\Big|}\!$  & $\varUpsilon$  & $\phantom{\Big|}\!\bar{\Theta}\phantom{\Big|}\!$  & $\Theta$  & $\phantom{\Big|}\!\bar{\Lambda}\phantom{\Big|}\!$  & $\Lambda$  & $\phantom{\Big|}\!\bar{\Pi}\phantom{\Big|}\!$  & $\Pi$\tabularnewline
\hline 
\hline 
Dimension  & 3  & 3  & $\frac{5}{2}$  & $\frac{5}{2}$  & $\frac{5}{2}$  & $\frac{5}{2}$  & $\frac{3}{2}$  & $\frac{3}{2}$  & $\frac{3}{2}$  & $\frac{3}{2}$\tabularnewline
\hline 
$c$-ghost number  & $-1$  & 0  & $-1$  & $-1$  & $0$  & $0$  & $0$  & $0$  & $-1$  & $1$\tabularnewline
\hline 
$\eta$-ghost number  & $-1$  & $-1$  & $0$  & $0$  & $0$  & $0$  & $0$  & $0$  & $0$  & $0$\tabularnewline
\hline 
$e$-charge  & 0  & 0  & $-1$  & $1$  & $-1$  & $1$  & $0$  & $0$  & $0$  & $0$\tabularnewline
\hline 
$U(4(N^{2}-1))$-charge  & $-1$  & $-1$  & 0  & 0  & 0  & 0  & $0$  & $0$  & $0$  & $0$\tabularnewline
\hline 
$U(4N)$-charge  & $0$  & $0$  & 0  & 0  & 0  & 0  & $-1$  & $1$  & $-1$  & $1$\tabularnewline
\hline 
Nature  & C  & A  & C  & C  & A  & A  & C  & C  & A  & A\tabularnewline
\hline 
\end{tabular}
\label{table2-1} 
}
\caption{Quantum numbers of the sources.}
\end{table}

\begin{table}[!ht]
{\small
\flushleft
\begin{tabular}{|c|c|c|c|c|c|c|c|c|c|}
\hline\hline  
Extra Sources  & $\phantom{\Big|}\!{J_{\lambda}}\phantom{\Big|}\!$  & $K_{\lambda}$  & $\phantom{\Big|}\!\bar{Z}\phantom{\Big|}\!$  & $\phantom{\Big|}\!\bar{W}\phantom{\Big|}\!$  & $Z$  & $W$  & $\phantom{\Big|}\!\bar{\Phi}\phantom{\Big|}\!$  & $\Phi$  & $\Gamma$\tabularnewline
\hline 
\hline
Dimension  & $2$  & $2$  & $3$  & $3$  & $3$  & $3$  & $\frac{5}{2}$  & $\frac{5}{2}$  & $4$\tabularnewline
\hline 
$c$-ghost number  & $0$  & $1$  & $0$  & $-1$  & $0$  & $1$  & $0$  & $0$  & $0$\tabularnewline
\hline 
$\eta$-ghost number  & $0$  & $0$  & $-1$  & $-1$  & $-1$  & $-1$  & $-1$  & $-1$  & $-2$\tabularnewline
\hline 
$e$-charge  & $0$  & $0$  & $-1$  & $-1$  & $1$  & $1$  & $-1$  & $1$  & $0$\tabularnewline
\hline 
$U(4(N^{2}-1))$-charge  & $0$  & $0$  & $0$  & $0$  & $0$  & $0$  & $0$  & $0$  & $0$\tabularnewline
\hline 
$U(4N)$-charge  & $0$  & $0$  & $-1$  & $-1$  & $1$  & $1$  & $0$  & $0$  & $0$\tabularnewline
\hline 
Nature  & C  & A  & C  & A  & C  & A  & A  & A  & C\tabularnewline
\hline 
\end{tabular}
}
\label{table60} 
\caption{Quantum numbers of the extra sources.}
\end{table}
\newpage

\section{Ward identities}
\label{Wardid} 

\hspace{0.5cm}The tree level extended action $\Sigma$ defined by \eqref{Sigma_full}
enjoys a large set of  Ward identities which we enlist below: 
\begin{itemize}
\item {The generalized Slavnov-Taylor identity translating in functional form the exact invariance \eqref{QinvSS}:}

\begin{equation}
\mathcal{B}(\Sigma)=0\,,\label{wi0}
\end{equation}
with

\begin{eqnarray}
\mathcal{B}(\Sigma) & = & \int d^{4}x\,\Bigg(\frac{\delta\Sigma}{\delta A_{\mu}^{a}}\frac{\delta\Sigma}{\delta\Omega_{\mu}^{a}}+\frac{\delta\Sigma}{\delta c^{a}}\frac{\delta\Sigma}{\delta L^{a}}+\frac{\delta\Sigma}{\delta\xi^{a}}\frac{\delta\Sigma}{\delta K^{a}}+\frac{\delta\Sigma}{\delta\varUpsilon_{\alpha}^{i}}\frac{\delta\Sigma}{\delta\bar{\psi}^{i,\alpha}}+\frac{\delta\Sigma}{\delta\bar{\varUpsilon}_{\alpha}^{i}}\frac{\delta\Sigma}{\delta\psi^{i,\alpha}}+ib^{a}\frac{\delta\Sigma}{\delta\bar{c}^{a}}\nonumber \\
 & + & \omega^{aI}\frac{\delta\Sigma}{\delta\varphi^{aI}}+\bar{\varphi}^{aI}\frac{\delta\Sigma}{\delta\bar{\omega}^{aI}}+M_{\mu}^{aI}\frac{\delta\Sigma}{\delta N_{\mu}^{aI}}+U_{\mu}^{aI}\frac{\delta\Sigma}{\delta V_{\mu}^{aI}}+K_{\varphi}\frac{\delta\Sigma}{\delta{J}_{\varphi}}+X^{I}\frac{\delta\Sigma}{\delta Y^{I}}-\bar{Y}^{abI}\frac{\delta\Sigma}{\delta\bar{X}^{abI}}\nonumber \\
 & + & \bar{\lambda}^{a\hat{I}}\frac{\delta\Sigma}{\delta\bar{\zeta}_{\hat{I}}^{a}}+\zeta^{a\hat{I}}\frac{\delta\Sigma}{\delta\lambda_{\hat{I}}^{a}}+\bar{\Lambda}^{i\alpha\hat{I}}\frac{\delta\Sigma}{\delta\bar{\Pi}_{\alpha\hat{I}}^{i}}+\Pi^{i\alpha\hat{I}}\frac{\delta\Sigma}{\delta\Lambda_{\alpha\hat{I}}^{i}}+K_{\lambda}\frac{\delta\Sigma}{\delta{G}_{\lambda}}-W^{\hat{I}}\frac{\delta\Sigma}{\delta Z_{\hat{I}}}+\bar{Z}^{ab\hat{I}}\frac{\delta\Sigma}{\delta\bar{W}_{\hat{I}}^{ab}}\Bigg)\nonumber \\
 & + & \chi\frac{\partial\Sigma}{\partial\alpha}\,.\label{sti}
\end{eqnarray}
For further use, let us introduce the so-called nilpotent linearized Slavnov-Taylor operator $\mathcal{B}_{\Sigma}$
\cite{Piguet:1995er}, 
\begin{eqnarray}
\mathcal{B}_{\Sigma} & = & \int d^{4}x\,\Bigg(\frac{\delta\Sigma}{\delta A_{\mu}^{a}}\frac{\delta}{\delta\Omega_{\mu}^{a}}+\frac{\delta\Sigma}{\delta\Omega_{\mu}^{a}}\frac{\delta}{\delta A_{\mu}^{a}}+\frac{\delta\Sigma}{\delta c^{a}}\frac{\delta}{\delta L^{a}}+\frac{\delta\Sigma}{\delta L^{a}}\frac{\delta}{\delta c^{a}}+\frac{\delta\Sigma}{\delta\xi^{a}}\frac{\delta}{\delta K^{a}}+\frac{\delta\Sigma}{\delta K^{a}}\frac{\delta}{\delta\xi^{a}}\nonumber \\
 & + & \frac{\delta\Sigma}{\delta\varUpsilon_{\alpha}^{i}}\frac{\delta}{\delta\bar{\psi}^{i,\alpha}}+\frac{\delta\Sigma}{\delta\bar{\psi}^{i,\alpha}}\frac{\delta}{\delta\varUpsilon_{\alpha}^{i}}+\frac{\delta\Sigma}{\delta\bar{\varUpsilon}_{\alpha}^{i}}\frac{\delta}{\delta\psi^{i,\alpha}}+\frac{\delta\Sigma}{\delta\psi^{i,\alpha}}\frac{\delta}{\delta\bar{\varUpsilon}_{\alpha}^{i}}+ib^{a}\frac{\delta}{\delta\bar{c}^{a}}+\omega^{aI}\frac{\delta}{\delta\varphi^{aI}}\nonumber 
 \end{eqnarray}
 \begin{eqnarray}
& + & \bar{\varphi}^{aI}\frac{\delta}{\delta\bar{\omega}^{aI}}+M_{\mu}^{aI}\frac{\delta}{\delta N_{\mu}^{aI}}+U_{\mu}^{aI}\frac{\delta}{\delta V_{\mu}^{aI}}+K_{\varphi}\frac{\delta}{\delta{J}_{\varphi}}+X^{I}\frac{\delta}{\delta Y^{I}}-\bar{Y}^{abI}\frac{\delta}{\delta\bar{X}^{abI}}+\bar{\lambda}^{a\hat{I}}\frac{\delta}{\delta\bar{\zeta}_{\hat{I}}^{a}}\nonumber \\
 & + & \zeta^{a\hat{I}}\frac{\delta}{\delta\lambda_{\hat{I}}^{a}}+\bar{\Lambda}^{i\alpha\hat{I}}\frac{\delta}{\delta\bar{\Pi}_{\alpha\hat{I}}^{i}}+\Pi^{i\alpha\hat{I}}\frac{\delta}{\delta\Lambda_{\alpha\hat{I}}^{i}}
 +K_{\lambda}\frac{\delta}{\delta{J}_{\lambda}}-W^{\hat{I}}\frac{\delta}{\delta Z_{\hat{I}}}+\bar{Z}^{ab\hat{I}}\frac{\delta}{\delta\bar{W}_{\hat{I}}^{ab}}\Bigg)+\chi\frac{\partial}{\partial\alpha}\,,\nonumber \\
\label{lsti}
\end{eqnarray}
with 
\begin{equation}
\mathcal{B}_{\Sigma}\mathcal{B}_{\Sigma}=0\;.\label{nln}
\end{equation}
The linearized nilpotent operator $\mathcal{B}_{\Sigma}$ plays a pivotal role in the analysis of the algebraic renormalization due to the fact that the most general local invariant counterterm which can be freely added to any loop order can be characterized in terms of  the cohomology of $\mathcal{B}_{\Sigma}$ in the space of the local integrated polynomials in the fields, sources and their derivatives \cite{Piguet:1995er}. 

\item {The anti-ghost Ward identity:}

\begin{equation}
\frac{\delta\Sigma}{\delta\bar{c}^{a}}+\partial_{\mu}\frac{\delta\Sigma}{\delta\Omega_{\mu}^{a}}=\frac{i}{2}\chi b^{a}\,.\label{wi2}
\end{equation}
Notice that the right-hand side of eq.\eqref{wi2} is a linear breaking, {\it i.e.} a breaking linear in the quantum fields. As such, it will be not affected by quantum corrections \cite{Piguet:1995er}.
\item The equation of motion of the Lagrange multiplier $b^{a}$:

\begin{equation}
\frac{\delta\Sigma}{\delta b^{a}}=i\partial_{\mu}A_{\mu}^{a}+\alpha b^{a}-\frac{i}{2}\chi\bar{c}^{a}\,,\label{wi3}
\end{equation}
expressing in functional form the linear covariant gauge-fixing condition adopted here. Again, the right-hand side of eq.\eqref{wi3} is a linear breaking,  not affected by the quantum corrections.  

\item The $\ensuremath{\tau^{a}}$ Ward identity:

Analogously to the anti-ghost Ward identity, the equation of motion of the $\tau^{a}$ field and the variation of the action with respect to the source $\mathcal{J}^{a}_{\mu}$, yield the following Ward identity:  
\begin{equation}
\frac{\delta\Sigma}{\delta\tau{a}}-\partial_{\mu}\frac{\delta\Sigma}{\delta\mathcal{J}_{\mu}^{a}}=0\,.\label{wi4}
\end{equation}

\item The  anti-ghost $\ensuremath{{\bar{\eta}}^{a}}$ Ward identity:

\begin{equation}
\frac{\delta\Sigma}{\delta\bar{\eta}^{a}}+\partial_{\mu}\frac{\delta\Sigma}{\delta\Xi_{\mu}^{a}}=0\,.\label{w15}
\end{equation}
Note that the presence of the composite field operator $D^{ab}_{\mu}(A^{h})\eta^{b}$, coupled to the source $\Xi^{a}_{\mu}$, is needed in order to establish this identity.

\item The integrated linearly broken  $\ensuremath{\eta^{a}}$-ghost Ward identity:

\begin{eqnarray}
\int d^{4}x\,\bigg(\frac{\delta\Sigma}{\delta\eta^{a}}+gf^{abc}\bar{\eta}^{b}\frac{\delta\Sigma}{\delta\tau^{c}}-gf^{abc}\Xi_{\mu}^{b}\frac{\delta\Sigma}{\delta\mathcal{J}_{\mu}^{c}}\bigg) & = & \int d^{4}x\,\bigg(\bar{X}^{abI}\omega^{bI}-\bar{Y}^{abI}\varphi^{bI}+X\bar{\omega}^{aI}\nonumber \\
 & - & Y^{I}\bar{\varphi}^{aI}+Z^{\hat{I}}\bar{\lambda}_{\hat{I}}^{a}-W^{\hat{I}}\bar{\zeta}_{\hat{I}}^{a}+\bar{Z}^{ab\hat{I}}\lambda_{\hat{I}}^{b}\nonumber \\
 & - & \bar{W}^{ab\hat{I}}\zeta_{\hat{I}}^{b}+\Gamma^{ab}\eta^{b}+\Phi_{\alpha}^{i}\bar{\psi}^{h,j\alpha}T^{a,ji}\nonumber \\
 & - & \bar{\Phi}^{i\alpha}T^{a,ij}\psi_{\alpha}^{h,j}\bigg)\,.
 \label{w16}
\end{eqnarray}

\item The global $U(4(N^{2}-1))$ symmetry: 
\begin{equation}
U_{IJ}(\Sigma)=0\,,\label{wi5}
\end{equation}
with 
\begin{eqnarray}
U_{IJ}(\Sigma) & = & \int d^{4}x\bigg(\varphi^{aI}\frac{\delta\Sigma}{\delta\varphi^{aJ}}-\bar{\varphi}^{aJ}\frac{\delta\Sigma}{\delta\bar{\varphi}^{aI}}+\omega^{aI}\frac{\delta\Sigma}{\delta\omega^{aJ}}-\bar{\omega}^{aJ}\frac{\delta\Sigma}{\delta\bar{\omega}^{aI}}\nonumber \\
 & - & M_{\mu}^{aJ}\frac{\delta\Sigma}{\delta M_{\mu}^{aI}}+V_{\mu}^{aI}\frac{\delta\Sigma}{\delta V_{\mu}^{aJ}}-N_{\mu}^{aJ}\frac{\delta\Sigma}{\delta N_{\mu}^{aI}}+U_{\mu}^{aJ}\frac{\delta\Sigma}{\delta U_{\mu}^{aI}}\nonumber \\
 & + &X^{J}\frac{\delta\Sigma}{\delta X^{I}}+ Y^{I}\frac{\delta\Sigma}{\delta Y^{J}}-\bar{X}^{abJ}\frac{\delta\Sigma}{\delta\bar{X}^{abI}}-\bar{Y}^{abJ}\frac{\delta\Sigma}{\delta\bar{Y}^{abI}}\bigg)\,.\label{wi6}
\end{eqnarray}
\item The $U(4N)$ symmetry: 
\begin{equation}
\hat{U}^{\hat{I}\hat{J}}(\Sigma)=0\,,\label{wi5-1}
\end{equation}
where 
\begin{eqnarray}
\hat{U}^{\hat{I}\hat{J}}(\Sigma) & = & \int d^{4}x\,\Bigg(\lambda^{a\hat{I}}\frac{\delta\Sigma}{\delta\lambda_{\hat{J}}^{a}}-\bar{\lambda}^{a\hat{J}}\frac{\delta\Sigma}{\delta\bar{\lambda}_{\hat{I}}^{a}}+\zeta^{a\hat{I}}\frac{\delta\Sigma}{\delta\zeta_{\hat{J}}^{a}}-\bar{\zeta}^{a\hat{J}}\frac{\delta\Sigma}{\delta\bar{\zeta}_{\hat{I}}^{a}}+\Lambda^{i\alpha\hat{I}}\frac{\delta\Sigma}{\delta\Lambda_{\alpha\hat{J}}^{i}}\nonumber \\
 & - & \bar{\Lambda}^{i\alpha\hat{J}}\frac{\delta\Sigma}{\delta\bar{\Lambda}_{\alpha\hat{I}}^{i}}+\Pi^{i\alpha\hat{I}}\frac{\delta\Sigma}{\delta\Pi_{\alpha\hat{J}}^{i}}-\bar{\Pi}^{i\alpha\hat{J}}\frac{\delta\Sigma}{\delta\bar{\Pi}_{\alpha\hat{I}}^{i}}+Z^{\hat{J}}\frac{\delta\Sigma}{\delta Z_{\hat{I}}}+W^{\hat{I}}\frac{\delta\Sigma}{\delta W_{\hat{J}}}\nonumber \\
 & - & \bar{Z}^{ab\hat{J}}\frac{\delta\Sigma}{\delta\bar{Z}_{\hat{I}}^{ab}}-\bar{W}^{ab\hat{J}}\frac{\delta\Sigma}{\delta\bar{W}_{\hat{I}}^{ab}}\Bigg)\,.\label{wi6-1}
\end{eqnarray}
\item The $e$-charge, or spinor number Ward identity : 
\begin{eqnarray*}
\mathcal{N}_{e}(\Sigma) & = & 0\,,
\end{eqnarray*}
\begin{eqnarray}
\mathcal{N}_{e}(\Sigma) & = & \int d^{4}x\,\Bigg(\psi^{i\alpha}\frac{\delta\Sigma}{\delta\psi_{\alpha}^{i}}-\bar{\psi}^{i\alpha}\frac{\delta\Sigma}{\delta\bar{\psi}_{\alpha}^{i}}+\varUpsilon^{i\alpha}\frac{\delta\Sigma}{\delta\varUpsilon_{\alpha}^{i}}-\bar{\varUpsilon}{}^{i\alpha}\frac{\delta\Sigma}{\delta\bar{\varUpsilon}_{\alpha}^{i}}+\Theta^{i\alpha}\frac{\delta\Sigma}{\delta\Theta_{\alpha}^{i}}-\bar{\Theta}{}^{i\alpha}\frac{\delta\Sigma}{\delta\bar{\Theta}_{\alpha}^{i}}\nonumber \\
 & + & \lambda^{a\hat{I}}\frac{\delta\Sigma}{\delta\lambda_{\hat{I}}^{a}}-\bar{\lambda}^{a\hat{I}}\frac{\delta\Sigma}{\delta\bar{\lambda}_{\hat{I}}^{a}}+\zeta^{a\hat{I}}\frac{\delta\Sigma}{\delta\zeta_{\hat{I}}^{a}}-\bar{\zeta}^{a\hat{I}}\frac{\delta\Sigma}{\delta\bar{\zeta}_{\hat{I}}^{a}}+\Phi^{i\alpha}\frac{\delta\Sigma}{\delta\Phi_{\alpha}^{i}}-\bar{\Phi}{}^{i\alpha}\frac{\delta\Sigma}{\delta\bar{\Phi}_{\alpha}^{i}}+Z^{\hat{I}}\frac{\delta\Sigma}{\delta Z_{\hat{I}}}\nonumber \\
 & + & W^{\hat{I}}\frac{\delta\Sigma}{\delta W_{\hat{I}}}-\bar{Z}^{ab\hat{I}}\frac{\delta\Sigma}{\delta\bar{Z}_{\hat{I}}^{ab}}-\bar{W}^{ab\hat{I}}\frac{\delta\Sigma}{\delta\bar{W}_{\hat{I}}^{ab}}\Bigg)\,.
\end{eqnarray}
\item The linearly broken identities:

\begin{equation}
\frac{\delta\Sigma}{\delta\bar{\varphi}^{aI}}+\partial_{\mu}\frac{\delta\Sigma}{\delta M_{\mu}^{aI}}+gf^{abc}V_{\mu}^{bI}\frac{\delta\Sigma}{\delta\mathcal{J}_{\mu}^{c}}=-{J}_{\varphi}\varphi^{aI}+Y^{I}\eta^{a}\,,\label{wi7}
\end{equation}

\begin{equation}
\frac{\delta\Sigma}{\delta\varphi^{aI}}+\partial_{\mu}\frac{\delta\Sigma}{\delta V_{\mu}^{aI}}-gf^{abc}\bar{\varphi}^{bI}\frac{\delta\Sigma}{\delta\tau^{c}}+gf^{abc}M_{\mu}^{bI}\frac{\delta\Sigma}{\delta\mathcal{J}_{\mu}^{c}}=-{J}_{\varphi}\bar{\varphi}^{aI}-K_{\varphi}\bar{\omega}^{aI}+\bar{Y}^{baI}\eta^{b}\,,\label{wi8}
\end{equation}

\begin{equation}
\frac{\delta\Sigma}{\delta\bar{\omega}^{aI}}+\partial_{\mu}\frac{\delta\Sigma}{\delta N_{\mu}^{aI}}-gf^{abc}U_{\mu}^{bI}\frac{\delta\Sigma}{\delta\mathcal{J}_{\mu}^{c}}={J}_{\varphi}\omega^{aI}-{K}_{\varphi}\varphi^{aI}-X^{I}\eta^{a}\,,\label{wi9}
\end{equation}

\begin{equation}
\frac{\delta\Sigma}{\delta\omega^{aI}}+\partial_{\mu}\frac{\delta\Sigma}{\delta U_{\mu}^{aI}}-gf^{abc}\bar{\omega}^{bI}\frac{\delta\Sigma}{\delta\tau^{c}}+gf^{abc}N_{\mu}^{bI}\frac{\delta\Sigma}{\delta\mathcal{J}_{\mu}^{c}}=-{J}_{\varphi}\bar{\omega}^{aI}-\bar{X}^{baI}\eta^{a}\,.\label{wi10}
\end{equation}

\item The linearly broken integrated Ward identities for the matter sector:

\begin{eqnarray}
\int d^{4}x\,\bigg(\frac{\delta\Sigma}{\delta\zeta^{a\hat{I}}}+gf^{abc}\bar{\zeta}_{\hat{I}}^{b}\frac{\delta\Sigma}{\delta\tau^{c}}-\bar{\Pi}_{\hat{I}}^{i\alpha}\,T^{a,ij}\frac{\delta\Sigma}{\delta\Theta_{\alpha}^{j}}\bigg) & = & \int d^{4}x\,\bigg({J}_{\lambda}\bar{\zeta}_{\hat{I}}^{a}-\bar{W}_{\hat{I}}^{ba}\,\eta^{b}\bigg)\,,\nonumber\\
\end{eqnarray}
\begin{eqnarray}
\int d^{4}x\,\bigg(\frac{\delta\Sigma}{\delta\lambda^{a\hat{I}}}+gf^{abc}\frac{\delta\Sigma}{\delta\tau^{c}}\bar{\lambda}_{\hat{I}}^{b}+T^{a,ij}\,\bar{\Lambda}_{\hat{I}}^{i\alpha}\frac{\delta\Sigma}{\delta\Theta_{\alpha}^{j}}\bigg) & = & \int d^{4}x\,\bigg(J_{\lambda}\bar{\lambda}_{\hat{I}}^{a}+K_{\lambda}\bar{\zeta}^{a}_{\hat{I}}
+\bar{Z}_{\hat{I}}^{ba}\,\eta^{b}\bigg)\,,\nonumber\\
\end{eqnarray}
\begin{eqnarray}
\int d^{4}x\,\bigg(\frac{\delta\Sigma}{\delta\bar{\zeta}^{a\hat{I}}}+\Pi_{\hat{I}}^{i\alpha}\,T^{a,ij}\,\frac{\delta\Sigma}{\delta\bar{\Theta}_{\alpha}^{j}}\bigg) & = & \int d^{4}x\,\bigg(J_{\lambda}\zeta_{\hat{I}}^{a}-W_{\hat{I}}\,\eta^{a}-K_{\lambda}\lambda^{a}_{\hat{I}}\bigg)\,,
\end{eqnarray}
\begin{eqnarray}
\int d^{4}x\,\bigg(\frac{\delta\Sigma}{\delta\bar{\lambda}^{a\hat{I}}}+\Lambda_{\hat{I}}^{i\alpha}\,T^{a,ij}\,\frac{\delta\Sigma}{\delta\bar{\Theta}_{\alpha}^{j}}\bigg) & = & \int d^{4}x\,\bigg(J_{\lambda}\lambda_{\hat{I}}^{a}+Z_{\hat{I}}\,\eta^{a}\bigg)\,.
\end{eqnarray}

\item The $\eta$-ghost number identity:

A ghost number can be assigned to the anti-commuting fields $(\bar \eta, \eta)$ and to the source $\Xi_{\mu}$, resulting in the following $\eta$-ghost number Ward identity
\begin{eqnarray}
\mathcal{N}_{\eta-ghost}(\Sigma) & = & \int d^{4}x\,\bigg(\eta^{a}\frac{\delta\Sigma}{\delta\eta^{a}}+-\bar{\eta}^{a}\frac{\delta\Sigma}{\delta\bar{\eta}^{a}}-\Xi_{\mu}^{a}\frac{\delta\Sigma}{\delta\Xi_{\mu}^{a}}-X^{I}\frac{\delta\Sigma}{\delta X^{I}}-Y^{I}\frac{\delta\Sigma}{\delta Y^{I}}\nonumber \\
 & - & \bar{X}^{abI}\frac{\delta\Sigma}{\delta\bar{X}^{abI}}-\bar{Y}^{abI}\frac{\delta\Sigma}{\delta\bar{Y}^{abI}}-Z^{\hat{I}}\frac{\delta\Sigma}{\delta Z_{\hat{I}}}-W^{\hat{I}}\frac{\delta\Sigma}{\delta W_{\hat{I}}}-\bar{Z}^{ab\hat{I}}\frac{\delta\Sigma}{\delta\bar{Z}_{\hat{I}}^{ab}}\nonumber \\
 & - & \bar{W}^{ab\hat{I}}\frac{\delta\Sigma}{\delta\bar{W}_{\hat{I}}^{ab}}-2\Gamma^{ab}\frac{\delta\Sigma}{\delta\Gamma^{ab}}\bigg)=0\,.
\end{eqnarray}

\item The $c$-ghost number identity:

Analogously, we have also the usual $c$-ghost number in the Faddeev-Popov sector, expressed by
\begin{eqnarray}
\mathcal{N}_{c-ghost}(\Sigma) & = & \int d^{4}x\,\bigg(c^{a}\frac{\delta\Sigma}{\delta c^{a}}-\bar{c}^{a}\frac{\delta\Sigma}{\delta\bar{c}^{a}}+\omega^{aI}\frac{\delta\Sigma}{\delta\omega^{aI}}-\bar{\omega}^{aI}\frac{\delta\Sigma}{\delta\bar{\omega}^{aI}}-\varUpsilon^{i\alpha}\frac{\delta\Sigma}{\delta\varUpsilon_{\alpha}^{i}}-\bar{\varUpsilon}{}^{i\alpha}\frac{\delta\Sigma}{\delta\bar{\varUpsilon}_{\alpha}^{i}}\nonumber \\
 & - & \Omega_{\mu}^{a}\frac{\delta\Sigma}{\delta\Omega_{\mu}^{a}}-2L^{a}\frac{\delta\Sigma}{\delta L^{a}}-K^{a}\frac{\delta\Sigma}{\delta K^{a}}+U_{\mu}^{aI}\frac{\delta\Sigma}{\delta U_{\mu}^{aI}}-N_{\mu}^{aI}\frac{\delta\Sigma}{\delta N_{\mu}^{aI}}+K_{\varphi}\frac{\partial\Sigma}{\delta K_{\varphi}}+X^{I}\frac{\delta\Sigma}{\delta X^{I}}\nonumber \\
 & - & \bar{X}^{abI}\frac{\delta\Sigma}{\delta\bar{X}^{abI}}+\zeta^{a\hat{I}}\frac{\delta\Sigma}{\delta\zeta_{\hat{I}}^{a}}-\bar{\zeta}^{a\hat{I}}\frac{\delta\Sigma}{\delta\bar{\zeta}_{\hat{I}}^{a}}+\Pi^{i\alpha\hat{I}}\frac{\delta\Sigma}{\delta\Pi_{\alpha\hat{I}}^{i}}-\bar{\Pi}^{i\alpha\hat{I}}\frac{\delta\Sigma}{\delta\bar{\Pi}_{\alpha\hat{I}}^{i}}+K_{\lambda}\frac{\delta\Sigma}{\delta K_{\lambda}}\nonumber \\
 & + & B^{\hat{I}}\frac{\delta\Sigma}{\delta B_{\hat{I}}}-\bar{B}^{ab\hat{I}}\frac{\delta\Sigma}{\delta\bar{B}_{\hat{I}}^{ab}}\bigg)+\chi\frac{\partial\Sigma}{\partial\chi}=0\,.\label{wi11}
\end{eqnarray}

\item The exactly $\mathcal{R}_{IJ}$ symmetry;

\begin{equation}
\mathcal{R}_{IJ}(\Sigma)=0\,,\label{wi13}
\end{equation}
with

\begin{equation}
\mathcal{R}_{IJ}(\Sigma)=\int d^{4}x\,\bigg(\varphi^{aI}\frac{\delta\Sigma}{\delta\omega^{aJ}}-\bar{\omega}^{aJ}\frac{\delta\Sigma}{\delta\bar{\varphi}^{aI}}+V_{\mu}^{aI}\frac{\delta\Sigma}{\delta U_{\mu}^{aJ}}-N_{\mu}^{aI}\frac{\delta\Sigma}{\delta M_{\mu}^{aJ}}+\bar{X}^{abJ}\frac{\delta\Sigma}{\delta\bar{Y}^{abI}}+Y^{I}\frac{\delta\Sigma}{\delta X^{J}}\bigg)\,.\label{wi14}
\end{equation}

\item Identities that mix the Zwanziger ghosts with the $(\eta,\bar{\eta})$
ghosts 
\begin{eqnarray}
W_{(1)}^{I}(\Sigma) & = & \int d^{4}x\,\bigg(\bar{\omega}^{aI}\frac{\delta\Sigma}{\delta\bar{\eta}^{a}}+\eta^{a}\frac{\delta\Sigma}{\delta\omega^{aI}}+N_{\mu}^{aI}\frac{\delta\Sigma}{\delta\Xi_{\mu}^{a}}+J_{\varphi}\frac{\delta\Sigma}{\delta X^{I}}+\bar{X}^{abI}\frac{\delta\Sigma}{\delta\Gamma^{ab}}\bigg)=0\,,\nonumber \\
\\
W_{(2)}^{I}(\Sigma) & = & \int d^{4}x\,\bigg(\bar{\varphi}^{aI}\frac{\delta\Sigma}{\delta\bar{\eta}^{a}}-\eta^{a}\frac{\delta\Sigma}{\delta\varphi^{aI}}+M_{\mu}^{aI}\frac{\delta\Sigma}{\delta\Xi_{\mu}^{a}}-J_{\varphi}\frac{\delta\Sigma}{\delta Y^{I}}+K_{\varphi}\frac{\delta\Sigma}{\delta X^{I}}\nonumber \\
 & - & \bar{Y}^{abI}\frac{\delta\Sigma}{\delta\Gamma^{ab}}\bigg)=0\,,\\
W_{(3)}^{I}(\Sigma) & = & \int d^{4}x\,\bigg(\varphi^{aI}\frac{\delta\Sigma}{\delta\bar{\eta}^{a}}-\eta^{a}\frac{\delta\Sigma}{\delta\bar{\varphi}^{aI}}-gf^{abc}\frac{\delta\Sigma}{\delta\bar{Y}^{abI}}\frac{\delta\Sigma}{\delta\tau^{c}}-V_{\mu}^{aI}\frac{\delta\Sigma}{\delta\Xi_{\mu}^{a}}+J_{\varphi}\frac{\delta\Sigma}{\delta\bar{Y}^{aaI}}\nonumber \\
 & + & Y^{I}\frac{\delta\Sigma}{\delta\Gamma^{aa}}\bigg)=0\,,\\
W_{(4)}^{I}(\Sigma) & = & \int d^{4}x\,\bigg(\omega^{aI}\frac{\delta\Sigma}{\delta\bar{\eta}^{a}}-\eta^{a}\frac{\delta\Sigma}{\delta\bar{\omega}^{aI}}+gf^{abc}\frac{\delta\Sigma}{\delta\bar{X}^{abI}}\frac{\delta\Sigma}{\delta\tau^{c}}+U_{\mu}^{aI}\frac{\delta\Sigma}{\delta\Xi_{\mu}^{a}}+J_{\varphi}\frac{\delta\Sigma}{\delta\bar{X}^{aaI}}\nonumber \\
 & + & K_{\varphi}\frac{\delta\Sigma}{\delta\bar{Y}^{aaI}}+X^{I}\frac{\delta\Sigma}{\delta\Gamma^{aa}}\bigg)=0\,.
\end{eqnarray}
\item Identities that mix the ghosts related to fermionic matter fields with the
$(\eta,\bar{\eta})$ ghosts: 
\begin{eqnarray}
W_{(1)}^{\hat{I}}(\Sigma) & = & \int d^{4}x\,\bigg(\eta^{a}\frac{\delta\Sigma}{\delta\lambda_{\hat{I}}^{a}}+\bar{\lambda}^{a\hat{I}}\frac{\delta\Sigma}{\delta\bar{\eta}^{a}}+\bar{\Lambda}^{i\alpha\hat{I}}\frac{\delta\Sigma}{\delta\Phi_{\alpha}^{i}}
+J_{\lambda}\frac{\delta\Sigma}{\delta Z_{\hat{I}}}-\bar{Z}^{ab\hat{I}}\frac{\delta\Sigma}{\delta\Gamma^{ab}}\bigg)=0\,,\nonumber \\
\\
W_{(2)}^{\hat{I}}(\Sigma) & = & \int d^{4}x\,\bigg(\bar{\zeta}{}^{a\hat{I}}\frac{\delta\Sigma}{\delta\bar{\eta}^{a}}-\eta^{a}\frac{\delta\Sigma}{\delta\zeta_{\hat{I}}^{a}}+\bar{\Pi}^{i\alpha\hat{I}}\frac{\delta\Sigma}{\delta\Phi_{\alpha}^{i}}-J_{\lambda}\frac{\delta\Sigma}{\delta W_{\hat{I}}}+K_{\lambda}\frac{\delta\Sigma}{\delta Z_{\hat{I}}}-\bar{W}^{ab\hat{I}}\frac{\delta\Sigma}{\delta\Gamma^{ab}}\bigg)=0\,,\nonumber \\
\\
W_{(3)}^{\hat{I}}(\Sigma) & = & \int d^{4}x\,\bigg(\eta^{a}\frac{\delta\Sigma}{\delta\bar{\lambda}_{\hat{I}}^{a}}+\lambda{}^{a\hat{I}}\frac{\delta\Sigma}{\delta\bar{\eta}^{a}}-\Lambda^{i\alpha\hat{I}}\frac{\delta\Sigma}{\delta\bar{\Phi}{}_{\alpha}^{i}}-J_{\lambda}\frac{\delta\Sigma}{\delta\bar{Z}_{\hat{I}}^{aa}}-K_{\lambda}\frac{\delta\Sigma}{\delta\bar{W}_{\hat{I}}^{aa}}-Z^{\hat{I}}\frac{\delta\Sigma}{\delta\Gamma^{aa}}\bigg)=0\,,\nonumber \\
\\
W_{(4)}^{\hat{I}}(\Sigma) & = & \int d^{4}x\,\bigg(\zeta{}^{a\hat{I}}\frac{\delta\Sigma}{\delta\bar{\eta}^{a}}+\eta^{a}\frac{\delta\Sigma}{\delta\bar{\zeta}_{\hat{I}}^{a}}+\Pi^{i\alpha\hat{I}}\frac{\delta\Sigma}{\delta\bar{\Phi}_{\alpha}^{i}}+J_{\lambda}\frac{\delta\Sigma}{\delta\bar{W}_{\hat{I}}^{aa}}-W^{\hat{I}}\frac{\delta\Sigma}{\delta\Gamma^{aa}}\bigg)=0\,.\nonumber \\
\end{eqnarray}
\end{itemize}

\section{Algebraic renormalization analysis}
\label{REn}
\hspace{0.5 cm} In section \eqref{Wardid}, we displayed all Ward Identities that
the tree level action $\Sigma$, eq\eqref{Sigma_full}, obeys. We can focus thus on the  search of
the most general counterterm in order to check out  
the renormalizability of our model. To that aim it turns out to be helpful to adopt a slightly different notation and reparametrize the fields and sources as  
\begin{equation}
    \left(\,A^{a}_{\mu}\,,\,\,b^{a}\,,\,\,\xi^{a}\,,\,\,\alpha\,,\,\,\tau^{a}\,,\,\,\mathcal{J}^{a}_{\mu}\,,\,\,J\,\right)\rightarrow
    \left(\,
    \frac{1}{g}A^{a}_{\mu}\,,\,\,
    gb^{a}\,,\,\,
    \frac{1}{g}\xi^{a}\,,\,\,
    \frac{1}{g^{2}}\alpha\,,\,\, 
    g\tau^{a}\,,\,\,
    g\mathcal{J}^{a}_{\mu}\,,\,\,
    g^{2}J\,\right)\,.
\end{equation}

\subsection{Algebraic characterization of the most general counterterm}

In order to characterize the most general local invariant counterterm which
can be freely added to all orders in perturbation theory, we follow
the setup of the algebraic renormalization \cite{Piguet:1995er} and
perturb the tree level action $\Sigma$ by adding an integrated local
quantity in the fields and sources, $\Sigma_{\mathrm{CT}}$, with
dimension bounded by four and vanishing $c$-ghost number. We demand
thus that the perturbed action, $( \Sigma+\epsilon\Sigma_{\mathrm{CT}})$, 
where $\epsilon$ stands for an expansion parameter, fulfills  to the first
order in $\epsilon$ the same Ward identities obeyed by the starting 
action $\Sigma$, \textit{i.e.} 
\begin{eqnarray*}
 &  & \mathcal{B}_{\Sigma}(\Sigma+\epsilon\Sigma_{\mathrm{CT}})=\mathcal{O}(\epsilon^{2})\,,\nonumber \\
 &  & \bigg(\frac{\delta}{\delta\bar{c}^{a}}+\partial_{\mu}\frac{\delta}{\delta\Omega_{\mu}^{a}}\bigg)(\Sigma+\epsilon\Sigma_{\mathrm{CT}})-\frac{i}{2}\chi b^{a}=\mathcal{O}(\epsilon^{2})\,,\nonumber \\
 &  & \frac{\delta}{\delta b^{a}}(\Sigma+\epsilon\Sigma_{\mathrm{CT}})=i\partial_{\mu}A_{\mu}^{a}+\alpha b^{a}-\frac{i}{2}\chi\bar{c}^{a}+\mathcal{O}(\epsilon^{2})\,,\nonumber \\
 &  & \bigg(\frac{\delta}{\delta\tau^{a}}-\partial_{\mu}\frac{\delta}{\delta\mathcal{J}_{\mu}^{a}}\bigg)(\Sigma+\epsilon\Sigma_{\mathrm{CT}})=\mathcal{O}(\epsilon^{2})\,,\nonumber \\
 &  & U_{IJ}(\Sigma+\epsilon\Sigma_{\mathrm{CT}})=\mathcal{O}(\epsilon^{2})\,,\nonumber \\
 &  & \hat{U}_{\hat{I}}^{\hat{J}}(\Sigma+\epsilon\Sigma_{\mathrm{CT}})=\mathcal{O}(\epsilon^{2})\,,\nonumber \\
 &  & \mathcal{N}_{e}(\Sigma+\epsilon\Sigma_{\mathrm{CT}})=\mathcal{O}(\epsilon^{2})\,,\nonumber \\
 &  & \mathcal{N}_{c-ghost}(\Sigma+\epsilon\Sigma_{\mathrm{CT}})=\mathcal{O}(\epsilon^{2})\,,\nonumber \\
 &  & \mathcal{N}_{\eta-ghost}(\Sigma+\epsilon\Sigma_{\mathrm{CT}})=\mathcal{O}(\epsilon^{2})\,,\nonumber \\
 \end{eqnarray*}
 \begin{eqnarray}
 &  & \bigg(\frac{\delta}{\delta\bar{\varphi}^{aI}}+\partial_{\mu}\frac{\delta}{\delta M_{\mu}^{ai}}+f^{abc}V_{\mu}^{bI}\frac{\delta}{\delta\mathcal{J}_{\mu}^{c}}\bigg)(\Sigma+\epsilon\Sigma_{\mathrm{CT}})=-\tilde{J}\varphi^{aI}+Y^{I}\eta^{a}+\mathcal{O}(\epsilon^{2})\,,\nonumber \\
 &  & \bigg(\frac{\delta}{\delta\varphi^{aI}}+\partial_{\mu}\frac{\delta}{\delta V_{\mu}^{aI}}-f^{abc}\bar{\varphi}^{bI}\frac{\delta}{\delta\tau^{c}}+f^{abc}M_{\mu}^{bI}\frac{\delta}{\delta\mathcal{J}_{\mu}^{c}}\bigg)(\Sigma+\epsilon\Sigma_{\mathrm{CT}})=-\tilde{J}\bar{\varphi}^{aI}-H\bar{\omega}^{aI}+\bar{Y}^{baI}\eta^{b}+\mathcal{O}(\epsilon^{2})\nonumber \\
 &  & \bigg(\frac{\delta}{\delta\bar{\omega}^{aI}}+\partial_{\mu}\frac{\delta}{\delta N_{\mu}^{aI}}-f^{abc}U_{\mu}^{bI}\frac{\delta}{\delta\mathcal{J}_{\mu}^{c}}\bigg)(\Sigma+\epsilon\Sigma_{\mathrm{CT}})=\tilde{J}\omega^{aI}-H\varphi^{aI}-X^{I}\eta^{a}+\mathcal{O}(\epsilon^{2})\,,\nonumber \\
 &  & \bigg(\frac{\delta}{\delta\omega^{aI}}+\partial_{\mu}\frac{\delta}{\delta U_{\mu}^{aI}}-f^{abc}\bar{\omega}^{bI}\frac{\delta}{\delta\tau^{c}}+f^{abc}N_{\mu}^{bI}\frac{\delta}{\delta\mathcal{J}_{\mu}^{c}}\bigg)(\Sigma+\epsilon\Sigma_{\mathrm{CT}})=-\tilde{J}\bar{\omega}^{aI}-\bar{X}^{baI}\eta^{b}+\mathcal{O}(\epsilon^{2})\,,\nonumber \\
 &  & \int d^{4}x\,\bigg(\frac{\delta}{\delta\zeta^{a\hat{I}}}+f^{abc}\bar{\zeta}_{\hat{I}}^{b}\frac{\delta}{\delta\tau^{c}}-\bar{\Pi}_{\hat{I}}^{i\alpha}(T^{a})^{ij}\frac{\delta}{\delta\Theta_{\alpha}^{j}}\bigg)(\Sigma+\epsilon\Sigma_{\mathrm{CT}})=\int d^{4}x\,\bigg(\tilde{G}\bar{\zeta}_{\hat{I}}^{a}+P\bar{\lambda}_{\hat{I}}^{a}-\bar{B}_{\hat{I}}^{ba}\,\eta^{b}\bigg)+\mathcal{O}(\epsilon^{2})\,,\nonumber \\
 &  & \int d^{4}x\,\bigg(\frac{\delta}{\delta\lambda^{a\hat{I}}}+f^{abc}\frac{\delta}{\delta\tau^{c}}\bar{\lambda}_{\hat{I}}^{b}+(T^{a})^{ij}\bar{\Lambda}_{\hat{I}}^{i\alpha}\frac{\delta}{\delta\Theta_{\alpha}^{j}}\bigg)(\Sigma+\epsilon\Sigma_{\mathrm{CT}})=\int d^{4}x\,\bigg(\tilde{G}\bar{\lambda}_{\hat{I}}^{a}+\bar{Z}_{\hat{I}}^{ba}\,\eta^{b}\bigg)+\mathcal{O}(\epsilon^{2})\,,\nonumber \\
 &  & \int d^{4}x\,\bigg(\frac{\delta}{\delta\bar{\zeta}^{a\hat{I}}}+\Pi_{\hat{I}}^{i\alpha}(T^{a})^{ij}\frac{\delta}{\delta\bar{\Theta}_{\alpha}^{j}}\bigg)(\Sigma+\epsilon\Sigma_{\mathrm{CT}})=\int d^{4}x\,\bigg(\tilde{G}\zeta_{\hat{I}}^{a}-B_{\hat{I}}\,\eta^{a}-\bar{B}_{\hat{I}}^{ba}\,\eta^{b}\bigg)+\mathcal{O}(\epsilon^{2})\,,\nonumber \\
 &  & \int d^{4}x\,\bigg(\frac{\delta}{\delta\bar{\lambda}^{a\hat{I}}}+\Lambda_{\hat{I}}^{i\alpha}(T^{a})^{ij}\frac{\delta}{\delta\bar{\Theta}_{\alpha}^{j}}\bigg)(\Sigma+\epsilon\Sigma_{\mathrm{CT}})=\int d^{4}x\,\bigg(\tilde{G}\lambda_{\hat{I}}^{a}+P\zeta_{\hat{I}}^{a}+Z_{\hat{I}}\,\eta^{a}\bigg)+\mathcal{O}(\epsilon^{2})\,,\nonumber \\
 &  & \mathcal{R}_{ij}(\Sigma+\epsilon\Sigma_{\mathrm{CT}})=\mathcal{O}(\epsilon^{2})\,,\nonumber \\
 &  & \bigg(\frac{\delta}{\delta\bar{\eta}^{a}}+\partial_{\mu}\frac{\delta}{\delta\Xi_{\mu}^{a}}\bigg)(\Sigma+\epsilon\Sigma_{\mathrm{CT}})=\mathcal{O}(\epsilon^{2})\,,\nonumber \\
 &  & \int d^{4}x\bigg(\frac{\delta}{\delta\eta^{a}}+f^{abc}\bar{\eta}^{b}\frac{\delta}{\delta\tau^{c}}-f^{abc}\Xi_{\mu}^{b}\frac{\delta}{\delta\mathcal{J}_{\mu}^{c}}\bigg)(\Sigma+\epsilon\Sigma_{\mathrm{CT}})=\int d^{4}x\bigg(-\bar{Y}^{abI}\varphi^{bI}+\bar{X}^{abI}\omega^{bI}+X\bar{\omega}^{aI}-Y^{I}\bar{\varphi}^{aI}\nonumber \\
 &  & +Z^{\hat{I}}\bar{\lambda}_{\hat{I}}^{a}-B^{\hat{I}}\bar{\zeta}_{\hat{I}}^{a}+\bar{Z}^{ab\hat{I}}\lambda_{\hat{I}}^{b}-\bar{B}^{ab\hat{I}}\zeta_{\hat{I}}^{b}+\Gamma^{ab}\eta^{b}+\Phi_{\alpha}^{i}\bar{\psi}^{(h)j\alpha}(T^{a})^{ji}-\bar{\Phi}^{i\alpha}(T^{a})^{ij}\psi_{\alpha}^{(h)j}\bigg)+\mathcal{O}(\epsilon^{2})\,,\nonumber \\
 &  & W_{(1,2,3,4)}^{I}(\Sigma+\epsilon\Sigma_{\mathrm{CT}})=\mathcal{O}(\epsilon^{2})\,.\nonumber \\
 &  & W_{(1,2,3,4)}^{\hat{I}}(\Sigma+\epsilon\Sigma_{\mathrm{CT}})=\mathcal{O}(\epsilon^{2})\,.\label{w17}
\end{eqnarray}
Looking at the first condition of eqs.\eqref{w17}, one gets 
\begin{equation}
\mathcal{B}_{\Sigma}\Sigma_{\mathrm{CT}}=0\,,\label{w18}
\end{equation}
where $\mathcal{B}_{\Sigma}$ is the linearized nilpotent Slavnov-Taylor operator  defined in \eqref{lsti}. The equation
\eqref{w18} means  that the invariant counterterm $\Sigma_{\mathrm{CT}}$
belongs to the cohomology of $\mathcal{B}_{\Sigma}$ in the space of
the integrated local polynomials in the fields and sources with $c$-ghost
number zero and bounded by dimension four. From the general results on the cohomology of Yang-Mills
theories,  see \cite{Piguet:1995er}, the most
general solution for $\ensuremath{\Sigma_{CT}}$ can be written as 
\begin{equation}
\Sigma_{\mathrm{CT}}=\Delta+\mathcal{B}_{\Sigma}\Delta^{(-1)}\,,\label{w20}
\end{equation}
with $\ensuremath{\Delta}$ and $\ensuremath{\Delta^{\left(-1\right)}}$ 
being the nontrivial and trivial solutions of  equation \eqref{w18}, respectively. In particular,  for the nontrivial term $\Delta$, we have: 
\begin{equation} 
\mathcal{B}_{\Sigma} \Delta=0 \;, \qquad  \Delta\neq{\cal B}_{\Sigma}T \;, \label{ntcoh} 
\end{equation} 
for some local integrated field polynomial $T$.
Note also that, according to the quantum numbers of the fields,
$\ensuremath{\Delta^{\left(-1\right)}}$ is an integrated polynomial
of dimension four, c-ghost number $-1$ and $\ensuremath{\eta}$-number
equals to zero. One can appreciate now the usefulness of the introduction of the generalized extended BRST operator $Q$. In fact, from eqs.\eqref{Q_nonlinear},\eqref{Q_doublets},\eqref{Q_singlets} one sees that the auxiliary fields and sources introduced to implement the restriction of the functional measure to the Gribov region in a local fashion transform as doublets\footnote{We remind here that a pair of variables $(\omega, \beta)$ is a doublet if 
\begin{equation} 
Q \omega = \beta\;, \qquad Q\beta=0 \;. \label{db}
\end{equation}
} under $Q$. As a consequence \cite{Piguet:1995er}, they can appear only in the  
 exact trivial part of the cohomology of ${\cal B}_{\Sigma}$,
that is they can enter  only in the term $\Delta^{(-1)}$. Therefore, excluding the doublet pairs, the most general expression  for $\Delta$
can be written as 
\begin{eqnarray}
\Delta & = & \int d^{4}x\,\bigg[\frac{a_{0}}{4g^{2}}F_{\mu\nu}^{a}F_{\mu\nu}^{a}+a_{1}J_{\psi}\,\bar{\psi}_{\alpha}^{i}\psi^{i,\alpha}+a_{2}(\partial_{\mu}(A^{h})_{\mu}^{a})(\partial_{\nu}(A^{h})_{\nu}^{a})+a_{3}(\partial_{\mu}(A^{h})_{\nu}^{a})(\partial_{\mu}(A^{h})_{\nu}^{a})\nonumber \\
 & + & a_{4}f^{abc}(A^{h})_{\mu}^{a}(A^{h})_{\nu}^{b}\partial_{\mu}(A^{h})_{\nu}^{c}+\pi^{abcd}(A^{h})_{\mu}^{a}(A^{h})_{\mu}^{b}(A^{h})_{\nu}^{c}(A^{h})_{\nu}^{d}+\hat{\mathcal{J}}_{\mu}^{a}\mathcal{O}_{\mu}^{a}(A,\xi)+(J+J_{\psi}^{2})\mathcal{O}(A,\xi)\nonumber \\
 & + & a_{5}(\partial_{\mu}\bar{\eta}^{a}+\Xi_{\mu}^{a})(\partial_{\mu}\eta^{a})+f^{abc}(\partial_{\mu}\bar{\eta}^{a}+\Xi_{\mu}^{a})\EuScript{P}_{\mu}^{b}(A,\xi)\eta^{c}+a_{6}\frac{\kappa_{1}}{2}J^{2}+\bar{\Theta}^{i,\alpha}\mathcal{F}_{\alpha}^{i}(\psi,\xi)\nonumber \\
 & + & \Theta^{i,\alpha}\bar{\mathcal{F}}_{\alpha}^{i}(\bar{\psi},\xi)+\bar{\Phi}^{i,\alpha}(T^{a})^{ij}\eta^{a}\mathcal{W}_{\alpha}^{j}(\psi,\xi)+\Phi^{i,\alpha}(T^{a})^{ij}\eta^{a}\bar{\mathcal{W}}_{\alpha}^{j}(\bar{\psi},\xi)+a_{7}\kappa_{2}J_{\psi}^{2}J+a_{8}\kappa_{3}J_{\psi}^{4}\bigg]\,,\label{w21}
\end{eqnarray}
where $(a_{0},a_{1},\ldots,a_{8},\pi)$ are arbitrary dimensionless
coefficients and  $\mathcal{O}_{\mu}^{a}(A,\xi)$, $\mathcal{O}(A,\xi)$,
$\EuScript{P}_{\mu}^{b}(A,\xi)$ are local expressions in the fields $A_{\mu}^{a}$ and $\xi^{a}$ with ghost number zero and
dimensions $(1, 1, 2)$, respectively. Moreover, $\mathcal{W}_{\alpha}^{i}(\psi,\xi)$
and $\bar{\mathcal{W}}_{\alpha}^{i}(\psi,\xi)$ are local functionals
of $\psi_{\alpha}^{i}$ and $\ensuremath{\xi^{a}}$ while 
$\bar{\mathcal{F}}_{\alpha}^{i}(\bar{\psi},\xi)$ and $\bar{\mathcal{W}}_{\alpha}^{i}(\bar{\psi},\xi)$
are local expressions of $\bar{\psi}_{\alpha}^{i}$ and $\ensuremath{\xi^{a}}$. Also, in expression \eqref{w21} use has been made of the fact that, from the Ward identity \eqref{wi4}, the variables $(\tau^{a},\mathcal{J}_{\mu}^{a})$
can  enter  only through the combination, 
\begin{equation}
\hat{\mathcal{J}}_{\mu}^{a}=\mathcal{J}_{\mu}^{a}-\partial_{\mu}\tau^{a}\;.\label{cb}
\end{equation}
It is worth observing that the quantities $(\mathcal{O}_{\mu}^{a}(A,\xi), \mathcal{O}(A,\xi),
\EuScript{P}_{\mu}^{b}(A,\xi))$ as well as $(\mathcal{W}_{\alpha}^{i}(\psi,\xi),
\bar{\mathcal{W}}_{\alpha}^{i}(\psi,\xi), \bar{\mathcal{F}}_{\alpha}^{i}(\bar{\psi},\xi),\bar{\mathcal{W}}_{\alpha}^{i}(\bar{\psi},\xi))$ are fully independent from the external sources. As a conseuqnece, the action of the linearized operator $\mathcal{B}_{\Sigma}$ reduces to that of the BRST operator $s$, namely, from eq.\eqref{db}, one gets 
\begin{eqnarray}
\mathcal{B}_{\Sigma}\mathcal{O}_{\mu}^{a}(A,\xi) & = & Q\mathcal{O}_{\mu}^{a}(A,\xi)=s\mathcal{O}_{\mu}^{a}(A,\xi)=0\,,\label{1}
\end{eqnarray}
\begin{eqnarray}
\mathcal{B}_{\Sigma}\mathcal{O}(A,\xi) & = & Q\mathcal{O}(A,\xi)=s\mathcal{O}(A,\xi)=0\,,\label{2}
\end{eqnarray}
\begin{eqnarray}
\mathcal{B}_{\Sigma}\EuScript{P}_{\mu}^{a}(A,\xi) & = & Q\EuScript{P}_{\mu}^{a}(A,\xi)=s\EuScript{P}_{\mu}^{a}(A,\xi)=0\,,\label{3}
\end{eqnarray}
\begin{eqnarray}
{\cal B}_{\Sigma}\mathcal{F}_{\alpha}^{i}(\psi,\xi) & = & Q\mathcal{F}_{\alpha}^{i}(\psi,\xi)=s\mathcal{F}_{\alpha}^{i}(\psi,\xi)=0\,,\label{4}
\end{eqnarray}
\begin{eqnarray}
{\cal B}_{\Sigma}\bar{\mathcal{F}}_{\alpha}^{i}(\bar{\psi},\xi) & = & Q\bar{\mathcal{F}}_{\alpha}^{i}(\bar{\psi},\xi)=s\bar{\mathcal{F}}_{\alpha}^{i}(\bar{\psi},\xi)=0\,,\label{5}
\end{eqnarray}
\begin{eqnarray}
{\cal B}_{\Sigma}\mathcal{W}_{\alpha}^{i}(\psi,\xi) & = & Q\mathcal{W}_{\alpha}^{i}(\psi,\xi)=s\mathcal{W}_{\alpha}^{i}(\psi,\xi)=0\,,\label{6}
\end{eqnarray}
\begin{eqnarray}
{\cal B}_{\Sigma}\bar{\mathcal{W}}_{\alpha}^{i}(\bar{\psi},\xi) & = & Q\bar{\mathcal{W}}_{\alpha}^{i}(\bar{\psi},\xi)=s\bar{\mathcal{W}}_{\alpha}^{i}(\bar{\psi},\xi)=0\,,\label{7}
\end{eqnarray}
which imply that $\mathcal{O}_{\mu}^{a}(A,\xi)$,
$\mathcal{O}(A,\xi)$, $\EuScript{P}_{\mu}^{b}(A,\xi)$, $\mathcal{F}_{\alpha}^{i}(\psi,\xi)$,
$\bar{\mathcal{F}}_{\alpha}^{i}(\bar{\psi},\xi)$, $\mathcal{W}_{\alpha}^{i}(\psi,\xi)$
and $\bar{\mathcal{W}}_{\alpha}^{i}(\bar{\psi},\xi)$ are BRST invariant. In 
\cite{Capri:2019drm}
and \cite{Fiorentini:2016rwx}, the general solution of the eqs. \eqref{1}-\eqref{5}
were obtained, yielding 
\begin{eqnarray}
\mathcal{O}_{\mu}^{a}(A,\xi) & = & b_{1}(A^{h})_{\mu}^{a}\,,
\end{eqnarray}
\begin{eqnarray}
\mathcal{O}(A,\xi) & = & \bigg(\frac{b_{2}+b_{2}^{\prime}}{2}\bigg)(A^{h})_{\mu}^{a}(A^{h})_{\mu}^{a}\,,
\end{eqnarray}
\begin{eqnarray}
\EuScript{P}_{\mu}^{a}(A,\xi) & = & b_{3}(A^{h})_{\mu}^{a}\,,
\end{eqnarray}
\begin{eqnarray}
\mathcal{F}_{\alpha}^{i}(\psi,\xi) & = & b_{4}\psi_{\alpha}^{(h)i}\,,
\end{eqnarray}
\begin{eqnarray}
\bar{\mathcal{F}}_{\alpha}^{i}(\bar{\psi},\xi) & = & b_{5}\bar{\psi}{}_{\alpha}^{(h)i}\,,
\end{eqnarray}
with $(b_{1},b_{2},b_{2}^{\prime},b_{3},b_{4},b_{5})$ free dimensionless
parameters. \\\\Let us consider thus the equations \eqref{6} and \eqref{7}, that
is 
\begin{eqnarray}
s\mathcal{W}_{\alpha}^{i}(\psi,\xi)=0\,,\label{17}
\end{eqnarray}
\begin{eqnarray}
s\bar{\mathcal{W}}_{\alpha}^{i}(\bar{\psi},\xi)=0\,,\label{18}
\end{eqnarray}
It is immediate to realize that they have the same structure of eqs.\eqref{4},\eqref{5}. As a consequence, they can be solved by repeating exactly the same analysis presented in \cite{Capri:2019drm}. Skipping the intermediate algebraic manipulations, for $(\mathcal{W}_{\alpha}^{i}(\psi,\xi), \bar{\mathcal{W}_{\alpha}^{i}}(\bar{\psi},\xi))$  we get 
\begin{eqnarray}
\mathcal{W}_{\alpha}^{i}(\psi,\xi) & = & b_{6}\psi_{\alpha}^{(h)i}\,; \nonumber \\
\bar{\mathcal{W}_{\alpha}^{i}}(\bar{\psi},\xi) & = & b_{7}\bar{\psi}{}_{\alpha}^{(h)i}    \;,\label{b6b7}
\end{eqnarray} 
with $(b_6,b_7)$ constant free parameters. Therefore, after imposing the constraints
\eqref{1}-\eqref{6}, for the  cohomological non-trivial term  $\Delta$ we get 
\begin{eqnarray}
\Delta & = & \int d^{4}x\bigg[\frac{a_{0}}{4g^{2}}F_{\mu\nu}^{a}F_{\mu\nu}^{a}+a_{1}J_{\psi}\,\bar{\psi}_{\alpha}^{i}\psi^{i,\alpha}+a_{2}(\partial_{\mu}(A^{h})_{\mu}^{a})(\partial_{\nu}(A^{h})_{\nu}^{a})+a_{3}(\partial_{\mu}(A^{h})_{\nu}^{a})(\partial_{\mu}(A^{h})_{\nu}^{a})\nonumber \\
 & + & a_{4}f^{abc}(A^{h})_{\mu}^{a}(A^{h})_{\nu}^{b}\partial_{\mu}(A^{h})_{\nu}^{c}+\pi^{abcd}(A^{h})_{\mu}^{a}(A^{h})_{\mu}^{b}(A^{h})_{\nu}^{c}(A^{h})_{\nu}^{d}+b_{1}\hat{\mathcal{J}}_{\mu}^{a}(A^{h})_{\mu}^{a}\nonumber \\
 & + & \bigg(b_{2}\frac{J}{2}+b_{2}^{\prime}\frac{J_{\psi}^{2}}{2}\bigg)(A^{h})_{\mu}^{a}(A^{h})_{\mu}^{a}+a_{5}(\partial_{\mu}\bar{\eta}^{a}+\Xi_{\mu}^{a})(\partial_{\mu}\eta^{a})+b_{3}f^{abc}(\partial_{\mu}\bar{\eta}^{a}+\Xi_{\mu}^{a})(A^{h})_{\nu}^{b}\eta^{c}\nonumber \\
 & + & a_{6}\frac{\kappa_{1}}{2}J^{2}+b_{4}\bar{\Theta}^{i,\alpha}\psi_{\alpha}^{(h)i}+b_{5}\Theta^{i,\alpha}\bar{\psi}{}_{\alpha}^{(h)i}+b_{6}\bar{\Phi}^{i,\alpha}(T^{a})^{ij}\eta^{a}\psi_{\alpha}^{(h)i}+b_{7}\Phi^{i,\alpha}(T^{a})^{ij}\eta^{a}\bar{\psi}{}_{\alpha}^{(h)i}\nonumber\\
 &+&a_{7}\kappa_{2}J_{\psi}^{2}J+a_{8}\kappa_{3}J_{\psi}^{4}\bigg]\,.\label{w25}
\end{eqnarray}
Let us also point out here  that the parameters $(\alpha,\chi)$ form a
$Q$-doublet, so that  they cannot appear in the nontrivial
sector of the $Q$-cohomology, meaning that these parameters are not
present in $\Delta$. A further reduction of the free parameters appearing in the expression \eqref{w25} is possible by making use of the following argument.  If the
values of the extra sources are set to zero as 
\begin{eqnarray}
J=\tilde{J}=M=N=V=U=H=\chi=K=\mathcal{J}=\Xi=\Gamma=X=Y=\bar{X}=\bar{Y} & = & 0\label{lim1}
\end{eqnarray}
as well as 
\begin{eqnarray}
\varUpsilon=\bar{\varUpsilon}=\Lambda=\bar{\Lambda}=\Theta=\bar{\Theta}=\Phi=\bar{\Phi}= L= \Omega & = & 0,\label{lim2}
\end{eqnarray}
the tree level complete action $\Sigma$ reduces to 
\begin{equation} 
\Sigma \rightarrow \Sigma_{\mathrm{LCG}M} \;, \label{redsig}
\end{equation} 
where 
\begin{eqnarray}
\Sigma_{\mathrm{LCG}M} & = & \int d^{4}x~\bigg[\frac{1}{4g^{2}}F_{\mu\nu}^{a}F_{\mu\nu}^{a}+i\bar{\psi}^{i,\alpha}(\gamma_{\mu})_{\alpha\beta}D_{\mu}^{ij}\psi^{j,\beta}-J_{\psi}\bar{\psi}_{\alpha}^{i}\psi^{i,\alpha}+ib^{a}\partial_{\mu}A_{\mu}^{a}+\frac{\alpha}{2}b^{a}b^{a}+\bar{c}^{a}\partial_{\mu}D_{\mu}^{ab}c^{b}\nonumber \\
 & - & \bar{\varphi}_{\mu}^{ac}\mathcal{M}(A^{h}){\varphi}_{\mu}^{bc}+\bar{\omega}_{\mu}^{ac}\mathcal{M}^{ab}(A^{h})\omega_{\mu}^{bc}+\tau^{a}\partial_{\mu}(A^{h})_{\mu}^{a}-\bar{\eta}^{a}\mathcal{M}^{ab}(A^{h})\eta^{b}+\bar{\lambda}_{\alpha}^{ai}\bigg(-\partial_{\mu}D_{\mu}^{ab}(A^{h})\bigg)\lambda^{\alpha,bi}\nonumber \\
 & + & \bar{\zeta}^{a\hat{I}}\bigg(-\partial_{\mu}D_{\mu}^{ab}(A^{h})\bigg)\zeta_{\hat{I}}^{b}+\sigma^{\frac{3}{2}}\bigg(\bar{\lambda}_{\alpha}^{ai}(T^{a})^{ij}\psi^{h,j\alpha}+\bar{\psi}_{\alpha}^{h,i}(T^{a})^{ij}\lambda^{aj\alpha}\bigg)\bigg]\,,
\end{eqnarray}
is nothing but the Yang-Mills action with the  fermionic matter fields 
 in the fundamental representation,  gauge-fixed in the  (LCG) linear covariant gauge,  with the addition of the following terms: 
\begin{equation}
-\int d^{4}x\bigg(\bar{\varphi}_{\mu}^{ac}\mathcal{M}^{ab}(A^{h}){\varphi}_{\mu}^{bc}-\bar{\omega}_{\mu}^{ac}\mathcal{M}^{ab}(A^{h})\omega_{\mu}^{bc}+\tau^{a}\partial_{\mu}(A^{h})_{\mu}^{a}-\bar{\eta}^{a}\mathcal{M}^{ab}(A^{h})\eta^{b}\bigg)\,,\label{w27}
\end{equation}
\begin{equation}
-\int d^{4}x\bigg(\bar{\lambda}_{\alpha}^{ai}\partial_{\mu}D_{\mu}^{ab}(A^{h})\lambda^{\alpha,bi}+\bar{\zeta}^{a\hat{I}}\partial_{\mu}D_{\mu}^{ab}(A^{h})\zeta_{\hat{I}}^{b}-\sigma^{\frac{3}{2}}\bar{\lambda}_{\alpha}^{ai}(T^{a})^{ij}\psi^{h,j\alpha}-\sigma^{\frac{3}{2}}\bar{\psi}_{\alpha}^{h,i}(T^{a})^{ij}\lambda^{aj\alpha}\bigg)\,.\label{w27-1}
\end{equation}
Nevertheless, upon integration over $(\bar{\varphi},\varphi,\bar{\omega},\omega,\tau,\bar{\eta},\eta,\lambda,\bar{\lambda},\zeta,\bar{\zeta})$,
the terms \eqref{w27} and \eqref{w27-1} give rise to  to a unity. As a consequence,
the correlation functions of the original fields $(A,\bar{c},c,b,\psi,{\bar \psi})$ are the same as those
computed with the standard Yang-Mills action in the linear covariant gauges, eq.\eqref{rgzlcg1}, supplemented with the usual  spinor matter term. From this remark, it follows that, in
the limits \eqref{lim1} and \eqref{lim2}, the counterterm \eqref{w25}
 reduces to the well-known standard Yang-Mills counterterm  in linear covariant gauges, see for instance \cite{Fiorentini:2016rwx,Capri:2018gpu,Capri:2019drm}. This implies that 
\begin{equation}
a_{2}=a_{3}=a_{4}=0\,,\qquad a_{5}=b_{3}\,,\qquad\pi^{abcd}=0\,,\label{w28}
\end{equation}
yielding 
\begin{eqnarray}
\Delta & = & \int d^{4}x\bigg[\frac{a_{0}}{4g^{2}}F_{\mu\nu}^{a}F_{\mu\nu}^{a}+b_{1}\hat{\mathcal{J}}_{\mu}^{a}(A^{h})_{\mu}^{a}+\bigg(b_{2}\frac{J}{2}+b_{2}^{\prime}\frac{J_{\psi}^{2}}{2}\bigg)(A^{h})_{\mu}^{a}(A^{h})_{\mu}^{a}+b_{3}(\partial_{\mu}\bar{\eta}^{a}+\Xi_{\mu}^{a})D_{\mu}^{ab}(A^{h})\eta^{b}\nonumber \\
 & + & a_{6}\frac{\kappa_{1}}{2}J^{2}+b_{4}\bar{\Theta}\psi_{\alpha}^{(h)i}+b_{5}\Theta\bar{\psi}{}_{\alpha}^{(h)i}+b_{6}\bar{\Phi}^{i,\alpha}(T^{a})^{ij}\eta^{a}\psi_{\alpha}^{(h)i}+b_{7}\Phi^{i,\alpha}(T^{a})^{ij}\eta^{a}\bar{\psi}{}_{\alpha}^{(h)i}+a_{7}\kappa_{2}J_{\psi}^{2}J
 \nonumber\\
 &+&a_{8}\kappa_{3}J_{\psi}^{4}\bigg]\,;\label{189}
\end{eqnarray}
where, from the Ward identity \eqref{w16}, use has been made of the  constraint
\begin{equation}
a_{5}=b_3=-b_{1}\,.\label{w30}
\end{equation}
Let us turn now our attention to the exact part, $\Delta^{(-1)}$, of the cohomology of the linearized operator 
$\mathcal{B}_{\Sigma}$. The term $\Delta^{(-1)}$  is a local integrated 
expression in the fields and sources with $c$-ghost number $(-1)$. Moreover,
by considering the quantum numbers of all fields and sources given
in  tables (I -- \ref{table60}) as well as  the
set of constraints \eqref{w17}, the most general  expression
for $\Delta^{(-1)}$ can be written down as 
\begin{eqnarray}
\Delta^{(-1)} & = & \int d^{4}x\,\bigg[f_{1}^{ab}(\xi,\alpha)(\Omega_{\mu}^{a}+\partial_{\mu}\bar{c}^{a})A_{\mu}^{b}+f_{2}^{ab}(\xi,\alpha)c^{a}L^{b}+K^{a}f^{ab}(\xi,\alpha)\xi^{b}\nonumber \\
 & - & b_{1}\left(V_{\mu}^{aI}N_{\mu}^{aI}+V_{\mu}^{aI}D_{\mu}^{ab}(A^{h})\bar{\omega}^{bI}+N_{\mu}^{aI}D_{\mu}^{ab}(A^{h})\varphi^{bI}+(\partial_{\mu}\bar{\omega}^{aI})D_{\mu}^{ab}(A^{h})\varphi^{bI}\right)\nonumber \\
 & + & f_{3}(\xi,\alpha)\bigg(\bar{\psi}^{i,\alpha}\varUpsilon{}_{\alpha}^{i}+\bar{\varUpsilon}{}^{i,\alpha}\psi_{\alpha}^{i}\bigg)+b_{4}\bigg((T^{a})^{ij}\psi^{h,i\alpha}\bar{\zeta}^{a\hat{I}}\Lambda_{\alpha\hat{I}}^{j}+(T^{a})^{ij}\bar{\psi}^{h,i\alpha}\lambda^{a\hat{I}}\bar{\Pi}_{\alpha\hat{I}}^{j}\bigg)\nonumber \\
 & + & f_{4}(\xi,\alpha)(\partial^{2}\lambda^{a\hat{I}})\bar{\zeta}_{\hat{I}}^{a}+f_{5}\left(\xi,\alpha\right)\varrho J_{\psi}\Lambda^{i\alpha\hat{I}}\bar{\Pi}_{\alpha\hat{I}}^{i}\bigg]\,,
\end{eqnarray}
with $f_{1}^{ab}(\xi,\alpha)$, $f_{2}^{ab}(\xi,\alpha)$, $f_{3}\left(\xi,\alpha\right),f_{4}\left(\xi,\alpha\right),f_{5}\left(\xi,\alpha\right)$
and $f^{ab}(\xi,\alpha)$ being arbitrary functions of $\xi^{a}$ and of the gauge parameter $\alpha$.
Invoking again the limits \eqref{lim1} and \eqref{lim2}, we are
able to conclude that 
\begin{equation}
f_{1}^{ab}(\xi,\alpha)=\delta^{ab}d_{1}\,,\qquad f_{2}^{ab}(\xi,\alpha)=\delta^{ab}d_{2}\,,\qquad f_{3}(\xi,\alpha)=d_{3},\qquad f_{4}(\xi,\alpha)=d_{4},\qquad f_{5}(\xi,\alpha)=d_{5},\label{w32}
\end{equation}
where $(d_{1},d_{2},d_{3},d_{4},d_{5})$ are free constant parameters which
might  depend on the gauge parameter $\alpha$. Acting  thus with the operator  $\mathcal{B}_{\Sigma}$ on $\Delta^{(-1)}$, we obtain 
\begin{eqnarray}
\mathcal{B}_{\Sigma}\Delta^{(-1)} & = & \int d^{4}x\,\bigg\{ d_{1}\bigg(\frac{\delta\Sigma}{\delta A_{\mu}^{a}}+i\partial_{\mu}b^{a}A_{\mu}^{a}\bigg)-d_{1}(\Omega_{\mu}^{a}+\partial_{\mu}\bar{c}^{a})\frac{\delta\Sigma}{\delta\Omega_{\mu}^{a}}+d_{2}\left(\frac{\delta\Sigma}{\delta L^{a}}L^{a}+\frac{\delta\Sigma}{\delta c^{a}}c^{a}\right)+\frac{\delta\Sigma}{\delta\xi^{a}}f^{ab}(\xi)\xi^{b}\nonumber \\
 & - & K^{b}\frac{\delta\Sigma}{\delta K^{a}}\bigg(\frac{\partial f^{bc}}{\partial\xi^{a}}\xi^{c}+f^{ba}(\xi)\bigg)+d_{3}\bigg(\bar{\psi}^{i,\alpha}\varUpsilon{}_{\alpha}^{i}+\bar{\varUpsilon}{}^{i,\alpha}\psi_{\alpha}^{i}\bigg)+b_{1}f^{abc}(A^{h})_{\mu}^{c}\bigg(U_{\mu}^{aI}\bar{\omega}^{bI}+V_{\mu}^{aI}\bar{\varphi}^{bI}+M_{\mu}^{aI}{\varphi}^{bI}\nonumber \\
 & - & N_{\mu}^{aI}{\omega}^{bI}-\omega^{aI}\partial_{\mu}\bar{\omega}^{bI}-\varphi^{aI}\partial_{\mu}\bar{\varphi}^{bI}\bigg)-b_{1}\bigg(U_{\mu}^{aI}N_{\mu}^{aI}+V_{\mu}^{aI}M_{\mu}^{aI}+U_{\mu}^{aI}\partial_{\mu}\bar{\omega}^{aI}+V_{\mu}^{aI}\partial_{\mu}\bar{\varphi}^{aI}+M_{\mu}^{aI}\partial_{\mu}{\varphi}^{aI}\nonumber \\
 & - & N_{\mu}^{aI}\partial_{\mu}{\omega}^{aI}+(\partial_{\mu}\bar{\varphi}^{aI})\partial_{\mu}\varphi^{aI}-(\partial_{\mu}\bar{\omega}^{aI})\partial_{\mu}\omega^{aI}+\bar{\lambda}^{a\hat{I}}\partial^{2}\lambda_{\hat{I}}^{a}+\zeta^{a\hat{I}}\partial^{2}\bar{\zeta}_{\hat{I}}^{a}\bigg)+b_{4}\bigg(\bar{\lambda}^{a\hat{I}}(T^{a})^{ij}\psi^{h,i\alpha}\Lambda_{\alpha\hat{I}}^{j}\nonumber \\
 & - & (T^{a})^{ij}\Pi^{j\alpha\hat{I}}\psi_{\alpha}^{h,i}\bar{\zeta}_{\hat{I}}^{a}-\zeta^{a\hat{I}}(T^{a})^{ij}\bar{\psi}^{h,i\alpha}\bar{\Pi}_{\alpha\hat{I}}^{j}+(T^{a})^{ij}\bar{\Lambda}^{j\alpha\hat{I}}\bar{\psi}_{\alpha}^{h,i}\lambda_{\hat{I}}^{a}\bigg)+d_{5}\varrho J_{\psi}\bigg(\bar{\Lambda}^{i\alpha\hat{I}}\Lambda_{\alpha\hat{I}}^{i}+\Pi^{i\alpha\hat{I}}\bar{\Pi}_{\alpha\hat{I}}^{i}\bigg)\nonumber \\
 & + & \chi\frac{\partial d_{1}}{\partial\alpha}(\Omega_{\mu}^{a}+\partial_{\mu}\bar{c}^{a})A_{\mu}^{a}+\chi\frac{\partial d_{2}}{\partial\alpha}L^{a}c^{a}+\chi K^{a}\frac{\partial f^{ab}}{\partial\alpha}\xi^{b}+\chi\bar{\zeta}_{\hat{I}}^{a}\frac{\partial d_{4}}{\partial\alpha}(\partial^{2}\lambda^{a\hat{I}})+\chi\varrho J_{\psi}\frac{\partial d_{5}}{\partial\alpha}\Lambda^{i\alpha\hat{I}}\bar{\Pi}_{\alpha\hat{I}}^{i}\bigg\}\,.\label{190}
\end{eqnarray}
Substituting  expressions \eqref{189} and \eqref{190} in equation
\eqref{w20}, the most general allowed local invariant counterterm turns out to be: 
\begin{eqnarray}
\Sigma_{\mathrm{CT}} & = & \Delta+\mathcal{B}_{\Sigma}\Delta^{(-1)}\nonumber \\
 & = & \int d^{4}x\bigg\{\frac{a_{0}}{4g^{2}}F_{\mu\nu}^{a}F_{\mu\nu}^{a}+a_{1}J_{\psi}\bar{\psi}^{i,\alpha}\psi_{\alpha}^{i}+b_{1}\bigg(\hat{\mathcal{J}}_{\mu}^{a}(A^{h})_{\mu}^{a}-(\partial_{\mu}\bar{\eta}^{a}+\Xi_{\mu}^{a})D_{\mu}^{ab}(A^{h})\eta^{b}\bigg)\nonumber \\
 & + & \bigg(b_{2}\frac{J}{2}+b_{2}^{\prime}\frac{J_{\psi}^{2}}{2}\bigg)(A^{h})_{\mu}^{a}(A^{h})_{\mu}^{a}+b_{1}\frac{\kappa_{1}}{2}J^{2}+a_{7}\kappa_{2}J_{\psi}^{2}J+a_{8}\kappa_{3}J_{\psi}^{4}
 +\frac{\delta\Sigma}{\delta\xi^{a}}f^{ab}(\xi)\xi^{b}\nonumber \\
 & + & d_{1}\bigg(\frac{\delta\Sigma}{\delta A_{\mu}^{a}}+i\partial_{\mu}b^{a}A_{\mu}^{a}-(\Omega_{\mu}^{a}+\partial_{\mu}\bar{c}^{a})\frac{\delta\Sigma}{\delta\Omega_{\mu}^{a}}\bigg)+d_{2}\left(\frac{\delta\Sigma}{\delta L^{a}}L^{a}+\frac{\delta\Sigma}{\delta c^{a}}c^{a}\right)\nonumber
\\
 & - & K^{b}\frac{\delta\Sigma}{\delta K^{a}}\bigg(\frac{\partial f^{bc}}{\partial\xi^{a}}\xi^{c}+f^{ba}(\xi)\bigg)+d_{3}\bigg(\bar{\psi}^{i,\alpha}\varUpsilon{}_{\alpha}^{i}+\bar{\varUpsilon}{}^{i,\alpha}\psi_{\alpha}^{i}\bigg)+b_{1}f^{abc}(A^{h})_{\mu}^{c}\bigg(U_{\mu}^{aI}\bar{\omega}^{bI}+\nonumber \\
 & + & V_{\mu}^{aI}\bar{\varphi}^{bI}+M_{\mu}^{aI}{\varphi}^{bI}-N_{\mu}^{aI}{\omega}^{bI}-\omega^{aI}\partial_{\mu}\bar{\omega}^{bI}-\varphi^{aI}\partial_{\mu}\bar{\varphi}^{bI}\bigg)\nonumber \\
 & - & b_{1}\bigg(U_{\mu}^{aI}N_{\mu}^{aI}+V_{\mu}^{aI}M_{\mu}^{aI}+U_{\mu}^{aI}\partial_{\mu}\bar{\omega}^{aI}+V_{\mu}^{aI}\partial_{\mu}\bar{\varphi}^{aI}+M_{\mu}^{aI}\partial_{\mu}{\varphi}^{aI}-N_{\mu}^{aI}\partial_{\mu}{\omega}^{aI}\nonumber \\
 & + & (\partial_{\mu}\bar{\varphi}^{aI})\partial_{\mu}\varphi^{aI}-(\partial_{\mu}\bar{\omega}^{aI})\partial_{\mu}\omega^{aI}+\bar{\lambda}^{a\hat{I}}\partial^{2}\lambda_{\hat{I}}^{a}+\zeta^{a\hat{I}}\partial^{2}\bar{\zeta}_{\hat{I}}^{a}\bigg)+b_{4}\bigg(\bar{\Theta}\psi_{\alpha}^{(h)i}+\Theta\bar{\psi}{}_{\alpha}^{(h)i}\nonumber \\
 & + & \bar{\lambda}^{a\hat{I}}(T^{a})^{ij}\psi^{h,i\alpha}\Lambda_{\alpha\hat{I}}^{j}-(T^{a})^{ji}\Pi^{i\alpha\hat{I}}\psi_{\alpha}^{h,j}\bar{\zeta}_{\hat{I}}^{a}-\zeta^{a\hat{I}}(T^{a})^{ij}\bar{\psi}^{h,i\alpha}\bar{\Pi}_{\alpha\hat{I}}^{j}+(T^{a})^{ji}\bar{\Lambda}^{i\alpha\hat{I}}\bar{\psi}_{\alpha}^{h,j}\lambda_{\hat{I}}^{a}\nonumber \\
 & + & \bar{\Phi}^{i,\alpha}(T^{a})^{ij}\eta^{a}\psi_{\alpha}^{(h)j}+\Phi^{i,\alpha}(T^{a})^{ij}\eta^{a}\bar{\psi}{}_{\alpha}^{(h)j}\bigg)+d_{5}\varrho J_{\psi}\bigg(\bar{\Lambda}^{i\alpha\hat{I}}\Lambda_{\alpha\hat{I}}^{i}+\Pi^{i\alpha\hat{I}}\bar{\Pi}_{\alpha\hat{I}}^{i}\bigg)\nonumber \\
 & + & \chi\frac{\partial d_{1}}{\partial\alpha}(\Omega_{\mu}^{a}+\partial_{\mu}\bar{c}^{a})A_{\mu}^{a}+\chi\frac{\partial d_{2}}{\partial\alpha}L^{a}c^{a}+\chi K^{a}\frac{\partial f^{ab}}{\partial\alpha}\xi^{b}+\chi\bar{\zeta}_{\hat{I}}^{a}\frac{\partial d_{4}}{\partial\alpha}(\partial^{2}\lambda^{a\hat{I}})\nonumber \\
 & + & \chi\varrho J_{\psi}\frac{\partial d_{5}}{\partial\alpha}\Lambda^{i\alpha\hat{I}}\bar{\Pi}_{\alpha\hat{I}}^{i}\bigg\}\,.\label{w34}
\end{eqnarray}
Having found the most general local counterterm, eq.\eqref{w34}, compatible with all Ward identities of the tree level action $\Sigma$, the final step in the proof of the renormalizability of $\Sigma$ is to check out the stability of $\Sigma$ \cite{Piguet:1995er}, that is to  show that the counterterm $\Sigma^{\mathrm{CT}}$ can be reabsorbed in the starting action $\Sigma$ through a redefinition of the fields, parameters and sources. This final step is greatly simplified by re-casting expression \eqref{w34} in the so-called parametric form \cite{Piguet:1995er}, namely
\begin{eqnarray}
\Sigma^{\mathrm{CT}} & = & -a_{0}g^{2}\frac{\partial\Sigma}{\partial g^{2}}+2d_{1}\alpha\frac{\partial\Sigma}{\partial\alpha}+(b_{1}+2b_{2})\kappa_{1}\frac{\partial\Sigma}{\partial\kappa_{1}}+[(a_{7}+2a_{1}+b_{2})\kappa_{2}+b_{2}^{\prime}\kappa_{1}]\frac{\partial\Sigma}{\partial\kappa_{2}}\nonumber \\
 & + & [(a_{8}+4a_{1})\kappa_{3}+b_{2}^{\prime}\kappa_{2}]\frac{\partial\Sigma}{\partial\kappa_{3}}+a_{1}\int d^{4}x\,J_{\psi}\frac{\partial\Sigma}{\partial J_{\psi}}+b_{1}\int d^{4}x\bigg[\mathcal{J}_{\mu}^{a}\frac{\delta\Sigma}{\delta\mathcal{J}_{\mu}^{a}}+\tau^{a}\frac{\delta\Sigma}{\delta\tau^{a}}\nonumber \\
 & + & \frac{1}{2}\bigg(\bar{\eta}^{a}\frac{\delta\Sigma}{\delta\bar{\eta}^{a}}+\eta^{a}\frac{\delta\Sigma}{\delta\eta^{a}}+\Xi_{\mu}^{a}\frac{\delta\Sigma}{\delta\Xi_{\mu}^{a}}-\Gamma^{ab}\frac{\delta\Sigma}{\delta\Gamma^{ab}}\bigg)\bigg]+d_{3}\int d^{4}x\bigg(\frac{\delta\Sigma}{\delta\varUpsilon_{\alpha}{}^{i}}\varUpsilon_{\alpha}{}^{i}+\bar{\varUpsilon}_{\alpha}^{i}\frac{\delta\Sigma}{\delta\bar{\varUpsilon}_{\alpha}^{i}}\nonumber \\
 & - & \frac{\delta\Sigma}{\delta\psi_{\alpha}{}^{i}}\psi_{\alpha}{}^{i}-\bar{\psi}_{\alpha}^{i}\frac{\delta\Sigma}{\delta\bar{\psi}_{\alpha}^{i}}\bigg)+\int d^{4}x~\bigg(b_{2}J+b_{2}^{\prime}J_{\psi}^{2}\bigg)\frac{\delta\Sigma}{\delta J}-d_{1}\int d^{4}x~b^{a}\frac{\delta\Sigma}{\delta b^{a}}-d_{1}\int d^{4}x~\bar{c}^{a}\frac{\delta\Sigma}{\delta\bar{c}^{a}}\nonumber \\
 & + & \int d^{4}x\bigg[d_{1}A_{\mu}^{a}\frac{\delta\Sigma}{\delta A_{\mu}^{a}}-d_{1}\Omega_{\mu}^{a}\frac{\delta\Sigma}{\delta\Omega_{\mu}^{a}}+d_{2}L^{a}\frac{\delta\Sigma}{\delta L^{a}}+d_{2}c^{a}\frac{\delta\Sigma}{\delta c^{a}}+f^{ab}(\xi)\xi^{b}\frac{\delta\Sigma}{\delta\xi^{a}}\nonumber \\
 & - & K^{a}\frac{\delta\Sigma}{\delta K^{a}}\bigg(\frac{\partial f^{bc}}{\partial\xi^{a}}\xi^{c}+f^{ba}(\xi)\bigg)\bigg]-\frac{b_{1}}{2}\int d^{4}x\bigg[\bar{\varphi}^{aI}\frac{\delta\Sigma}{\delta\bar{\varphi}^{aI}}+\varphi^{aI}\frac{\delta\Sigma}{\delta\varphi^{aI}}+\bar{\omega}^{aI}\frac{\delta\Sigma}{\delta\bar{\omega}^{aI}}+\omega^{aI}\frac{\delta\Sigma}{\delta\omega^{aI}}\nonumber \\
 & + & M_{\mu}^{aI}\frac{\delta\Sigma}{\delta M_{\mu}^{aI}}+V_{\mu}^{aI}\frac{\delta\Sigma}{\delta V_{\mu}^{aI}}+N_{\mu}^{aI}\frac{\delta\Sigma}{\delta N_{\mu}^{aI}}+U_{\mu}^{aI}\frac{\delta\Sigma}{\delta U_{\mu}^{aI}}+2\tilde{J}\frac{\delta\Sigma}{\delta\tilde{J}}+2H\frac{\delta\Sigma}{\delta H}+2\tilde{G}\frac{\delta\Sigma}{\delta\tilde{G}}\nonumber \\
 & + & 2P\frac{\delta\Sigma}{\delta P}+2\varrho\frac{\partial\Sigma}{\partial\varrho}\bigg]+\int d^{4}x\bigg[\bigg(\frac{d_{5}+b_{4}-b_{1}}{2}\bigg)\bar{\lambda}^{a\hat{I}}\frac{\delta\Sigma}{\delta\bar{\lambda}_{\hat{I}}^{a}}+\bigg(\frac{d_{5}-b_{1}-b_{4}}{2}\bigg)\lambda^{a\hat{I}}\frac{\delta\Sigma}{\delta\lambda_{\hat{I}}^{a}}\nonumber 
  \end{eqnarray}
 \begin{eqnarray}
 & + & \bigg(\frac{b_{1}+b_{4}+d_{5}}{2}\bigg)\bigg(\bar{\zeta}^{a\hat{I}}\frac{\delta\Sigma}{\delta\bar{\zeta}_{\hat{I}}^{a}}+\bar{\Pi}^{i\alpha\hat{I}}\frac{\delta\Sigma}{\delta\bar{\Pi}_{\alpha\hat{I}}^{i}}+\Lambda^{i\alpha\hat{I}}\frac{\delta\Sigma}{\delta\Lambda_{\alpha\hat{I}}^{i}}+\bar{\Lambda}^{i\alpha\hat{I}}\frac{\delta\Sigma}{\delta\bar{\Lambda}_{\alpha\hat{I}}^{i}}\bigg)\nonumber \\
 & + & \bigg(\frac{b_{1}+d_{5}-b_{4}}{2}\bigg)\bigg(\zeta^{a\hat{I}}\frac{\delta\Sigma}{\delta\zeta_{\hat{I}}^{a}}+\Pi^{i\alpha\hat{I}}\frac{\delta\Sigma}{\delta\Pi_{\alpha\hat{I}}^{i}}\bigg)-\bigg(\frac{b_{4}+d_{5}}{2}\bigg)\bigg(Z^{\hat{I}}\frac{\delta\Sigma}{\delta Z_{\hat{I}}}+B^{\hat{I}}\frac{\delta\Sigma}{\delta B_{\hat{I}}}\bigg)\nonumber \\
 & + & \bigg(\frac{b_{4}-d_{5}}{2}\bigg)\bigg(\bar{Z}^{ab\hat{I}}\frac{\delta\Sigma}{\delta\bar{Z}_{\hat{I}}^{ab}}+\bar{B}^{ab\hat{I}}\frac{\delta\Sigma}{\delta\bar{B}_{\hat{I}}^{ab}}\bigg)+b_{4}\int d^{4}x\,\bigg[\Theta^{i\alpha}\frac{\delta\Sigma}{\delta\Theta_{\alpha}^{i}}+\bar{\Theta}^{i\alpha}\frac{\delta\Sigma}{\delta\bar{\Theta}_{\alpha}^{i}}+\Phi^{i\alpha}\frac{\delta\Sigma}{\delta\Phi_{\alpha}^{i}}+\bar{\Phi}^{i\alpha}\frac{\delta\Sigma}{\delta\bar{\Phi}_{\alpha}^{i}}\bigg]\,,\nonumber \\ \label{w48-1-1-1}
\end{eqnarray}
or, equivalently: 
\begin{equation}
\Sigma^{\mathrm{CT}}=\mathcal{R}\Sigma\,,\label{w49}
\end{equation}
with $\mathcal{R}$ being the operator
\begin{eqnarray}
\mathcal{R} & = & -a_{0}g^{2}\frac{\partial}{\partial g^{2}}+2d_{1}\alpha\frac{\partial}{\partial\alpha}+(b_{1}+2b_{2})\kappa_{1}\frac{\partial}{\partial\kappa_{1}}+[(a_{7}+2a_{1}+b_{2})\kappa_{2}+b_{2}^{\prime}\kappa_{1}]\frac{\partial}{\partial\kappa_{2}}\nonumber \\
 & + & [(a_{8}+4a_{1})\kappa_{3}+b_{2}^{\prime}\kappa_{2}]\frac{\partial}{\partial\kappa_{3}}+a_{1}\int d^{4}x\,J_{\psi}\frac{\partial}{\partial J_{\psi}}+b_{1}\int d^{4}x\bigg[\mathcal{J}_{\mu}^{a}\frac{\delta}{\delta\mathcal{J}_{\mu}^{a}}+\tau^{a}\frac{\delta}{\delta\tau^{a}}\nonumber \\
 & + & \frac{1}{2}\bigg(\bar{\eta}^{a}\frac{\delta}{\delta\bar{\eta}^{a}}+\eta^{a}\frac{\delta}{\delta\eta^{a}}+\Xi_{\mu}^{a}\frac{\delta}{\delta\Xi_{\mu}^{a}}-\Gamma^{ab}\frac{\delta}{\delta\Gamma^{ab}}\bigg)\bigg]+d_{3}\int d^{4}x\bigg(\frac{\delta}{\delta\varUpsilon_{\alpha}{}^{i}}\varUpsilon_{\alpha}{}^{i}+\bar{\varUpsilon}_{\alpha}^{i}\frac{\delta}{\delta\bar{\varUpsilon}_{\alpha}^{i}}\nonumber \\
 & - & \frac{\delta}{\delta\psi_{\alpha}{}^{i}}\psi_{\alpha}{}^{i}-\bar{\psi}_{\alpha}^{i}\frac{\delta}{\delta\bar{\psi}_{\alpha}^{i}}\bigg)+\int d^{4}x~\bigg(b_{2}J+b_{2}^{\prime}J_{\psi}^{2}\bigg)\frac{\delta}{\delta J}-d_{1}\int d^{4}x~b^{a}\frac{\delta}{\delta b^{a}}-d_{1}\int d^{4}x~\bar{c}^{a}\frac{\delta}{\delta\bar{c}^{a}}\nonumber \\
 & + & \int d^{4}x\bigg[d_{1}A_{\mu}^{a}\frac{\delta}{\delta A_{\mu}^{a}}-d_{1}\Omega_{\mu}^{a}\frac{\delta}{\delta\Omega_{\mu}^{a}}+d_{2}L^{a}\frac{\delta}{\delta L^{a}}+d_{2}c^{a}\frac{\delta}{\delta c^{a}}+f^{ab}(\xi)\xi^{b}\frac{\delta}{\delta\xi^{a}}\nonumber \\
 & - & K^{a}\frac{\delta}{\delta K^{a}}\bigg(\frac{\partial f^{bc}}{\partial\xi^{a}}\xi^{c}+f^{ba}(\xi)\bigg)\bigg]-\frac{b_{1}}{2}\int d^{4}x\bigg[\bar{\varphi}^{aI}\frac{\delta}{\delta\bar{\varphi}^{aI}}+\varphi^{aI}\frac{\delta}{\delta\varphi^{aI}}+\bar{\omega}^{aI}\frac{\delta}{\delta\bar{\omega}^{aI}}+\omega^{aI}\frac{\delta}{\delta\omega^{aI}}\nonumber \\
 & + & M_{\mu}^{aI}\frac{\delta}{\delta M_{\mu}^{aI}}+V_{\mu}^{aI}\frac{\delta}{\delta V_{\mu}^{aI}}+N_{\mu}^{aI}\frac{\delta}{\delta N_{\mu}^{aI}}+U_{\mu}^{aI}\frac{\delta}{\delta U_{\mu}^{aI}}+2\tilde{J}\frac{\delta}{\delta\tilde{J}}+2H\frac{\delta}{\delta H}+2\tilde{G}\frac{\delta}{\delta\tilde{G}}\nonumber \\
 & + & 2P\frac{\delta}{\delta P}+2\varrho\frac{\partial}{\partial\varrho}\bigg]+\int d^{4}x\bigg[\bigg(\frac{d_{5}+b_{4}-b_{1}}{2}\bigg)\bar{\lambda}^{a\hat{I}}\frac{\delta}{\delta\bar{\lambda}_{\hat{I}}^{a}}+\bigg(\frac{d_{5}-b_{1}-b_{4}}{2}\bigg)\lambda^{a\hat{I}}\frac{\delta}{\delta\lambda_{\hat{I}}^{a}}\nonumber \\
 & + & \bigg(\frac{b_{1}+b_{4}+d_{5}}{2}\bigg)\bigg(\bar{\zeta}^{a\hat{I}}\frac{\delta}{\delta\bar{\zeta}_{\hat{I}}^{a}}+\bar{\Pi}^{i\alpha\hat{I}}\frac{\delta}{\delta\bar{\Pi}_{\alpha\hat{I}}^{i}}+\Lambda^{i\alpha\hat{I}}\frac{\delta}{\delta\Lambda_{\alpha\hat{I}}^{i}}+\bar{\Lambda}^{i\alpha\hat{I}}\frac{\delta}{\delta\bar{\Lambda}_{\alpha\hat{I}}^{i}}\bigg)\nonumber 
 \end{eqnarray}
 \begin{eqnarray}
 & + & \bigg(\frac{b_{1}+d_{5}-b_{4}}{2}\bigg)\bigg(\zeta^{a\hat{I}}\frac{\delta}{\delta\zeta_{\hat{I}}^{a}}+\Pi^{i\alpha\hat{I}}\frac{\delta}{\delta\Pi_{\alpha\hat{I}}^{i}}\bigg)-\bigg(\frac{b_{4}+d_{5}}{2}\bigg)\bigg(Z^{\hat{I}}\frac{\delta}{\delta Z_{\hat{I}}}+B^{\hat{I}}\frac{\delta}{\delta B_{\hat{I}}}\bigg)\nonumber \\
 & + & \bigg(\frac{b_{4}-d_{5}}{2}\bigg)\bigg(\bar{Z}^{ab\hat{I}}\frac{\delta}{\delta\bar{Z}_{\hat{I}}^{ab}}+\bar{B}^{ab\hat{I}}\frac{\delta}{\delta\bar{B}_{\hat{I}}^{ab}}\bigg)+b_{4}\int d^{4}x\,\bigg[\Theta^{i\alpha}\frac{\delta}{\delta\Theta_{\alpha}^{i}}+\bar{\Theta}^{i\alpha}\frac{\delta}{\delta\bar{\Theta}_{\alpha}^{i}}+\Phi^{i\alpha}\frac{\delta}{\delta\Phi_{\alpha}^{i}}+\bar{\Phi}^{i\alpha}\frac{\delta}{\delta\bar{\Phi}_{\alpha}^{i}}\bigg] \,.\label{w50}
\end{eqnarray}

\subsection{Stability of the tree level action action $\Sigma$ and the renormalization factors}

As said before, in order to complete the all order algebraic renormalization analysis of our model, we have to show that the
counterterm \eqref{w48-1-1-1} can be reabsorbed into the starting action \eqref{Sigma_full} by redefining the fields, sources and parameters. This procedure can be done in a quick way with the help of the parametric form
obtained in \eqref{w48-1-1-1}. According to the general setup of the algebraic renormalization \cite{Piguet:1995er},  we have to show that  
\begin{equation}
\Sigma[\Phi_{0}]=\Sigma[\Phi]+\epsilon\Sigma^{\mathrm{CT}}[\Phi]+O(\epsilon^2)\,,\label{w51}
\end{equation}
where $\epsilon$ is the expansion parameter and $\{\Phi\}$ stands for all fields, sources and parameters of the theory. As usual, the notation $\{\Phi_0\}$ refers to the bare or redefined quantities. To find out the explicit form of the redefined quantities $\{\Phi_0\}$ we make use of the parametric equation \eqref{w49}, from which we immediately get 
\begin{equation}
\Sigma[\Phi_{0}]=\Sigma[\Phi]+\epsilon\mathcal{R}\Sigma[\Phi]+O(\epsilon^2)\,,\label{w52}
\end{equation}
where the redefined  quantities   $\{\Phi_0\}$ are given by 
\begin{equation}
\Phi_{0}=(1+\epsilon\mathcal{R})\Phi\,.\label{w53}
\end{equation}
From eq.\eqref{w53} we can read off the whole set of redefinitions for all fields, parameters and sources, as given below: 
\begin{eqnarray}
A_{0} & = & Z_{A}^{1/2}A\,,\qquad b_{0}=Z_{b}^{1/2}b\,,\qquad c_{0}=Z_{c}^{1/2}c\,,\qquad\bar{c}_{0}=Z_{\bar{c}}^{1/2}\bar{c}\,,\nonumber \\
\xi_{0}^{a} & = & Z_{\xi}^{ab}(\xi)\xi^{b}\,,\qquad\tau_{0}=Z_{\tau}^{1/2}\tau\,,\qquad\eta_{0}=Z_{\eta}^{1/2}\eta\,,\qquad\bar{\eta}_{0}=Z_{\bar{\eta}}^{1/2}\bar{\eta}\,,\nonumber \\
\bar{\psi}_{0} & = & Z_{\bar{\psi}}^{1/2}\bar{\psi}\,,\qquad\psi_{0}=Z_{\psi}^{1/2}\psi\,,\qquad\bar{\Theta}_{0}=Z_{\bar{\Theta}}\bar{\Theta}\,,\qquad\Theta_{0}=Z_{\Theta}\Theta\,,\nonumber \\
\bar{\varphi}_{0} & = & Z_{\bar{\varphi}}^{1/2}\bar{\varphi}\,,\qquad\varphi_{0}=Z_{\varphi}^{1/2}\varphi\,,\qquad\bar{\omega}_{0}=Z_{\bar{\omega}}^{1/2}\bar{\omega}\,,\qquad\omega_{0}=Z_{\omega}^{1/2}\omega\,,\nonumber \\
\Omega_{0} & = & Z_{\Omega}\Omega\,,\qquad L_{0}=Z_{L}L\,,\qquad K_{0}^{a}=Z_{K}^{ab}(\xi)K^{b}\,,\qquad\mathcal{J}_{0}=Z_{\mathcal{J}}\mathcal{J}\,,\nonumber \\
\alpha_{0} & = & Z_{\alpha}\alpha\,,\qquad M_{0}=Z_{M}M\,,\qquad V_{0}=Z_{V}V\,,\qquad H_{0}=Z_{H}H\nonumber \\
N_{0} & = & Z_{N}N\,,\qquad U_{0}=Z_{U}U\,,\qquad\Xi_{0}=Z_{\Xi}\,\Xi\,,\qquad X_{0}=Z_{X}X\,,\nonumber \\
Y_{0} & = & Z_{Y}Y\,,\qquad\bar{X}_{0}=Z_{\bar{X}}\bar{X}\,,\qquad\bar{Y}_{0}=Z_{\bar{Y}}\bar{Y}\,,\qquad\Gamma_{0}=Z_{\Gamma}\Gamma\,,\nonumber \\
\bar{\lambda}_{0} & = & Z_{\bar{\lambda}}^{1/2}\bar{\lambda}\,,\qquad\lambda_{0}=Z_{\lambda}^{1/2}\lambda\,,\qquad\bar{\zeta}_{0}=Z_{\bar{\zeta}}^{1/2}\bar{\zeta}\,,\qquad\zeta_{0}=Z_{\zeta}^{1/2}\zeta\,,\nonumber \\
\tilde{G}_{0} & = & Z_{\tilde{G}}\tilde{G}\,,\qquad P_{0}=Z_{P}P\,,\qquad{\varrho}_{0}=Z_{\varrho}\varrho\,,\qquad J_{\psi0}=Z_{J_{\psi}}J_{\psi}\,,\nonumber \\
\Pi_{0} & = & Z_{\Pi}\Pi\,,\qquad\bar{\Pi}_{0}=Z_{\bar{\Pi}}\bar{\Pi}\,,\qquad\bar{\Lambda}_{0}=Z_{\bar{\Lambda}}\bar{\Lambda}\,,\qquad\Lambda_{0}=Z_{\Lambda}\Lambda\,.,\nonumber \\
Z_{0} & = & Z_{Z}Z\,,\qquad\bar{Z}_{0}=Z_{\bar{Z}}\bar{Z}\,,\qquad\bar{B}_{0}=Z_{\bar{B}}\bar{B}\,,\qquad B_{0}=Z_{B}B\,,\nonumber \\
\Phi_{0} & = & Z_{\Phi}\Phi\,,\qquad\bar{\Phi}_{0}=Z_{\bar{\Phi}}\bar{\Phi}\,,\qquad\bar{\varUpsilon}_{0}=Z_{\bar{\varUpsilon}}\bar{\varUpsilon}\,,\qquad\varUpsilon_{0}=Z_{\varUpsilon}\varUpsilon\,,\nonumber \\
\tilde{J}_{0} & = & Z_{\tilde{J}}\tilde{J}\,,\qquad g_{0}=Z_{g}g\,,\label{w54}
\end{eqnarray}
and
\begin{equation}
    \left(
    \begin{tabular}{c}
    $J_{0}$\cr
    $J^{2}_{\psi,0}$
    \end{tabular}
    \right)
    =
    \left(
    \begin{tabular}{c|c}
    $Z_{J}$&$Z_{J,J_{\psi}}$\cr
    \hline
    0&$Z^{2}_{J_{\psi}}$
    \end{tabular}
    \right)
    \left(
    \begin{tabular}{c}
    $J$\cr
    $J^{2}_{\psi}$
    \end{tabular}
    \right)\,,\qquad
    \left(
    \begin{tabular}{c}
    $\kappa_{1,0}$\cr
    $\kappa_{2,0}$\cr
    $\kappa_{3,0}$
    \end{tabular}
    \right)=
        \left(
    \begin{tabular}{c|c|c}
    $Z_{\kappa_{1}}$&$0$&$0$\cr
    \hline
    $Z_{\kappa_{2}\kappa_{1}}$&$Z_{\kappa_{2}}$&$0$\cr
    \hline
    $0$&$Z_{\kappa_{3}\kappa_{2}}$&$Z_{\kappa_{3}}$
    \end{tabular}
    \right)
    \left(
    \begin{tabular}{c}
    $\kappa_{1}$\cr
    $\kappa_{2}$\cr
    $\kappa_{3}$
    \end{tabular}
    \right)\,,
\end{equation}
with 
\begin{eqnarray}
&\displaystyle
Z_{J}=1+\epsilon(b_{2}+b_{2}^{\prime})\,,\qquad
Z_{J,J_{\psi}}=\epsilon\,b_{2}^{\prime}\,,\qquad
Z_{J_{\psi}}=1+\epsilon\,a_{1}\,,\nonumber\\
&
Z_{\kappa_{1}}=1+\epsilon\,(b_{1}+2b_{2})\,,\qquad
Z_{\kappa_{2}}=1+\epsilon\,(a_{7}+2a_1+b_{2})\,,\qquad
Z_{\kappa_{3}}=1+\epsilon(a_{8}+4a_{1})
&
   \label{1564}
\end{eqnarray}
and
\begin{eqnarray}
Z_{A}^{1/2} & = & 1+\epsilon d_{1}\,,\qquad Z_{c}^{1/2}=1+\epsilon d_{2}\,,\qquad Z_{g}=1-\epsilon\frac{a_{0}}{2}\,,\qquad Z_{\tau}^{1/2}=1+\epsilon b_{1}\,,\nonumber \\
Z_{\bar{\Theta}}^{1/2} & = & Z_{\Theta}^{1/2}=Z_{\Phi}=Z_{\bar{\Phi}}=1+\epsilon b_{4}\,,\qquad Z_{\varUpsilon}=Z_{\bar{\varUpsilon}}=Z_{\psi}^{-\frac{1}{2}}=Z_{\bar{\psi}}^{-\frac{1}{2}}=1+\epsilon d_{3}\,,\nonumber \\ Z_{\xi}^{ab}(\xi)&=&\delta^{ab}+\epsilon f^{ab}(\xi)\,,\qquad Z_{K}^{ab}(\xi)=\delta^{ab}-\epsilon\left(f^{ba}(\xi)+\frac{\partial f^{bc}}{\partial\xi^{a}}\xi^{c}\right)\,.\label{w55}
\end{eqnarray}
For the other fields, sources and parameters, the following relations
hold 
\begin{eqnarray}
Z_{A}^{1/2} & = & Z_{\Omega}^{-1}=Z_{\bar{c}}^{-1/2}=Z_{b}^{-1/2}=Z_{\alpha}^{1/2}\,,\nonumber \\
Z_{\tau}^{1/2} & = & Z_{\bar{\eta}}=Z_{\eta}=Z_{\Xi}^{2}=Z_{\Gamma}^{2}=Z_{\mathcal{J}}\,,\nonumber \\
Z_{c}^{1/2} & = & Z_{L}\,,\qquad Z_{X}=Z_{Y}=Z_{\bar{X}}=Z_{\bar{Y}}=1\,.\label{w56}
\end{eqnarray}
Looking at  the renormalization factors of  fields and sources introduced in order to implement the Gribov horizon related to the gauge-invariant composite bosonic field sector, we have 
\begin{eqnarray}
Z_{\tau}^{-1/4}=Z_{\bar{\varphi}}^{1/2}=Z_{\varphi}^{1/2}=Z_{\bar{\omega}}^{1/2}=Z_{\omega}^{1/2}=Z_{M}=Z_{V}=Z_{N}=Z_{U}=Z_{\tilde{J}}^{1/2}=Z_{H}^{1/2}=Z_{\tilde{G}}^{1/2}=Z_{P}^{1/2}=Z^{1/2}_{\varrho}\,,\label{w57}
\end{eqnarray}
while the renormalization factors of fields and sources associated to the horizon-like term of the fermionic gauge-invariant composite fields are given by
\begin{eqnarray}
Z_{\lambda}^{1/2} & = & 1+\epsilon\bigg(\frac{d_{5}-b_{1}-b_{4}}{2}\bigg)\,,\nonumber \\
Z_{\bar{\lambda}}^{1/2} & = & 1+\epsilon\bigg(\frac{d_{5}+b_{4}-b_{1}}{2}\bigg)\,,\nonumber \\
Z_{\bar{\zeta}} & = & Z_{\bar{\Pi}}=Z_{\bar{\Lambda}}=Z_{\Lambda}=1+\epsilon\bigg(\frac{b_{1}+b_{4}+d_{5}}{2}\bigg)\,,\nonumber \\
Z_{\zeta} & = & Z_{\Pi}=1+\epsilon\bigg(\frac{b_{1}+d_{5}-b_{4}}{2}\bigg)\,,\nonumber \\
Z_{Z} & = & Z_{B}=1-\epsilon\bigg(\frac{b_{4}+d_{5}}{2}\bigg)\,,\nonumber \\
Z_{\bar{Z}} & = & Z_{\bar{B}}=1+\epsilon\bigg(\frac{b_{4}-d_{5}}{2}\bigg)\,,\label{w60}
\end{eqnarray}
In particular, from  the physical values of the sources $J$ and $J_{\psi}$, eq.\eqref{J_phys}, we obtain the renormalization of the corresponding mass parameters, i.e. 
\begin{eqnarray}
m^{2}_{0}&=&m^{2}+\epsilon\,(b_{2}\,m^{2}+b_{2}^{\prime}\,m^{2}_{\psi})\,,\nonumber\\
m_{\psi,0}&=&m_{\psi}+\epsilon\,a_{1}\,m_{\psi}\,.
\label{178}
\end{eqnarray}
As one can observe, there is a mixing between the mass parameters, indicating    that even if the gauge-invariant operator $A^{h}_\mu A^{h}_\mu $ would have not been introduced from the beginning, {\it i.e.} $(m^{2}=0)$, it shows up through quantum corrections if the theory contains a fermionic mass parameter like $m_{\psi}$. Finally, we notice the  relationship 
\begin{equation}
Z_{\mathcal{J}}Z_{\varphi}^{1/2}Z_{\bar{\varphi}}^{1/2}=1\,,\label{w59}
\end{equation}
telling us that $Z_{\mathcal{J}}$ can be obtained from the knowledge of $(Z_{\varphi},Z_{\bar{\varphi}})$. \\\\Therefore, 
after performing  a proper redefinition of the fields, sources and parameters, eqs.\eqref{1564}, \eqref{w55}, \eqref{w56}, \eqref{w57}, \eqref{w60}, \eqref{178}, \eqref{w59},  the most general local invariant counterterm $\Sigma^{\mathrm{CT}}$ compatible with all Ward identities can be fully reabsorbed in the tree level 
 action \eqref{Sigma_full}, providing thus an all order purely algebraic proof of the renormalizability of $\Sigma$. \\\\Let us end this section by observing that, due to the dimensionless character of the Stueckelberg field $\xi^a$, the renormalization factors $(Z_{\xi}^{ab}(\xi), Z_{K}^{ab}(\xi))$ are nonlinear in  $\xi^{a}$, a well known feature of dimensionless fields, see  \cite{Fiorentini:2016rwx,Capri:2019drm, Capri:2018gpu, Capri:2018riz}. It is  worth to point out that the renormalization factors related to the massive Gribov parameters $\gamma^{2}$ and $\sigma^{3}$ are not independent quantities of the model, {\it i.e.} they are expressed in terms of other renormalization factors. Moreover, the renormalization factor of the source $J$, eq.\eqref{1564}, gives rise to a mixing between the coefficients related to the bosonic and fermionic gauge-invariant composite fields encoded in the mixing matrix  for $Z_{J}$. Finally, we observe that equation \eqref{w59} implies that the three vertex  $(A^h\varphi\bar{\varphi})$ is not renormalized,   as already noticed  in \cite{Capri:2016aif,Capri:2017bfd}.

\section{Conclusion }
\label{conclu}

In this work we have studied the BRST symmetry content and the all orders algebraic renormalization of an effective fermionic model  model in which the inverse of the Faddeev-Popov operator, $[\mathcal{M}^{ab}(A^{h})]^{-1}=[-\partial_{\mu}D_{\mu}^{ab}(A^{h})]^{-1}$, where $A^h_\mu$ is the dressed gauge invariant field of eq.\eqref{intro13-1}, has been coupled in a gauge invariant fashion to the gauge invariant spinor quantities $(\bar{\psi}^h,\psi^h)$, eqs.\eqref{local_psih},\eqref{psih_expansion}, giving rise to the effective horizon matter term $S_{\sigma}$ of eq\eqref{horizon}. \\\\After a suitable localization procedure, the gauge invariant nature of $(A^h_\mu, {\bar \psi}^h,\psi^h)$ has allowed us to quantize the theory in the class of the linear covariant gauges, as observed from the gauge fixing term of the local action $S^{local}$, eq.\eqref{S_local}. \\\\The proof of the all orders algebraic renormalization of $S^{local}$ has been achieved by embedding it into a  more general action $\Sigma$, eq.\eqref{Sigma_full}, which fulfills a very rich set of Ward identities.  The action $S^{local}$ is in fact re-obtained from the generalized action $\Sigma$  when the external sources attain  the  values
\eqref{phys_values_gamma},\eqref{phys_values_sigma},\eqref{J_phys}, while setting the remaining sources and the Grassmannian parameter $\chi$ to zero. \\\\As already mentioned in the first Sections of the work, the main reason behind the construction of the renormalizable local action $S^{local}$ relies on the resulting fermion propagator, whose tree level expression turns out to be 
\begin{equation}
    \langle\bar{\psi}^{i}_{\alpha}(p)\psi^{j}_{\beta}(-p)\rangle =\frac{-ip_{\mu}(\gamma_{\mu})_{\alpha\beta}+\mathcal{A}(p^{2})\,\delta_{\alpha\beta}}{p^{2}+\mathcal{A}^{2}(p^{2})}\,\,\delta^{ij}\,, \label{two-point-quark}
\end{equation}
where $\mathcal{A}(p^{2})$ is the so-called mass function,   given by
\begin{equation}
    \mathcal{A}(p^{2})=m_{\psi}+g^{2}\,\frac{N^{2}-1}{2N}\,\frac{\sigma^{3}}{p^{2}+w^{2}}\,. \label{mass-function}
\end{equation}
In particular, in the Landau gauge, corresponding to set the gauge parameter $\alpha$ to $0$,  the mass function $\mathcal{A}(p^{2}) $ turns out to be in rather good agreement with the numerical lattice simulations, see for instance  \cite{Parappilly:2005ei,Oliveira:2018lln} and refs. therein.\\\\ As such, the action $S^{local}$ can be seen as the first step towards the non-perturbative investigation of the possible dependence of the quark propagator from the gauge parameter $\alpha$ in the class of the linear covariant gauges, a topic under current intensive studies. \\\\Further relevant  topics which could be exploited by means of $S^{local}$ are the study of the gauge dependence of the quark-gluon vertex in the linear covariant gauges as well as the related applications within the framework of the  Bethe-Salpeter equations for the  bound states. \\\\ Let us end by mentioning the issue of the possible extension of the whole setup presented here to the case of finite temperature, a topic which we aim at investigating sometime in the near future. Such an extension would enable us to investigate aspects related to the chiral symmetry restoration as well as to the confinement-deconfinement transition and respective critical temperatures.

\section*{Acknowledgments}
The National Council for Scientific and Technological Development (CNPq/MCTI) and the SR2-UERJ are gratefully acknowledged for financial support. R. C. Terin is supported through the Junior Postdoctoral fellowship program (PDJ/CNPq), Finance Code - 151397/2020-1; S. P. Sorella is a level PQ-1 researcher
under the program Produtividade em Pesquisa (CNPq), Finance Code - 301030/2019-7; M. A. L. Capri is a level PQ-2 researcher under
the program Produtividade em Pesquisa (CNPq), Finance Code - 313068/2020-8.



\end{document}